\documentclass{aa}

\usepackage[utf8]{inputenx}
\usepackage{graphicx}
\usepackage{txfonts}

\usepackage{url}
\usepackage{grffile}

\newcommand{\esothanks}{Based on observations made with ESO Telescopes
  at the La Silla Paranal Observatory under programme ID 094.A-0605,
  programme ID 095.A-0570, and programme ID 097.A-0831.}

\begin{document}

\title{Deciphering Lyman $\alpha$ blob 1 with deep MUSE observations\thanks{\esothanks}}

\author{
  Edmund~Christian~Herenz\inst{\ref{inst2},\ref{inst1}} \and
  Matthew~Hayes\inst{\ref{inst1}} \and
  Claudia~Scarlata\inst{\ref{inst3}}
}

\institute{
  European Southern Observatory,
  Av. Alonso de C\'ordova 3107,
  763 0355 Vitacura,
  Santiago, Chile \\ e-mail: \url{eherenz@eso.org}
  \label{inst2}
  \and
  Department of Astronomy, Stockholm University,
  AlbaNova University Centre, SE-106 91,
  Stockholm, Sweden
  \label{inst1}
  \and
  Minnesota Institute for Astrophysics,
  School of Physics and Astronomy,
  University of Minnesota,
  316 Church Str. SE,
  Minneapolis,
  MN 55455, USA
  \label{inst3}
}

\abstract{
  Lyman $\alpha$ blobs (LABs) are large-scale radio-quiet Lyman $\alpha$ (Ly$\alpha$)
  nebula at high-$z$ that occur predominantly in overdense proto-cluster regions.
  Especially the prototypical SSA22a-LAB1 at $z=3.1$ has become an observational reference
  for LABs across the electromagnetic spectrum.  } {%
  We want to understand the powering mechanisms that drive the LAB to gain empirical
  insights into galaxy formation processes within a rare dense environment at
  high-$z$. Thus we need to infer the distribution, the dynamics, and the ionisation
  state of  LAB\,1's Ly$\alpha$ emitting gas.}
  %
{ LAB\,1 was observed for 17.2\,h with the VLT/MUSE integral-field spectrograph.  We
  produced optimally extracted narrow band images in Ly$\alpha$ $\lambda1216$,
  \ion{He}{ii} $\lambda1640$, and we tried to detect \ion{C}{iv} $\lambda1549$ emission.
  By utilising a moment based analysis we mapped the kinematics and the line profile
  characteristics of the blob.  We also linked the inferences from the line profile
  analysis to previous results from imaging polarimetry. } {%
  We map Ly$\alpha$ emission from the blob down to surface-brightness limits of
  $\approx 6 \times 10^{-19}$erg\,s$^{-1}$cm$^{-2}$arcsec$^{-2}$.  At this depth we reveal
  a bridge between LAB\,1 and its northern neighbour LAB\,8, as well as a shell-like
  filament towards the south of LAB\,1.  Complexity and morphology of the Ly$\alpha$
  profile vary strongly throughout the blob.  Despite the complexity, we find a coherent
  large scale east-west $\sim$1000\,km\,s$^{-1}$ velocity gradient that is aligned
  perpendicular to the major axis of the blob. Moreover, we observe a negative correlation
  of Ly$\alpha$ polarisation fraction with Ly$\alpha$ line width and a positive
  correlation with absolute line-of-sight velocity. Finally, we reveal \ion{He}{ii}
  emission in three distinct regions within the blob, but we can only provide upper limits
  for \ion{C}{iv}.  }{
  Various gas excitation mechanisms are at play in LAB\,1: Ionising radiation and feedback
  effects dominate near the embedded galaxies, while Ly$\alpha$ scattering is contributing
  at larger distances.  However, \ion{He}{ii}/Ly$\alpha$ ratios combined with upper limits
  on \ion{C}{iv}/Ly$\alpha$ can not discriminate between AGN ionisation and feedback
  driven shocks.  The alignment of the angular momentum vector parallel to the
  morphological principal axis appears odds with the predicted norm for high-mass halos,
  but likely reflects that LAB\,1 resides at a node of multiple intersecting filaments of
  the cosmic web. LAB\,1 can thus be thought of as a progenitor of present day massive
  elliptical within a galaxy cluster.  }


\maketitle

\section{Introduction}
\label{sec:intr}

Lyman $\alpha$ (Ly$\alpha$) blobs (LABs) are very luminous
($L_\mathrm{Ly\alpha} \gtrsim 10^{43.5}$\,erg\,s$^{-1}$) and very extended
($\gtrsim 10^2$\,kpc in projection) Ly$\alpha$ emitting nebulae.  They were unexpectedly
revealed in narrow-band imaging campaigns targeting Lyman $\alpha$ emitting galaxies
(LAEs) at $z\gtrsim 3$ \citep{Francis1996,Steidel2000}.  LABs have now been found
in numerous, sometimes LAB-dedicated, high-$z$ galaxy surveys
\citep[e.g.][]{Matsuda2004,Nilsson2006,Prescott2012,Prescott2013}.  Their presence is
confirmed from $z\sim1$ \citep{Barger2012} up to $z\sim7$
\citep[][]{Ouchi2009,Sobral2015,Shibuya2018a,Zhang2019}.  Moreover, a very rare class of
extended $z \sim 0.3$ [\ion{O}{iii}] nebulae have been proposed to share similarities with
high-redshift LABs \citep{Schirmer2016}.

The distinctive observational feature of LABs with respect to similarly extended and
luminous high-$z$ Ly$\alpha$ nebulae around radio-galaxies
\citep[e.g.][]{Morais2017,Vernet2017,Marques-Chaves2019}, radio-loud quasars
\citep[e.g.][]{Smith2009,Roche2014}, or radio-quiet quasars
\citep[e.g.][]{Christensen2006,Borisova2016,Ginolfi2018,Husemann2018,Arrigoni-Battaia2019,Drake2019,Farina2019,Travascio2020}
is that the primary powering source driving their Ly$\alpha$ emission is usually not
detected or not obvious from the rest-frame UV and rest-frame optical discovery data
\citep[see also review by][and references therein]{Cantalupo2017}.  However, the defining
physical characteristic of LABs is their preferential occurrence within overdense high-$z$
proto-cluster regions.  In fact, the first LABs were found in narrow-band searches
targeting known or presumed high-density structures \citep{Francis1996,Steidel2000}.
Following these initial discoveries, other narrow-band surveys targeting redshift
overdensities were able to replicate the success in unveiling LABs
\citep[e.g.][]{Palunas2004,Erb2011,Mawatari2012,Cai2017,Kikuta2019}.  Conversely, LABs
found in blind searches could be linked to over-densities
\citep{Prescott2008,Yang2009,Yang2010,Badescu2017}.  Given that their preferred habitats
are proto-cluster regions and their sizes are enormous, it appears natural to suspect LABs
as the progenitors of extremely massive, if not the most-massive galaxies in present day
cluster environments \citep[see also review by][]{Overzier2016}.

The required amount of hydrogen ionising photons to drive the observed Ly$\alpha$ output
of the blobs via recombinations is
$\dot{Q}(h\nu \geq 13.6\,\mathrm{eV}) \gtrsim 10^{55}$\,s$^{-1}$ for a
$L_\mathrm{Ly\alpha} \gtrsim 10^{44}$\,erg\,s$^{-1}$ LAB in a standard case-B
recombination scenario.  This $\dot{Q}$ would correspond to star-formation rates
$\gtrsim 100$M$_\odot$yr$^{-1}$ and absolute UV magnitudes $M_\mathrm{UV} < -23$ using
canonical conversion factors \citep[e.g.][]{Kennicutt1998}.  The absence of such bright UV
galaxies in the vicinity of the nebulae may indicate that the powering sources are heavily
dust-obscured along the line of sight.  Moreover, it might also hint at an additional
source of Ly$\alpha$ photons in LABs: collisional excitations of neutral hydrogen by free
electrons, a process which also cools the heated electron gas.  As a coolant, Ly$\alpha$
is most effective for gas temperatures around $10^4$\,K.  In the case of LABs, potential
heating sources could be star-burst driven super-winds from the heavily obscured central
galaxies \citep[e.g.][]{Taniguchi2000,Mori2004} or the gravitational potential of the halo
hosting the blob \citep[``gravitational cooling'', see e.g.][]{Haiman2000,Rosdahl2012}.

The gravitational cooling mechanism was initially deemed the dominant powering source for
driving Ly$\alpha$ emission from LABs \citep[][]{Haiman2000,Dijkstra2009}.  This idea is
especially intriguing, as theoretical models predict that gas accretion onto galaxies
forms dense cold flow filaments \citep[e.g.,
][]{Keres2005,Dekel2006,Brooks2009,Stewart2017}.  Despite its theoretical importance,
empirical evidence for this processes in high-$z$ galaxies remains circumstantial
\citep[e.g.][]{Rauch2016}.  The filamentary Ly$\alpha$ morphology of LABs as well
as the alignment of their major axes with the surrounding large scale structure are
regarded as observational support for the gravitational cooling scenario
\citep{Erb2011,Matsuda2011}.

The polarisation of the observed Ly$\alpha$ emission from the blob could potentially
distinguish between the central engine hypothesis or the in-situ powering by gravitational
cooling \citep{Dijkstra2008,Eide2018}.  In the former case the polarisation fraction is
expected to increase with distance from the embedded sources.  And indeed, for Lyman
$\alpha$ blob 1, the object examined in detail in this study, such a polarisation
characteristic was detected \citep[][]{Hayes2011,Beck2016}.  These observations
were thought to rule out the gravitational cooling scenario, but
\cite{Trebitsch2016} showed that also gravitational cooling may be
  responsible for the observed polarisation pattern.

Regardless, follow-up campaigns with X-ray, sub-mm, IR, and radio-facilities
revealed that a significant fraction of LABs indeed harbour highly obscured
star-bursts with star-formation rates $\sim 10^2 - 10^3$\,M$_\odot$yr$^{-1}$ or AGNs
\citep[e.g.][]{Yang2011,Ao2015,Ao2017}.  Thus, mechanical heating and/or ionising
radiation from these buried systems are definitely contributing and possibly dominating
the energy budget that powers Ly$\alpha$ in LABs.  In such a scenario the
filamentary morphology can still be reconciled with the cooling flow
interpretation, as those flows are expected to fuel star-formation and AGN
activity in the first place.  But rather than being lit-up in Ly$\alpha$ by gravitational
cooling, these filamentary flows could simply be illuminated from the systems that they
are feeding \citep{Prescott2015}.  However, a potential counterargument to this scenario
is that the gas in the cold flows is self-shielded from ionising radiation given the
expected typical densities ($n_\mathrm{H} > 1$cm$^{-3}$; \citealt{Dijkstra2009}).
Moreover, it appears counter-intuitive that the heavily obscured embedded sources have
high escape fractions of ionising photons into large enough solid angles to power the
blobs.  As yet, there is still no consensus on the importance of the possible
Ly$\alpha$ powering mechanisms in LABs.

Integral-field spectroscopic \citep[IFS; ][]{Bacon2017a} observations are an especially
promising observational line of attack for such studies.  Disentangling the different
Ly$\alpha$ powering mechanisms at work in LABs is warranted, as the relevant Ly$\alpha$
emission processes are linked to the physical processes that regulate the build-up of
stellar mass and growth of super-massive black holes in the most-massive galaxies of the
present day universe.  The modern integral-field spectrographs on 10\,m class telescopes,
i.e. the ``Keck Cosmic Web Imager'' \citep[][]{Morrissey2018} at the Keck II telescope and
the ``Multi Unit Spectroscopic Explorer'' \citep[MUSE,][]{Bacon2010,Bacon2014} at ESO's
Very Large Telescope UT4, are ideally suited to cover the projected sizes of LABs.  Analyses of combined spectral and spatial information from IFS can provide a detailed view
of the Ly$\alpha$ morpho-kinematics.  Specifically, cold-mode accretion filaments are
expected to leave imprints on the velocity fields compared to simple Keplarian motions
\citep[]{Arrigoni-Battaia2018,Martin2019}.  Moreover, gas which is not affected by
star-formation or AGN driven feedback is expected to be kinematically more quiescent than
feedback heated gas.  Thus, these processes can potentially be distinguished by spatially
mapping the observed Ly$\alpha$ line width.

A difficultly in interpreting line-of-sight velocities and line-widths from Ly$\alpha$
emission, is that resonant scattering diffuses the intrinsic Ly$\alpha$ radiation field in
real and frequency-space \citep[see review by][]{Dijkstra2017}.  The spatial diffusion can
be envisioned as a projected smoothing processes \citep[][]{Bridge2018}, that can enhance
the apparent size of the Ly$\alpha$ blobs by reducing the steepness of their surface
brightness profiles \citep{Zheng2011,Gronke2017c}.  Moreover, it can also ``wash-out''
cold-accretion features of small angular size \citep[][]{Smith2019}.  The diffusion in
frequency space, which is dependent on the kinematics and column densities of the
scattering medium, can lead to significant modulations of the spectral profile
\citep[e.g.,][]{Laursen2009}.  Additionally, the transmission of the Ly$\alpha$ photons
through the intergalactic medium will also modify the line profile
\citep[e.g.,][]{Laursen2011}.  However, the observed low-velocity shifts between
  Ly$\alpha$ and optically thin emission lines for galaxies within LABs appear
at odds with the expectations from Ly$\alpha$ radiative transfer theory
\citep[e.g.][]{McLinden2013}.  This might indicate that Ly$\alpha$ scattering does not
significantly modulate the observable velocity field of LABs.  Thus, it is possible to
obtain a measure the angular momentum of the gas in the early formation stage of a massive
halo, which directly relates to the action of tidal torques from the surrounding large
scale structure and the cosmic web \citep[e.g.][]{Forero-Romero2014,Lee2018}.

Further insights into the thermodynamic properties of the emitting gas of $z\sim3$
blobs can be gained from ground based IFS data due to the potential detectability of other
rest-frame UV emission lines.  Especially \ion{He}{ii} $\lambda 1640$ and \ion{C}{iv}
$\lambda 1550$ emission lines have been used as diagnostics for LABs
\citep{Prescott2009,Scarlata2009b,Arrigoni-Battaia2015}.  Both lines act as gas-coolants,
but for a higher temperature ($T \approx 10^5$\,K) gas phase compared to Ly$\alpha$ which
cools the $10^4$K phase \citep{Yang2006}.  Heating sources driving this phase could be
feedback effects from the embedded galaxies \citep{Mori2004,Cabot2016} or the
gravitational potential of the halo hosting the blob \citep{Yang2006,Dijkstra2009}.  Both
lines can also be powered via photo-ionisation, but require the abundance of
higher energy photons to produce the recombining species.  Such hard ionising radiation is
only expected in the vicinity of extreme low metallicity stellar populations
\citep[e.g. ][]{Schaerer2013} or in the surroundings of an AGN
\citep[e.g.][]{Humphrey2019}.  Analysing relative line strengths and comparing the spatial
distribution of the \ion{He}{ii} and \ion{C}{iv} emitting gas to the positions of
potential ionising sources within blob may help to distinguish between photo-ionisation
and cooling-radiation scenarios.

With the aim to decipher the physical processes at work in LABs we present the deepest IFS
observations of a giant $z\sim3$ LAB obtained so far.  Our target is the prototypical
giant \citeauthor{Steidel2000} LAB -- SSA-22a Lyman $\alpha$ blob 1 (LAB1).  LAB\,1 lives
in one of the most overdense regions known at $z=3.1$.  This region, referred to as the
SSA 22 proto-cluster, shows a significant density peak of Lyman break galaxies
\citep{Steidel1998,Steidel2003,Saez2015}, Lyman $\alpha$ emitting galaxies
\citep[LAEs,][]{Hayashino2004,Yamada2012}, sub-mm galaxies \citep{Tamura2009}, AGNs
\citep{Lehmer2009b,Alexander2016}, and also LABs \citep{Matsuda2004,Matsuda2011}.
Interestingly, with an estimated cluster mass of 2-4$\times10^{14}$M$_\odot$, the SSA\,22
proto-cluster may actually be a unique structure within the horizon \citep{Kubo2015}.

Since its discovery LAB\,1 became the target of numerous follow-up observations and is
thus the most well studied LAB to-date.  We will provide an overview of those results in
the following Sect.~\ref{sec:known-sources} before describing our new 17.2\,h MUSE
observations in Sect.~\ref{sec:obs}.  In Sect.~\ref{sec:data-reduction} we
detail how we reduced the MUSE data.  We present our analysis and results in
Sect.~\ref{sec:ana}, and in Sect.~\ref{sec:disc} we discuss the interpretations of our
findings.  Lastly, we summarise and present our conclusions in
Sect.~\ref{sec:summary-conclusions}.

Throughout the paper we assume a canonical 737-cosmology, i.e.  $\Omega_\Lambda = 0.7$,
$\Omega_M = 0.3$, and H$_0=70$\,km\,s$^{-1}$Mpc$^{-1}$.  Adopted reference line
wavelengths stem from the atomic line list compiled by \cite{van-Hoof2018}, all
wavelengths $< 2000$\,\AA{} refer to vacuum wavelengths, and for conversions between air-
and vacuum wavelengths we follow the prescriptions adopted in the Vienna Atomic Line
Database \citep{Ryabchikova2015}.

\section{Summary of previous results on LAB\,1}
\label{sec:known-sources}

\begin{table*}
  \caption{Overview of known sources within the Ly$\alpha$ blob.} 
  \label{tab:sources}
  \centering
  \begin{tabular}{ccccc}
    \hline \hline 
    Name &
    RA &
    Dec &
    $z$ &
    Refs. \\ 
    \hline \noalign{\smallskip}
    \object{SSA22a-C11} & 22$^\mathrm{h}$17$^\mathrm{m}$25.70$^\mathrm{s}$ &
    +00$^\circ$12\arcmin{}34.4\arcsec{} &
    3.0999$\pm$0.0004
    & (1),(2)
    \\
    \object{SSA22a-C15} &
    22$^\mathrm{h}$17$^\mathrm{m}$26.15$^\mathrm{s}$ &
    +00$^\circ$12\arcmin{}54.7\arcsec{} &
    3.0986$\pm$0.0003 &
    (1),(2) \\
    \object{LAB1-ALMA1} &
    22$^\mathrm{h}$17$^\mathrm{m}$25.94$^\mathrm{s}$ &
    +00$^\circ$12\arcmin{}36.6\arcsec{} &
    $\dots$ &
    (3) \\
    \object{LAB1-ALMA2} &
    22$^\mathrm{h}$17$^\mathrm{m}$26.01$^\mathrm{s}$ &
    +00$^\circ$12\arcmin{}36.4\arcsec{} &
    $\dots$ &
    (3) \\
    \object{LAB1-ALMA3} &
    22$^\mathrm{h}$17$^\mathrm{m}$26.11$^\mathrm{s}$ &
    +00$^\circ$12\arcmin{}36.4\arcsec{} &
    3.0993$\pm$0.0004 &
    (3) \\
    \object{VLA-LAB1a} & 22$^\mathrm{h}$17$^\mathrm{m}$25.95$^\mathrm{s}$ &
    +00$^\circ$12\arcmin{}35.3\arcsec{} &
    $\dots$ &
    (4) \\
    \object{LAB01-K15b} & 22$^\mathrm{h}$17$^\mathrm{m}$25.70$^\mathrm{s}$ &
    +00$^\circ$12\arcmin{}38.7\arcsec{} &
    3.1007$\pm$0.0002 &
    (5) \\
    S1 &
    22$^\mathrm{h}$17$^\mathrm{m}$26.08$^\mathrm{s}$ &
    +00$^\circ$12\arcmin{}34.2\arcsec{} &
    3.0968 &
    (5),(6) \\
    \hline \hline
  \end{tabular}
  \tablebib{ (1)~\citet{Steidel2003}; (2)~\citet{McLinden2013}; (3)~\citet{Umehata2017};
    (4)~\citet{Ao2017}; (5)~\citet{Kubo2016}; (6)~\citet{Geach2016}}
  \tablefoot{Coordinates from \cite{Steidel2003} have been adjusted to the here used 2MASS
    astrometric reference frame.}
\end{table*}

\begin{figure}
  \centering
  \includegraphics[width=0.5\textwidth,trim=10 0 20 0,clip=True]{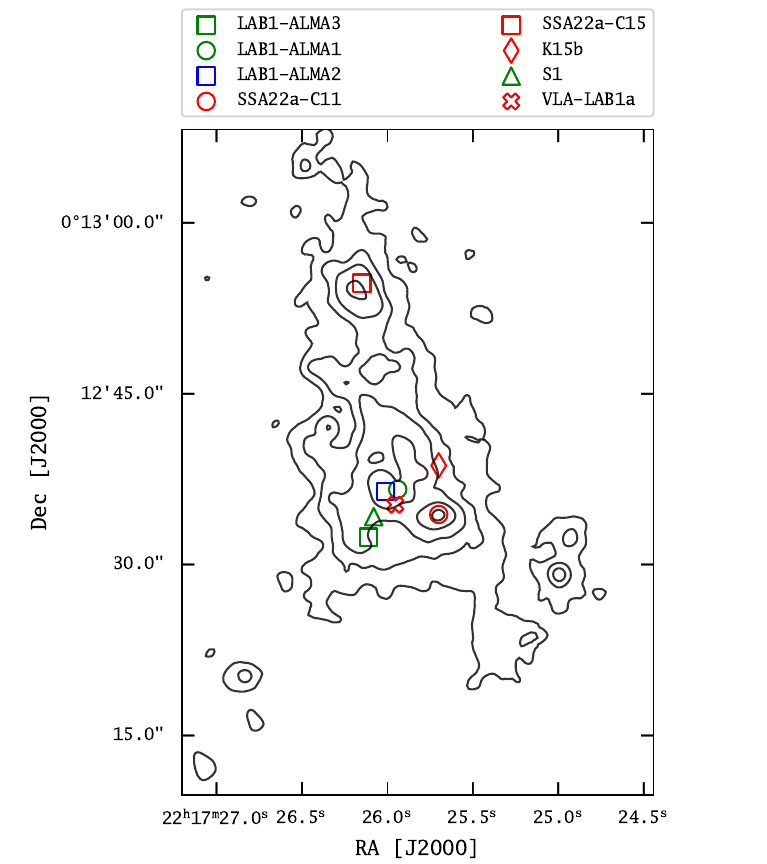}
  \caption{Positions of confirmed galaxies within the LAB\,1 / LAB\,8 structure
    (Table~\ref{tab:sources}) shown alongside Ly$\alpha$ surface brightness contours from
    our MUSE adaptive narrow band image (see Sect.~\ref{sec:adapt-lyalpha-image}).  The
    \emph{red circle} and \emph{red square} mark the Lyman Break Galaxies SSA22a-C11 and
    SSA22a-C15 from \cite{Steidel2003}, respectively.  The \emph{green circle} and
    \emph{blue square} mark the ALMA 850\,$\mu{}$m detections LAB1-ALMA1 and LAB1-ALMA2
    \citep{Geach2016,Umehata2017}, respectively.  The \emph{red cross} marks the position
    of the 3 GHz radio-continuum detection VLA-LAB1a \citep{Ao2017}.  The \emph{red
      diamond} indicates the K-band selected and spectroscopically confirmed galaxy
    LAB01-K15b from \cite{Kubo2015,Kubo2016}.  The \emph{green square} marks the ALMA
    850\,$\mu$m plus [\ion{C}{ii}] 158\,$\mu$m detected source
    \citep{Geach2016,Umehata2017}.  This source is also detected at 3 GHz with the VLA
    \citep[$S_\mathrm{10cm} = 7.3\,\mu$Jy;][]{Ao2017} and it has a tentative X-Ray
    counterpart in the 400\,ks deep Chandra data from \cite{Lehmer2009}.  The
    serendipitously found $z=3.1$ [\ion{O}{iii}] emitter S1 from \cite{Geach2016} is
    marked with a \emph{green triangle}. The symbols used to indicate the embedded
    galaxies within the blob will be used throughout the paper. North is to the top and
    east is to the left. The contours correspond to Ly$\alpha$
    surface-brightness levels
    $\mathrm{SB}_\mathrm{Ly\alpha} = [200 , 100 , 50 , 25 , 8.75]\times
    10^{-19}$\,erg\,s$^{-1}$cm$^{-2}$\,arcsec$^{-2}$.  }
  \label{fig:ocas}
\end{figure}

\object{LAB 1} (RA: 22h17m26.0s, Dec: +00$^\circ$12\arcmin{}36\arcsec{}) was discovered in
a narrow-band image by \cite{Steidel2000}.  These observations targeted Ly$\alpha$
emission of a previously identified redshift overdensity at $z=3.1$ that was revealed as a
peak in the redshift distribution of a spectroscopic follow-up campaign for Lyman break
selected galaxies in the SSA-22 field \citep{Steidel1998}.  With an isophotal area of
181\,arcsec$^2$ (10523\,kpc$^{2}$ in projection) and a total Ly$\alpha$ luminosity of
$8.1\times10^{43}$\,erg\,s$^{-1}$ LAB\,1 is one of the largest and one of most luminous
LABs known \citep{Matsuda2011}.  Directly north of LAB\,1, offset by $\approx 12$\arcsec{}
from its photometric centre, \cite{Matsuda2004} identified a companion blob: LAB\,8 (RA:
22h17m26.1s, Dec: +00$^\circ$12\arcmin{}55\arcsec{},
$L_\mathrm{Ly\alpha} = 1.5 \times 10^{43}$ \,erg\,s$^{-1}$, isophotal area
40\,arcsec$^2$).  As will be demonstrated in the present analysis LAB\,1 and LAB\,8 are,
in fact, a contiguous structure (see Figure~\ref{fig:ocas}).

Two of the Lyman break galaxies from \cite{Steidel1998} are within the combined LAB\,1 /
LAB\,8 structure.  Adopting the nomenclature of \cite{Steidel2003}, these are SSA22a-C11
and SSA22a-C15.  We list the coordinates of both galaxies in
Table~\ref{tab:sources}\footnote{We note the coordinates for these Lyman break galaxies
  provided in \cite{Steidel2003} --
  RA$=$22$^\mathrm{h}$17$^\mathrm{m}$25.67$^\mathrm{s}$/Dec$=$+00$^\circ$12\arcmin{}35.2\arcsec{}
  for SSA22a-C11, and
  RA$=$22$^\mathrm{h}$17$^\mathrm{m}$26.127$^\mathrm{s}$/Dec$=$+00$^\circ$12\arcmin{}55.3\arcsec{}
  for SSA22a-C15 -- appear to be offset by $\sim 1$\arcsec{} to the north-west with
  respect to where these galaxies are located in our data.  As described in
  Sect.~\ref{sec:astr-alignm}, we aligned our data with the 2MASS reference frame, while
  \cite{Steidel2003} tied their astrometry the HST guide star catalogue.  Earlier versions
  of the HST guide star catalogue were known to contain systematic errors of up to
  $\sim$1\arcsec{} \citep{Morrison2001}.  We thus speculate that this is the reason for
  the coordinate offsets.  Here we revise the coordinates of SSA22a-C11 and SSA22a-C15
  according to our adopted reference frame (Table~\ref{tab:sources}). \label{fn:1} }.
SSA22a-C11 is located in the south-west of LAB\,1, while SSA22a-C15 is found close to the
Ly$\alpha$ photometric centre of LAB\,8.  Both galaxies have their redshift confirmed via
near-IR detections of their [\ion{O}{iii}] $\lambda\lambda 4963,5007$ lines
\citep[][]{McLinden2013}.  Interestingly, \cite{McLinden2013} find no offset
  between Ly$\alpha$ emission line redshifts and [\ion{O}{III}] redshifts.  This is not
commonly the case for high-$z$ Ly$\alpha$ emitting galaxies where the Ly$\alpha$ line is
typically found to be offset by $\gtrsim 200$\,km\,s$^{-1}$ with respect to the systemic
redshift \citep[e.g.][]{Rakic2011,Chonis2013,Trainor2015}.  Since positive Ly$\alpha$
redshift offsets are usually interpreted as signs of outflowing expanding gas,
\cite{McLinden2013} argue that C11 and C15 have no such prominent outflows.  Moreover, C11
has an inferred star-formation rate of $\lesssim 10$\,M$_\odot{}$yr$^{-1}$
\citep{Steidel1998,Steidel2003}, thus falls short by more than a factor of ten in
producing the required amount of ionising photons to power LAB\,1.

Given the potential of obscured star-formation or AGN activity in the blob, it became
naturally the target of several sub-mm and radio campaigns
\citep{Chapman2001,Chapman2003,Matsuda2007,Yang2012,Geach2014,Geach2016,Umehata2017,Ao2017}.
Initial
studies provided a confusing picture with purported detections by some that were vastly
incommensurate with upper limits reached by others \citep[see Sect.~2 of][]{Geach2014}.
Nevertheless, advances in sub-mm detector technology and collecting area lead to a
significant detection of three 850$\mu$m sources within the blob
\citep{Geach2014,Geach2016,Umehata2017}.  Adopting the nomenclature of \cite{Umehata2017},
these ALMA detected systems are denoted LAB1-ALMA1, LAB1-ALMA2, and LAB1-ALMA3.  We list
their coordinates in Table~\ref{tab:sources}.  LAB1-ALMA1 and LAB1-ALMA2 are in close
vicinity to each other and close to the peak of Ly$\alpha$ surface brightness.  LAB1-ALMA3
is located in the south-eastern part of LAB\,1.  The total measured 850\,$\mu$m flux
density from these three resolved ALMA sources is $S_{850\mu\mathrm{m}} = 1.7$\,mJy.
This corresponds to star-formation rates of $\sim 150$\,M$_\odot$yr$^{-1}$ under
  standard dust-heating assumptions.  Moreover, hints for an extended low-surface
brightness dust-component which is not detected by ALMA are seen in the fact that the
single-dish SCUBA-2 observations \citep{Geach2014} yield a factor of $2.7$ higher flux
compared to the interferometric measurement.

\cite{Ao2017} report 3\,GHz radio-continuum detection slightly south of LAB\,1-ALMA1 and
LAB1-ALMA2.  This $S_{\mathrm{10cm}}=7.3\pm2.2 \,\mu$Jy radio source is denoted VLA-LAB1a
and we list its coordinates in Table~\ref{tab:sources}.  Given the proximity to
LAB\,1-ALMA1 and LAB1-ALMA2 the radio source is believed to be physically associated with
the $S_{850\mu\mathrm{m}} \approx 1$\,mJy sub-mm galaxies.  According to \cite{Ao2017} the
$S_{850\,\mu\mathrm{m}} / S_{\mathrm{10cm}}$ ratio is atypical for a purely star-forming
system and thus could be indicative of AGN activity.  Both systems have no spectroscopic
redshift confirmation independent of Ly$\alpha$.  However, sources at the positions of
LAB1-ALMA1 and LAB1-ALMA2 are reported as K-Band selected galaxies with photometric
redshifts in the range $2.6 < z < 3.6$ \citep[][their Figure~10]{Uchimoto2012}.

LAB1-ALMA3 is spectroscopically confirmed as a [\ion{C}{II}] 158\,$\mu$m emitting source
with ALMA \citep[$I_{\mathrm{[CII]}}= 16.8 \pm 2.1$Jy\,km\,s$^{-1}$,
$z_{\mathrm{[CII]}} = 3.0993\pm0.0004$;][]{Umehata2017}. It is also detected as a 3\,GHz
radio-continuum source with the VLA
\citep[$S_{\mathrm{10cm}}=7.3\pm2.2 \,\mu$Jy;][]{Ao2017}.  As the coordinates of this
radio counterpart are identical with LAB1-ALMA3 we do not list it as a separate source.
Moreover, \cite{Ao2017} report a tentative X-Ray signal at the position of LAB1-ALMA3
using the deep (400\,ks) \emph{Chandra} full-band (0.5 - 8 keV) image from
\cite{Lehmer2009}.  From these observations \cite{Ao2017} suggest the potential existence
of an AGN in this source.  Moreover, LAB1-ALMA3 appears as a K-Band selected galaxy in the
sample of \cite{Uchimoto2012} and has been spectroscopically confirmed via H$\beta$ and
[OIII] $\lambda5007$ emission with MOIRCS on the Subaru telescope
\citep{Kubo2015,Kubo2016}.  The redshift derived from those rest-frame optical lines
($z=3.1000\pm0.004$) is commensurate with the [\ion{C}{II}]-based redshift.

Two more galaxies are spectroscopically confirmed members of the blob.  LAB01-K15b is a
K-Band selected galaxy detected in [\ion{O}{iii}] emission \citep{Kubo2015,Kubo2016} and
S1 is a serendipitous [\ion{O}{iii}] detection from \cite{Geach2016}.  The coordinates and
redshifts of both sources were presented by \cite{Umehata2017} and are reproduced in
Table~\ref{tab:sources}.  Lastly, \cite{Kubo2016} display a faint K-Band selected galaxy
slightly south-west of SSA22a-C11 at a compatible photometric redshift (their Figure 2),
however, no coordinates for this potential member of LAB\,1 are provided.

In summary, the LAB\,1/LAB\,8 system contains 5 spectroscopically confirmed galaxies.
Guided by the systemic redshifts of the galaxies associated with the blob
(Table~\ref{tab:sources}) we fix $z=3.1$ as its systemic redshift in the following.  One
of the spectroscopically confirmed systems, LAB1-ALMA3, is detected in 850$\mu$m
dust-continuum, [\ion{C}{ii}] 158$\mu$m emission, 3GHz radio continuum, and tentatively in
X-Rays \citep{Ao2017}.  Additionally, two 850$\mu$m sources (LAB1-ALMA1 and LAB1-ALMA2)
which are accompanied by a 3GHz sources (VLA-LAB1a) are detected in the centre of LAB\,1.
While these sources are not spectroscopically confirmed members of the blob, their
physical association with the system appears likely, especially given their prominent
central position within the blobs structure.  To provide a visual overview of the with the
LAB\,1/LAB\,8 system associated galaxies we plot their positions with respect to the
Ly$\alpha$ surface brightness contours in Figure~\ref{fig:ocas}.  For the latter we made
already use of the MUSE data discussed in the remainder of the paper.

The interpretation that LAB1-ALMA1 and LAB1-ALMA2 are physically associated with the blob
is further supported by results from imaging-polarimetry with FORS2 on the VLT by
\cite{Hayes2011}.  The radial polarisation profile, as well as the orientation of the
polarisation vectors from those observations are consistent with predictions from
Ly$\alpha$ radiative-transfer theory for Ly$\alpha$ scattering from a central Ly$\alpha$
powering source \citep{Dijkstra2008}.  Interestingly, LAB1-ALMA1 and LAB1-ALMA2 are at the
centre of the circular pattern denoted by the polarisation vectors from \cite{Hayes2011}.
More recent spectro-polarimetry with FORS2 by \cite{Beck2016} appears consistent with this
scenario, although the interpretation of spectro-polarimetric Ly$\alpha$ data is more
complex \citep{Lee1998}, as different scattering geometries and kinematics are degenerate
with respect to observable polarisation signal \citep{Eide2018}.  Especially,
\cite{Trebitsch2016} challenged the interpretation of \cite{Hayes2011} by showing that the
observed polarisation signal can be reproduced in a pure cooling-flow scenario. Thus,
despite significant observational efforts and numerous counterpart identifications, there
is still considerable debate regarding the mechanisms that power the Ly$\alpha$ emission
of the blob.

Previous IFS observations of LAB\,1 were obtained with the SAURON instrument at the WHT
\citep{Bower2004,Weijmans2010}.  These observations revealed a complex morpho-kinematic
structure of the system in Ly$\alpha$.  The identification of multiple clump-like features
in these data was seen as evidence for the presence of multiple galaxies in the system,
while the chaotic motions where interpreted as signs for galaxy-galaxy interactions.
Moreover, \cite{Weijmans2010} found signatures of coherent velocity shear at the positions
of the Lyman break galaxies C11 and C15.  While these observations provided first insights
into the complex kinematic structure of the system, they were limited in depth and spatial
resolution.  Here we will present the new deep MUSE observations of the system, that will
allow us to map the spatial and kinematic structure of blob in unprecedented detail.

\section{ESO VLT/MUSE Observations of LAB\,1}
\label{sec:obs}

\begin{table}
  \caption{MUSE observations of LAB\,1 present in the ESO archive.}
  \begin{tabular}{lllllll} \hline \hline
    Date-Time          & AM      & DS & SGS  & $t_\mathrm{exp}$  & Sky & PID \\
     {[yy/mm/dd-UT]}            &         & [\arcsec{}]  & [\arcsec{}] & [s] &  & \\ \hline
14/11/13-00:33:06    & 1.14    & 0.61 & 0.69 & 1498             &   2    & a$^\dagger$  \\
14/11/13-00:59:29    & 1.19    & 0.82 & 0.73 & 1498             &   2    & a           \\
15/05/22-08:15:40    & 1.47    & 1.26 & 1.00 & 1500             &   3    & b           \\
15/05/22-08:42:34    & 1.34    & 0.94 & 0.94 & 1500             &   3    & b           \\
15/05/29-07:55:36    & 1.43    & ---  & 0.72 & 1500             &   3    & b           \\
15/05/29-08:22:31    & 1.31    & 0.59 & 0.71 & 1500             &   3    & b           \\
15/05/29-08:53:20    & 1.22    & 0.51 & 0.68 & 1500             &   3    & b           \\
15/05/29-09:20:15    & 1.16    & 0.54 & 0.65 & 1500             &   3    & b           \\
15/05/30-08:58:17    & 1.20    & 0.92 & 0.84 & 1500             &   3    & b           \\
15/05/30-09:25:11    & 1.15    & 0.67 & 0.82 & 1500             &   3    & b           \\
15/06/12-08:25:02    & 1.16    & 0.94 & 0.75 & 1500             &   2    & b           \\
15/06/12-08:51:58    & 1.13    & 1.21 & 0.78 & 1500             &   2    & b$^\dagger$  \\
15/06/12-09:21:56    & 1.11    & 1.06 & 0.79 & 1500             &   2    & b           \\
15/06/12-09:48:52    & 1.10    & 0.92 & 0.79 & 1500             &  2     & b$^\dagger$  \\
15/06/19-08:32:25    & 1.12    & 0.77 & 0.66 & 1498             &   2    & a           \\
15/06/19-08:59:09    & 1.10    & 0.80 & 0.68 & 1498             &   2    & a$^\dagger$  \\
15/06/19-09:37:28    & 1.11    & 0.88 & 0.63 & 1498             &   2    & a           \\
15/06/19-10:04:12    & 1.13    & 0.95 & 0.65 & 1498             &    2   & a           \\
15/06/20-07:40:02    & 1.19    & 0.78 & 0.70 & 1498             &   2    & a           \\
15/06/20-08:06:46    & 1.14    & 0.82 & 0.68 & 1498             &   2    & a           \\
15/06/20-08:38:08    & 1.11    & 0.93 & 0.66 & 1498             &   2    & a           \\
15/06/20-09:04:52    & 1.10    & 0.83 & 0.70 & 1498             &   2    & a           \\
15/06/22-08:01:52    & 1.14    & 1.14 & 0.63 & 1498             &   2    & a$^\ddagger$ \\
15/06/22-08:28:34    & 1.11    & 1.69 & 0.77 & 1498             &    2   & a           \\
15/06/24-06:39:13    & 1.32    & 0.60 & 0.68 & 1498             &   2    & a           \\
15/06/24-07:05:55    & 1.23    & 0.55 & 0.69 & 1498             &   2    & a           \\
15/09/17-02:35:33    & 1.12    & 1.10 & 0.71 & 1498             &   1    & a           \\
15/09/17-03:02:18    & 1.11    & 0.66 & 0.67 & 1498             &   1    & a           \\
16/07/15-09:09:06    & 1.23    & 1.10 & 1.14 & 1495             &   2    & c           \\
16/07/15-09:35:50    & 1.31    & 0.83 & 1.14 & 1495             &   2    & c           \\
16/08/12-06:47:59    & 1.16    & 1.12 & 0.89 & 1510             &   1    & c           \\
16/08/12-07:14:59    & 1.22    & 1.03 & 0.93 & 1510             &   1    & c           \\
16/09/02-02:33:13    & 1.22    & 0.84 & 1.00 & 1485             &   2    & c           \\
16/09/02-03:00:08    & 1.16    & 0.76 & 0.93 & 1485             &   2    & c           \\
16/09/05-01:57:24    & 1.29    & 0.87 & 1.00 & 1495             &   2    & c           \\
16/09/05-02:24:08    & 1.21    & 0.71 & 0.90 & 1495             &   2    & c           \\
16/09/25-00:19:41    & 1.36    & 0.91 & 1.01 & 1485             &   2    & c           \\
16/09/25-00:46:14    & 1.26    & 0.95 & 0.89 & 1485             &   2    & c           \\
16/09/30-00:11:15    & 1.32    & 1.76 & 1.02 & 1495             &   2    & c           \\
16/09/30-00:38:10    & 1.23    & 0.89 & 0.96 & 1495             &   2    & c           \\
16/10/04-00:43:00    & 1.18    & 1.56 & 1.39 & 510              &   2    & c           \\
16/10/04-01:55:53    & 1.10    & 0.85 & 1.02 & 1495             &    2   & c           \\
16/10/04-02:22:48    & 1.10    & 0.94 & 0.92 & 1495             &   2    & c           \\ \hline \hline
  \end{tabular}
  \label{tab:obslog}
  \tablefoot{AM = airmass, DS = Differential Image Motion Monitor Seeing measurement
    (FWHM), SGS = Slow Guiding System seeing measurement, Sky = sky transparency (1 =
    photometric, 2 = clear, 3 = thin cirrus), PID = ESO Programme ID (a = 094.A-0605, b =
    095.A-0570, c = 097.A-0831; with ${\dagger}$ indicating an exposure affected by a
    $\sim 2$\arcsec wide continuum bright trail and ${\ddagger}$ marking the exposure that
    could not be used due to a telescope tracking error).}
\end{table}

LAB 1 was observed with MUSE \citep[][]{Bacon2010} in its wide field mode configuration
without adaptive optics at the European Southern Observatories Unit Telescope 4 (Antu) in
three service-mode programmes from 2014 to 2016 (ESO Programme IDs 094.A-0605, 095.A-0570,
and 097.A-0831 with principal investigators Hayes, Bower, and Hayes, respectively).  A log
of these observations is given in Table~\ref{tab:obslog}.  The individual exposure times
are typically 1500\,s, only with one exposure being significantly shorter (510\,s).  In
total the three programmes accumulated a total open-shutter time of 63390\,s (17.6 h) on
the target.

The three different programmes centred the instrument on different regions of the blob.
While programmes 094.A-0605 and 095.A-0570 centred the 1\arcmin{}$\times$1\arcmin{} MUSE
field of view slightly north of LAB 1 to encompass also the northern neighbouring
Ly$\alpha$ blob LAB 8, programme 097.A-0831 centred on a region south-west of the
brightest blob structure.  This off-set was motivated by a detection of a previously
unknown low-surface brightness extension of the blob in a reduction of the data from
programme 094.A-0605 \citep[see][]{Geach2016}.  Each programme used the dithering and
rotation pattern that is recommended in the MUSE observing
manual\footnote{\url{http://www.eso.org/sci/facilities/paranal/instruments/muse/doc.html}}. Unfortunately,
one exposure suffered from a severe tracking error and could not be used in the final
analysis.  Thus the total integration time of the analysable dataset is 17.2\,h.

The DIMM seeing reported by ESO for the observation ranges from 0.5\arcsec{} to
1.7\arcsec{}, but with the majority of exposures taken at sub-arcsecond seeing (mean:
0.87\arcsec{}, median: 0.88\arcsec{}).  A potentially more accurate measurement of the
seeing is provided by the FHWM measurements of stars in the meteorology fields recorded by
MUSE's slow-guiding system (SGS column of Table~\ref{tab:obslog}).  Albeit having a
slightly larger scatter, the measured image quality by the SGS is often a bit better than
the site-wide values provided by the DIMM (mean: 0.83\arcsec{}, median: 0.78\arcsec{}),

Programmes 094.A-0605 and 097.A-0831 were taken without the blue cut-off filter in the
fore-optics (extended wavelength mode), thus allowing the wavelength range from 465\,nm to
930\,nm to be sampled (although at the cost of second-order contaminations at
$\lambda > 900$\,nm), while programme 095.A-0570 was taken with the blue cut-off filter
within the light path (nominal wavelength mode), thus sampling the wavelength range from
480\,nm to 930\,nm.

All observational raw data and the associated calibration frames where retrieved from the
ESO Science Archive Facility using the raw-data query
form\footnote{\url{http://archive.eso.org/cms/eso-data.html}} and the cal-selector tool.
For all exposures the associated calibration frames (bias frames, arc lamp frames,
continuum lamp frames, twilight flats, standard star exposures, and astrometric standard
fields) were taken as part of the standard calibration plan for MUSE observations.  This
means especially, that twilight flats were taken typically once a week, while
standard-star exposures are usually obtained daily.  Nevertheless, some retrieved
exposures were associated to standard star\footnote{The to our science observations
  associated spectrophotometric standards are Feige110, GJ754.1A, GD153, EG274, LTT3218,
  GD108, and LTT7987.} exposure that were taken one night before or after the science
observations.

\section{Data reduction}
\label{sec:data-reduction}

\begin{figure}
  \centering
  \includegraphics[width=0.5\textwidth]{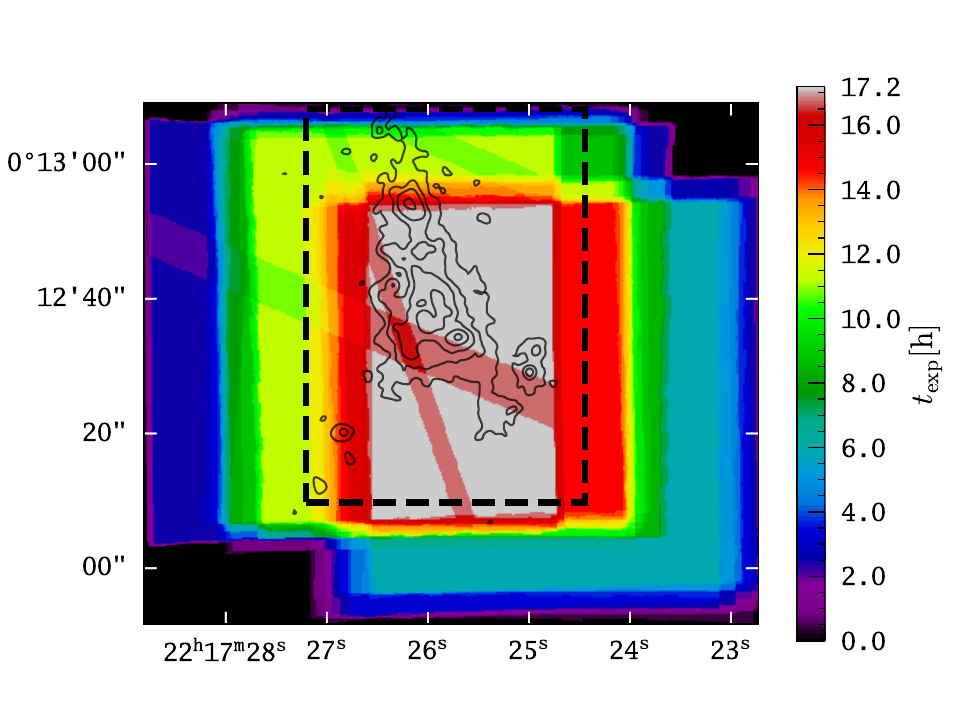}
  \vspace{-2em}
  \caption{Exposure map image for MUSE observations of LAB1 in the SSA22 field.
    The dashed rectangle indicates the zoomed-in region displayed in the
    spectral sequence shown in Figure~\ref{fig:channel_maps}.  Diagonal bands of
    lower exposure times are a result of masked out regions in the final cube
    stack due to contamination by bright satellite tracks or meteor trails in
    the individual exposures.  To indicate the position and morphology of LAB\,1
    we also overlay the Ly$\alpha$ surface brightness contours from our MUSE
    adaptive narrow-band image (see Sect.~\ref{sec:adapt-lyalpha-image}). }
  \label{fig:exp_map}
\end{figure}

\subsection{Production of the datacube}
\label{sec:production-datacube}

For the reduction of the 17.6\,h of MUSE observations into a science-ready datacube we
used version 2.4.2 of the MUSE Instrument Pipeline \citep[MUSE DRS -- ][]{Weilbacher2016}
provided by
ESO\footnote{\url{https://www.eso.org/sci/software/pipelines/muse/muse-pipe-recipes.html}}
and version 3.0 of the MUSE Python Data Analysis Framework \citep[MPDAF
--][]{Bacon2016,Piqueras2017} provided by the MUSE
consortium\footnote{\url{https://mpdaf.readthedocs.io/}}.  The here MUSE DRS version
used here incorporates the so-called self-calibration procedure to improve the
flat-fielding accuracy for deep datasets \citep[]{Conseil2016}.  For our data reduction we
adopted a similar strategy as the one used for the reduction of the MUSE Hubble Ultra Deep
Field \citep{Bacon2017}.

We first ran the MUSE DRS calibration recipes \texttt{muse\_bias}, \texttt{muse\_flat},
\texttt{muse\_wavecal}, and \texttt{muse\_lsf} on the calibration frames that are
associated to each science and standard star exposure. We also used the
\texttt{muse\_twilight} recipe on the twilight frames. Next, we applied the resulting
calibration data products to each science and standard star exposure using the recipe
\texttt{muse\_scibasic}.  The resulting standard-star pixtables where then fed into
\texttt{muse\_standard} to create response curves.  Those were the applied to each science
exposure with the \texttt{muse\_scipost} task, using its self-calibration feature, but not
using its sky-subtraction capabilities.  When running \texttt{muse\_scipost} we used the
associated astrometric calibrations provided from the ESO archive instead of the
astrometric calibrations shipped with the pipeline.  This approach was necessary,
as the instrument underwent several interventions during the long period over which
observations were taken.  Not using the correct astrometric calibrations would result in
uncorrected geometric distortions within the pixtables.

In order for self-calibration to work optimally, regions containing sources that are
bright in the continuum needed to be masked.  While in principle the DRS can
automatically detect such regions, we manually masked out bright continuum objects a bit
more conservatively compared to the DRS generated mask.  Manual masking was performed by
visual inspection of the datacubes with the ds9 software \citep{Joye2003} using
polygon-shaped and circular regions. These regions where then converted into datacube
masks using the \texttt{pyregion} python
package\footnote{\url{https://pyregion.readthedocs.io/}}.  Additionally, four science
exposures contained continuum bright linear trails from moving objects (likely satellite
flares, meteors or aeroplanes -- affected exposures are marked by a $\dagger$ in
Table~\ref{tab:obslog}). These trails were also masked manually for the self-calibration.

We then removed night sky-emission with \texttt{muse\_create\_sky} and
\texttt{muse\_subtract\_sky}.  During this step we iteratively masked out regions that
contained emission from the Ly$\alpha$ blob, so that the blob's emission is not subtracted
from the final data.

Next we resampled the individual sky-subtracted and flux-calibrated exposure tables to the
final 3D grid with \texttt{muse\_scipost\_make\_cube}.  We defined this final grid via an
initial run of \texttt{muse\_expcombine} on a subset of pointings from each observing
programme.  Using MPDAF's \texttt{combine} method we then produced an unweighted mean
stack of those individual datacubes to produce the final science-ready datacube.  Prior to
this stack we masked out bright linear trails that were present in some observations
(marked by a $\dagger$ in Table~\ref{tab:obslog}).  We also ensured that remaining
cosmic-ray residuals in the individual datacubes were filtered out in the final stack by
using the $\kappa$-$\sigma$ clipping algorithm in the \texttt{combine} task (2 maximum
iterations with $\sigma_\mathrm{clip} = 5$).

The resulting final datacube has $456 \times 378$ spatial elements (spaxels) that sample
the sky parallel to right ascension and declination at 0.2\arcsec{}$\times$0.2\arcsec{}.
Each spaxel consists of 3802 spectral elements that sample the spectral domain from
4599.6\,\AA{} to 9350.8\,\AA{} linearly with steps of 1.25\,\AA{}.

Throughout the above procedure a formal variance propagation is also carried out by each
of the used routines, thus a second datacube containing the variance for each volume pixel
(voxel) is also produced.  However, the resampling procedure in
\texttt{muse\_scipost\_make\_cube} produces correlated noise between neighbouring voxels
\citep[see Figure 5 in ][]{Bacon2017}.  Since this co-variance term between neighbouring
voxels can not easily be accounted for in the final reduction, the formal
variance cube contains underestimates of the true variances.  By processing artificial
pixtables filled only with Gaussian noise, \cite{Bacon2017} demonstrate that the variance
for a voxel in an individual exposure datacube must be corrected by a factor of
$(1/0.6)^2$ \citep[see Figure 6 in ][]{Bacon2017}.  Following \cite{Bacon2017} we apply
this correction to our final variance cube.

We display a map of the integration time for each spaxel in the MUSE datacube of LAB1 in
Figure~\ref{fig:exp_map}.  The maximum integration time of 61920\,s (17.2h) is reached in
a 32\arcsec{}$\times$48\arcsec{} central region of our datacube.  In this region all three
ESO observing programmes overlap.  This region completely encompassed the known extend of
LAB\,1 and LAB\,8 prior to the here presented observations.  Moreover,
Figure~\ref{fig:exp_map} also illustrates the location of the masked out regions due to
continuum-bright meteor trails or satellite tracks.

Lastly, we produce an emission line only datacube by subtracting a running median filter
in spectral direction.  This simple method for continuum removal has been proven effective
in previous MUSE studies for isolating emission line signals
\citep[e.g.][]{Borisova2016,Herenz2017,Herenz2017b,Herenz2017a,Arrigoni-Battaia2019}.
Here we set the width of the median filter conservatively to 301 spectral layers
(376.25\,\AA{}).

\subsection{Astrometric alignment}
\label{sec:astr-alignm}

We register the MUSE datacube to an absolute astrometric frame by using the
2MASS point-source catalogue \citep{Skrutskie2006}.  Unfortunately, there are
only two 2MASS sources within the limited FoV of the MUSE and STIS observations,
with one of those sources actually being an extended galaxy.  We therefore use
the only real 2MASS point source (2MASS J22172397+0012359) to anchor our
astrometric reference frame in both observational datasets.

By tying the astrometric reference frame only to one source there remain in
principle still several degrees of freedom with respect to geometrical
distortions and rotation.  However, rotation geometrical distortion are
corrected for MUSE data at the pipeline level, and the accuracy of this
procedure is constantly monitored by ESO using astrometric calibration fields.

We verified our absolute astrometry against an archival HST/STIS 50CCD clear-filter image
that is partially overlapping with our MUSE data \citep[Proposal ID: 9174, presented and
analysed in][]{Chapman2001,Chapman2004}.  We tied the absolute astrometry of this image
also to the 2MASS point source.  By visual inspection via blinking in \texttt{ds9} we then
ensured that no offsets exist between both datasets.  We conservatively estimate that the
absolute astrometric accuracy of our data to be on the order of one MUSE pixel
(0.2\arcsec{}).

\section{Analysis and results}
\label{sec:ana}

\subsection{Velocity sliced Ly$\alpha$ emission maps}
\label{sec:veloc-slic-lyalpha}

\begin{figure*}
  \centering
  \includegraphics[width=\textwidth]{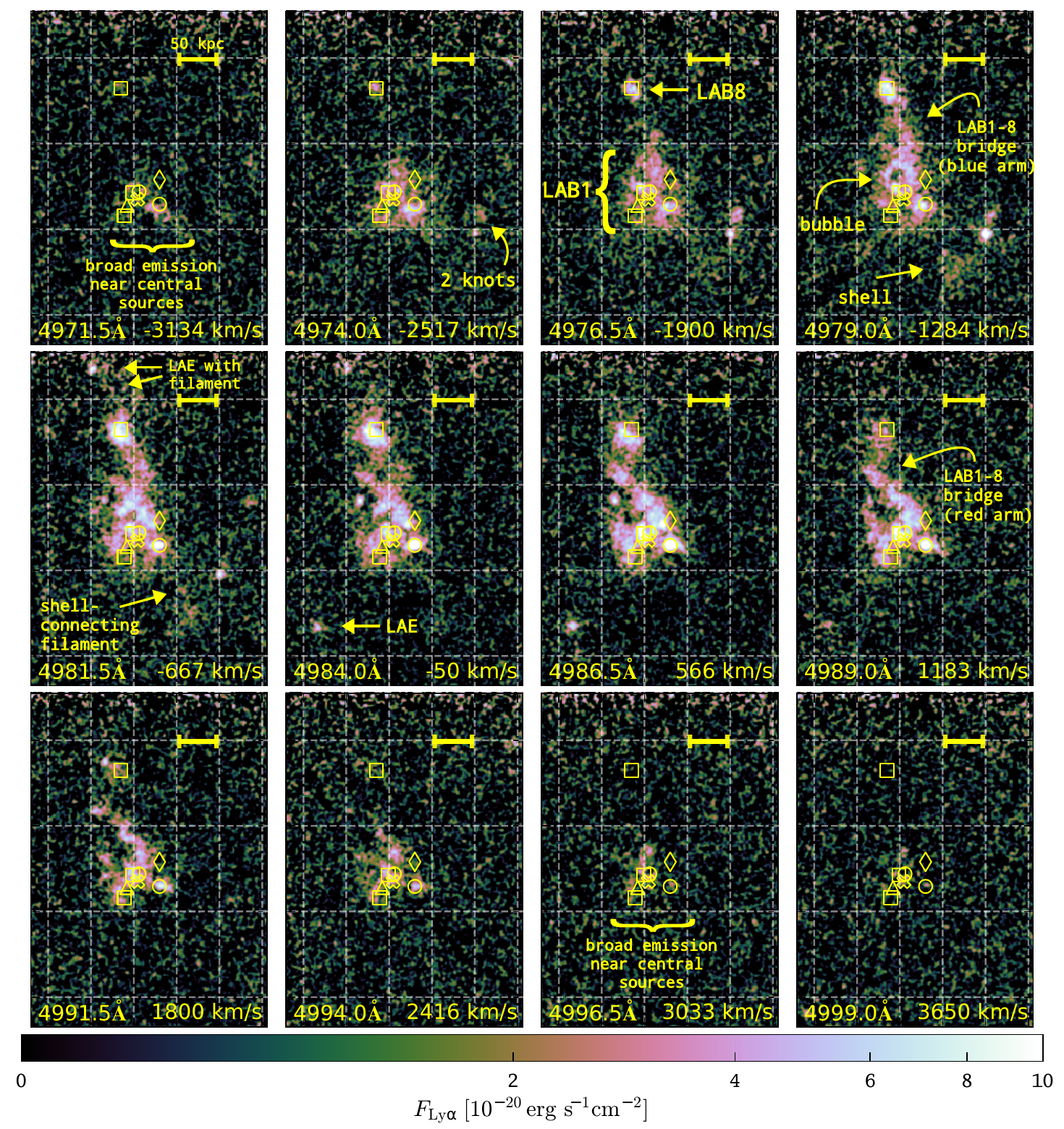}
  \caption{Spectral sequence of pseudo-narrowband images of LAB\,1 from 4971.5\,\AA{} to
    4999.0\,\AA{} created from the continuum subtracted MUSE datacube.  Each image has a
    band-width of 2.5\,\AA{} and in order to enhance low-surface brightness features the
    images have been smoothed with a $\sigma=1$px (0.2\arcsec{}) Gaussian kernel.  In each
    panel we indicate the positions of known galaxies within the blob (see
    Sect.~\ref{sec:known-sources} and Figure~\ref{fig:ocas}). Moreover, we show in each
    panel a scale that indicates 50 kpc in projection at $z=3.1$ (6.49\arcsec{}).  The
    morphological features described in Sect.~\ref{sec:veloc-slic-lyalpha} are indicated
    in the bluest image where they become visible.  We also indicate that we trace
    Ly$\alpha$ emission $\pm3000$km\,s$^{-1}$ around the central sources of LAB\,1.  North
    is up and east is to the left.}
  \label{fig:channel_maps}
\end{figure*}

We present a spectral sequence of pseudo-narrowband images over the Ly$\alpha$ line in
Figure~\ref{fig:channel_maps}.  These images show that we can trace Ly$\alpha$ emission
from LAB\,1 over a bandwidth of $\approx 28$\AA{} ($\pm 3000$km\,s$^{-1}$ around
$z_\mathrm{Ly}=3.1$).  As can be seen from Figure~\ref{fig:channel_maps}, the highest
velocity gas emitting Ly$\alpha$ is located in vicinity of the central sources LAB1-ALMA1,
LAB1-ALMA2, and VLA-LAB1a.  However, numerous other features with narrower spectral width
appear only in a few velocity slices.  Overall the velocity slices reveal a complex
spectral and morphological structure of Ly$\alpha$ emission throughout different parts of
the blob. We labelled notable features in Figure~\ref{fig:channel_maps}, where we point
at:
\begin{enumerate}
\item A circular structure devoid of strong Ly$\alpha$ emission towards
  the north of the sub-mm sources. This feature is labelled as ``bubble'' in
    Figure~\ref{fig:channel_maps}.  It is most clearly seen in the slice around
  4979\,\AA{}, where we indicate this feature with an arrow.  In subsequent redder slices
  this ``bubble'' appears with less contrast.  It appears as if its radius increases from
  $\sim 20$\,kpc at 4979\,\AA{} to up to $\sim 40$ kpc towards redder wavelengths.
\item A filamentary narrow bridge connecting LAB\,1 to LAB\,8 in the north.  This bridge
  becomes visible in the slice at 4979\,\AA{} and can be clearly followed until the
  4989\,\AA{} slice.  For LAB\,8 the flux shears from the north-west to the south-east
  with increasing wavelength and the filament shows the same velocity shearing.  We
  indicate this by labelling it as ``LAB1-8 bridge (blue arm)'' and
  ``LAB1-8 bridge (red arm)'' of the bridge in the panel displaying the
  4979\,\AA{} slice and 4989\,\AA{} slice, respectively.
\item A compact emission knot towards the north of LAB\,8, that connects to LAB\,8 via a
  faint low-surface brightness filament.  This feature is seen in the slices at
  4981.5\,\AA{} and 4984\,\AA{}, and we label it as ``LAE with filament'' in the
    4984\,\AA{} slice.  This newly discovered LAE and filament is close to the edge of
  the field of view of the observations, and thus only exposed at 8\,h to 12\,h, thus the
  noise in this part of the datacube is significantly larger.  Nevertheless, as we will
  discuss in the following Sect.~\ref{sec:newly-disc-isol}, the compact emitter and the
  filament are significant detections.
\item A faint extended, slightly curved, shell-like region towards the south-west of
  LAB\,1.  Along the major axis the extent of the shell is $\sim120$\,kpc, and its
  projected thickness is $\sim 30$\,kpc.  This low-surface brightness emission is visible
  in the slices from 4976.5\,\AA{} to 4981.5\,\AA{} where we label it simply as ``shell''.
  The indicated previously undetected Ly$\alpha$ emitting region appears to be connected
  to the main area of LAB\,1 via a low-surface brightness filament trailing from the
  north-east to the south-west.  The filament is most prominently visible in the
  4981.5\AA{} slice where we label it ``shell connecting filament''. In the northern part
  of this shell-like region a compact high-surface brightness knot is visible. This knot
  is accompanied by another slightly more diffuse knot.  These two knots in the northern
  part of the shell can be traced in the velocity slices from 4974\AA{} to 4981.5\AA{},
  and we label both features as ``2 knots'' in the 4974\AA{} slice.
\item Another compact isolated Ly$\alpha$ line emitter towards the south-east of LAB\,1
  seen in the slices from 4984\,\AA{} to 4989\,\AA{}.  We will show in the following
  Sect.~\ref{sec:adapt-lyalpha-image}, that this emitter is one of four newly detected
  Ly$\alpha$ emitters in the proximity of the LAB\,1/LAB\,8.  We label this
    emitter ``LAE'' in the 4984\AA{} slice.
\end{enumerate}

A closer inspection of the velocity slices reveals that there appears to be an overall
velocity shear.  The bluer $v<0$\,km\,s$^{-1}$ slices show predominantly emission towards
the west, while the redder $v>0$\,km\,s$^{-1}$ slices are dominated by emission
from the east.  This structure of the velocity shear becomes more clear
in our moment based analysis of the Ly$\alpha$ line profiles
(Sect.~\ref{sec:spectr-morph-lyalpha}).

\subsection{Detection and photometric measurements of Ly$\alpha$ emission}
\label{sec:adapt-lyalpha-image}

\begin{figure*}
  \sidecaption \includegraphics[width=12cm,trim=0 10 20
  30,clip=True]{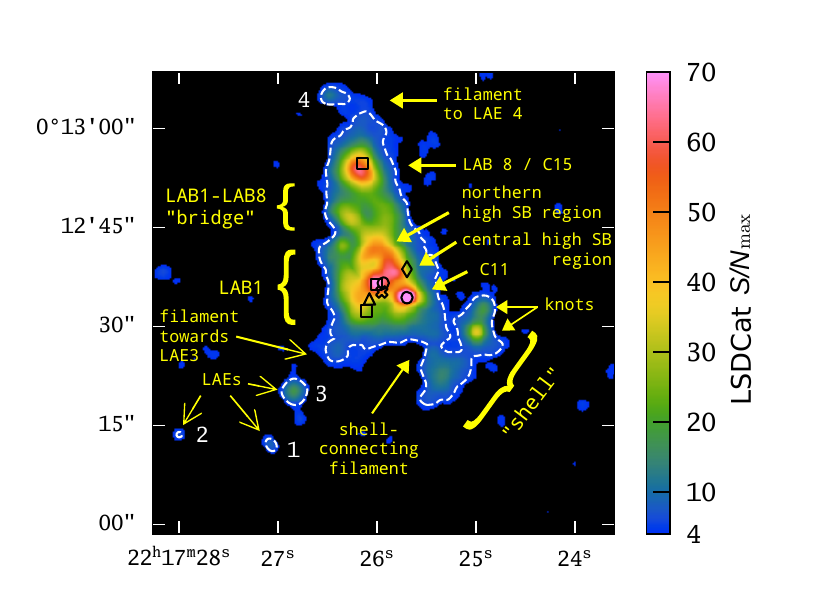}
  \caption{Map of the maximum signal-to-noise ratio after cross-correlating the datacube
    with a 3D Gaussian template (see Sect.~\ref{sec:adapt-lyalpha-image} for details on
    the construction of this image).  Pixels with $S/N_\mathrm{max}<4$ and contaminating
    foreground sources are masked (regions in black).  Thus, pixels shown in colour are a
    2D projection of the 3D mask utilised to construct the adaptive narrow band image
    displayed in Figure~\ref{fig:adapt_nb}. The dashed contour demarcates region of
    connected pixels with $S/N_\mathrm{max} \geq 6$. This highlights that LAB\,1, LAB\,8
    and the newly detected shell comprise a significantly detected contiguous region.  The
    four enumerated features in this image are regions that contain pixels with
    $S/N_\mathrm{max} \geq 6$ that have no pixel connectivity at $S/N_\mathrm{max} \geq 6$
    with the LAB1 / LAB8 + shell structure; these detections are LAEs in the vicinity of
    the blob and are further discussed in Sect.~\ref{sec:newly-disc-isol}. Previously
    known galaxies at $z=3.1$ are indicated using the symbols in Figure~\ref{fig:ocas} and
    interesting features are annotated.  }
  \label{fig:maxsn}
\end{figure*}

\begin{figure*}
  \sidecaption \includegraphics[width=12cm,trim=0 10 20
  30,clip=True]{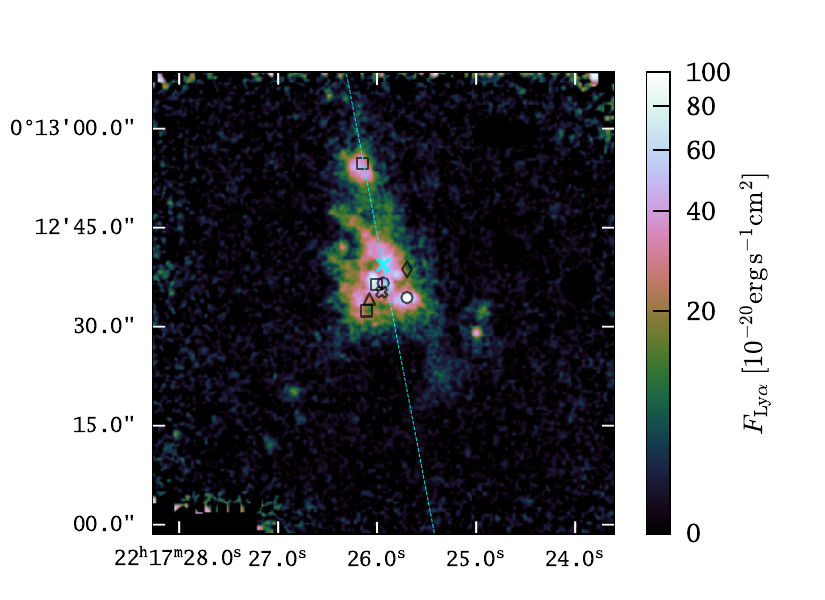}
  \caption{Adaptive narrow-band image of LAB1.  The image is the result of
    summing only over voxels in the continuum-subtracted datacube that have a
    S/N$>4$ in the \texttt{LSDCat} cross-correlated datacube.  For spaxels that
    do not contain voxels above this threshold we simply sum over 5\AA{} (four
    datacube layers) around 4985.6\AA{}
    ($=(1+z_\mathrm{LAB1})\times\lambda_\mathrm{Ly\alpha}$).  As in
    Figure~\ref{fig:maxsn}, we masked sources where the continuum subtraction
    with a running median filter failed.  In order to further enhance low-SB
    Ly$\alpha$ features, we smoothed the final image with a
    $\sigma=0.2$\arcsec{} Gaussian kernel.  The photometric centre and the
    principal axis of the blob are indicated by a cyan cross and a cyan dashed line,
    respectively. Previously known galaxies at $z=3.1$ are indicated using the
    same symbols as in Figure~\ref{fig:ocas}.}
  \label{fig:adapt_nb}
\end{figure*}

\subsubsection{Method}
\label{sec:method}

To determine the overall Ly$\alpha$ surface-brightness morphology of LAB\,1 from our
MUSE datacube it would not be optimal to create a simple narrow-band image by summing over
a certain number of datacube layers.  Choosing a single bandwidth for such an image would
either decrease the signal-to-noise ratio (S/N) for regions where the Ly$\alpha$ profiles
are very broad if the bandwidth is chosen too narrow or, conversely, decrease the S/N in
regions where Ly$\alpha$ is narrow if the adopted bandwidth would optimally
encompass the broader profiles.  Furthermore, such a simple summation would also not
account for the presence of velocity shear.  We thus create an adaptive narrow-band image.
For this image we sum for only over voxels that contain Ly$\alpha$ flux.  Our method is
similar to the creation of the narrow-band images in the analysis of extended Ly$\alpha$
emission around QSOs from MUSE data \citep[e.g.][]{Borisova2016,Arrigoni-Battaia2019}.

In order to find the spectral pixels over which we need to sum we utilised the 3D
cross-correlation procedure of the \texttt{LSDCat} software\footnote{\texttt{LSDCat} is
  publicly available via the Astrophysics Source Code Library:
  \url{http://ascl.net/1612.002} \citep{Herenz2016a}.}
\citep{Herenz2017}. \texttt{LSDCat} produces an S/N datacube by cross-correlating the
continuum subtracted datacube with a 3D Gaussian template.  The parameters of the template
are the amount of spatial and spectral dispersion of the 3D Gaussian.  Cross-correlation
suppresses high-frequency noise while maximising the S/N of signals within the data that
match the template.  Hence, the method is commonly called ``matched filtering'' \citep[e.g.,][]{Vio2016,Loomis2018}.

\texttt{LSDCat} was originally developed for the detection of Ly$\alpha$ emitting galaxies
in blind MUSE surveys \citep[see][]{Herenz2017a,Urrutia2018}.  For this application the
parameters of the template are optimally chosen when they match the width of the seeing
point spread function (PSF) and the average line width of LAEs \citep[see Sects. 4.2 and
4.3 in][]{Herenz2017}.  However, our goal here is not to optimise the template for compact
emission line sources, but to maximise the detectability of faint low-surface brightness
filaments in the outskirts of the blob.  Simultaneously, we want to preserve the
morphology of small scale surface-brightness variations.  As there is no optimal a priori
solution to this problem, the final set of adopted parameters had to be chosen by
parameter variation and visual inspection of the resulting images.  By experimenting with
different spatial filter widths, we found that a spatial FWHM of 1.8\arcsec{} preserved
most of the contrast of compact features and significantly enhanced the S/N of the
extended filamentary features in the outskirts of the blob.  The adopted filter FWHM is
roughly twice the seeing PSF FWHM\footnote{We determined the PSF FWHM via fitting a 2D
  Gaussian profile to the bright star within our field of view in a 45\,\AA{} wide
  narrow-band image centred at $z_\mathrm{Ly\alpha} = 3.1$.} of 0.95\arcsec{}.  As derived
by \cite{Zackay2017}, a filter width of twice of the seeing FWHM will reduce\footnote{In
  general for compact point-sources the relation
  $\mathrm{S/N}_\mathrm{max} \propto 2 \kappa / (\kappa + 1)^2$, with
  $FWHM_\mathrm{filt} = \kappa \times FWHM_\mathrm{PSF}$ holds, where $FWHM_\mathrm{filt}$
  and $FWHM_\mathrm{PSF}$ denote the adopted width of the filter and the width of the
  point-spread function, respectively \citep{Herenz2017}.}  the maximum S/N of compact
sources only by 20\%.  Similarly, we varied the FWHM of the spectral part and found that
300\,km/s is well suited for enhancing the detectability of the blobs low-surface
brightness features.

\subsubsection{Maximum S/N image}
\label{sec:maximum-sn-image}

In Figure~\ref{fig:maxsn} we show the resulting map when taking the maximum S/N in from
the \texttt{LSDCat} S/N datacube around $z_\mathrm{Ly\alpha}=3.1$.  We adopt an S/N
threshold of six as detection threshold to identify reliable regions from which Ly$\alpha$
emission is detected.  These regions are demarcated by a white dashed line in
Figure~\ref{fig:maxsn}.  An S/N threshold of six has previously been proven effective to
maximise the ratio of real- to spurious detections in blind emission line searches with
MUSE \citep{Herenz2017,Herenz2017a,Urrutia2018}.  This is also visualised in
Figure~\ref{fig:maxsn}, where we include spaxels in the display down to a S/N of four.
Inspecting spectra extracted in the $4 < \mathrm{S/N} < 6$ regions revealed that in all
those cases a possible emission line signature is at most marginal, while the $S/N>6$
regions are confident detections.

While our goal is to use the S/N datacube from the 3D cross-correlation as a mask to
produce an optimally extracted narrow-band image, its 2D representation in the form of
maximum S/N map in Figure~\ref{fig:maxsn} provides us with a schematic visualisation of
the main morphological features of the system.  We annotate these in
Figure~\ref{fig:maxsn}.  Most of the features were already hinted at in the display of the
velocity slices in Figure~\ref{fig:channel_maps}

Marked S/N peaks are found at the position of the LBG SSA22a-C11 in LAB\,1 and near the
LBG SSA22a-C15 in LAB\,8.  We point out that the LAB\,8 peak shows a slight offset towards
the south of SSA22a-C15.  The centre of LAB\,1 shows an extended region of high S/N, that
exhibits its peak values at LAB1-ALMA2.  This area is labelled ``central high SB
  region'' in Figure~\ref{fig:maxsn}. Interestingly, the sub-mm, [\ion{CII}] and
potentially X-Ray detected source LAB1-ALMA3 do not have an associated prominent
peak in the S/N map, and neither have the spectroscopically confirmed sources S1 and K15b
associated peaks.  But, these three sources (LAB1-ALMA3, S1, and K15b) demarcate the
central high surface-brightness region from the south-east (LAB1-ALMA3, S1) to the
north-west (K15b).  In the north-west another high surface-brightness region then curves
back to the north-east. This feature is labelled ``northern high surface-brightness
  region'' in Figure~\ref{fig:maxsn}  and does not contain known sources.  It is,
however, spatially coincident with the northern edge of the ``bubble'' that we
pointed out in Section \ref{sec:veloc-slic-lyalpha} (see Figure~\ref{fig:channel_maps}).

Our maximum S/N map also accentuates the two filamentary features that form a bridge
between LAB\,1 and LAB\,8. Moreover, the newly detected ``shell'' region in the south west
is clearly connected via a significantly detected filament to the central region of the
blob. This shell harbours a compact high-S/N knot, accompanied by more diffuse emission
knots towards the north and the south.  We also find four isolated $S/N>6$ peaks (labelled
1 to 4 in the figure) that are not connected to the central large $S/N>6$ region.  These
isolated $S/N>6$ peaks represent newly identified Ly$\alpha$ emitters in close vicinity to
the blob, and they will be analysed separately in Sect.~\ref{sec:newly-disc-isol}.  LAE 4,
to the north of LAB 8, appears to be connected by a filamentary structure to the main body
of the blob, but this potential filament is detected at lower significance than our
adopted detection threshold.  The map also hints a potential filament pointing towards LAE
3.

\subsubsection{Adaptive narrow-band image}
\label{sec:adaptive-narrow-band}

Equipped with the S/N datacube from \texttt{LSDCat} we constructed the optimal 3D
extraction mask for our adaptive narrow-band image.  We do so by summing the flux datacube
in the spectral direction only over voxels that contain at least an S/N value above 4 in
the S/N datacube. We note, that spaxels that do not contain a single voxel with $S/N > 4$
are blacked out in Figure~\ref{fig:maxsn}, thus the displayed spaxels in
Figure~\ref{fig:maxsn} can be interpreted as a 2D projection of the 3D extraction mask.
The choice of this analysis threshold is motivated by our observation that some 
$S/N > 4$ regions in Figure \ref{fig:maxsn} may contain a marginal Ly$\alpha$ signal.  The
use of a second S/N threshold that is lower than the detection threshold is also core
principle of the \texttt{LSDCat} software, which uses a detection threshold for finding
emission lines and an analysis threshold for performing measurements on the detected lines
\citep{Herenz2017}.  In order to provide a visual representation of the background noise
in source free regions we adopt the strategy from \cite{Borisova2016} and sum over 4
spectral bins.  We centre the summation around 4983.4\,\AA{} (the wavelength of Ly$\alpha$
at $z=3.1$).  The so constructed adaptive narrow-band image is displayed in
Figure~\ref{fig:adapt_nb}.  The $1\sigma$ noise of the image estimated from placing random
apertures in empty sky regions in the central parts of the image is
$4\times10^{-20}$\,erg\,s$^{-1}$cm$^{-2}$arcsec$^{-2}$.

The adaptive narrow-band image allows us to characterise the features that were pointed
out above (Figures~\ref{fig:channel_maps}) and \ref{fig:maxsn}) by Ly$\alpha$ surface
brightness\footnote{All $\mathrm{SB}_\mathrm{Ly\alpha}$ measurements were obtained within
  an circular aperture of 2\arcsec diameter.}  ($\mathrm{SB}_\mathrm{Ly\alpha}$).  We
distinguish three fragments in the shell: (1) A bright compact knot with
$\mathrm{SB}_\mathrm{Ly\alpha}\approx5\times10^{-18}$\,erg\,s$^{-1}$cm$^{-2}$arcsec$^{-2}$,
(2) a fainter more diffuse knot to the north of the bright knot with
$\mathrm{SB}_\mathrm{Ly\alpha}\approx2.8\times10^{-18}$\,erg\,s$^{-1}$cm$^{-2}$arcsec$^{-2}$,
and (3) an even more diffuse extended fragment in the south of the shell with
$\mathrm{SB}_\mathrm{Ly\alpha}\approx1.6\times10^{-18}$\,erg\,s$^{-1}$cm$^{-2}$arcsec$^{-2}$.
The filament connecting the diffuse part of the shell (labelled as shell-connecting
filament in Figure~\ref{fig:adapt_nb}) is characterised by
$\mathrm{SB}_\mathrm{Ly\alpha}\approx1\times10^{-18}$\,erg\,s$^{-1}$cm$^{-2}$arcsec$^{-2}$
emission, while the central- and northern high-surface brightness regions of LAB\,1 show
$\mathrm{SB}_\mathrm{Ly\alpha}\gtrsim1\times10^{-17}$\,erg\,s$^{-1}$cm$^{-2}$arcsec$^{-2}$.
The high-$\mathrm{SB}_\mathrm{Ly\alpha}$ regions clearly demarcate a central circular
region of lower $\mathrm{SB}_\mathrm{Ly\alpha}$
($\mathrm{SB}_\mathrm{Ly\alpha}\approx5\times10^{-18}$\,erg\,s$^{-1}$cm$^{-2}$arcsec$^{-2}$).
This apparent cavity, first seen by \cite{Bower2004}, was labelled ``bubble'' in
Figure~\ref{fig:channel_maps}.  To the north of LAB\,1 we find the two filaments
connecting LAB\,1 with LAB\,8, with the eastern one showing higher
$\mathrm{SB}_\mathrm{Ly\alpha}$
($\mathrm{SB}_\mathrm{Ly\alpha}\approx4.5\times10^{-18}$\,erg\,s$^{-1}$cm$^{-2}$arcsec$^{-2}$)
than the western one
($\mathrm{SB}_\mathrm{Ly\alpha}\approx2.5\times10^{-18}$\,erg\,s$^{-1}$cm$^{-2}$arcsec$^{-2}$).
LAB\,8 is characterised by
$\mathrm{SB}_\mathrm{Ly\alpha}\gtrsim1\times10^{-17}$\,erg\,s$^{-1}$cm$^{-2}$arcsec$^{-2}$,
and for the filament towards LAE\,4 in the north we measure
$\mathrm{SB}_\mathrm{Ly\alpha}\approx1\times10^{-18}$\,erg\,s$^{-1}$cm$^{-2}$arcsec$^{-2}$.
However, in the northern part of the image, due to the lower number of contributing
exposures (see Figure~\ref{fig:exp_map}), the background noise of the image is higher
($\sigma \approx 6\times10^{-20}$\,erg\,s$^{-1}$cm$^{-2}$arcsec$^{-2}$).  With
$\mathrm{S/N}_\mathrm{max}$ values $\sim 5 - 5.5$ after the 3D cross-correlation procedure
it missed the adopted detection threshold and we regard this filament conservatively only
as a tentative feature.

\subsubsection{Size and total Ly$\alpha$ luminosity}
\label{sec:size-total-lyalpha}

We define the size of the unveiled LAB\,1 + LAB\,8 + shell structure as the area of the
region above the adopted detection threshold of $\mathrm{S/N}_\mathrm{max} = 6$ (white
contour in Figure~\ref{fig:maxsn}), excluding the isolated LAEs.  In terms of surface
brightness this threshold corresponds to a limit of
$\approx 6 \times 10^{-19}$\,erg\,s$^{-1}$cm$^{-2}$arcsec$^{-2}$. The corresponding limit
in surface luminosity is $8.7 \times 10^{38}$erg\,s$^{-1}$kpc$^{-2}$.  At this threshold
Ly$\alpha$ emission from the LAB\,1/LAB\,8 structures covers an area of
553\,arcsec$^2$. This corresponds to a projected surface of $3.2\times10^4$kpc$^2$ at
$z=3.1$.  The total measured Ly$\alpha$ flux from the structure is
$F_\mathrm{Ly\alpha} = 1.73\times10^{-15}$\,erg\,s$^{-1}$cm$^{-2}$, which corresponds to a
Ly$\alpha$ luminosity of $L_\mathrm{Ly\alpha} = 1.45\times10^{44}$erg\,s$^{-1}$.

While the Ly$\alpha$ structure revealed here is enormous, two even larger Ly$\alpha$
nebulae, namely the ``Slug'' nebulae with an extent of $\approx500$\,kpc
\citep{Cantalupo2014} and the MAMMOTH-1 nebula with an extent of $\approx440$\,kpc
\citep{Cai2017}, are known.  While the ``Slug'' nebulae surrounds a luminous
($L_\mathrm{bol} = 10^{47.3}$\,erg\,s$^{-1}$) type-I quasar, the MAMMOTH-1 nebulae
surrounds a relatively faint broad-band source whose emission line spectrum appears to be
consistent with a quasar.  Nevertheless, both the ``Slug'' and the MAMMOTH-1 nebulae are
also characterised by a factor of 3.4 and 1.5 higher luminosity than the LAB\,1 + LAB\,8 +
shell structure, respectively.  Both nebulae are at $z\approx2.3$, and thus the effect of
cosmological surface brightness dimming is reduced by a factor of $2.4$ compared to our
observations.  While no limiting surface brightness for the ``Slug'' observations is
published, \cite{Cai2017} report a surface brightness detection limit of
$4.8\times10^{-18}$\,erg\,s$^{-1}$cm$^{-2}$arcsec$^{-2}$ for MAMMOTH-1. This translates to
limiting surface luminosity of $6.8 \times 10^{38}$erg\,s$^{-1}$kpc$^{-2}$ , which is
comparable to our physical limit.  The projected maximum extend of our structure, measured
from the northernmost tip in LAB\,8 to the southernmost point in the shell, is
45.4\arcsec{} or, correspondingly, 346.3\,kpc in projection, which is a factor of
$\approx 0.8$ smaller that the extent of the MAMMOTH-1 nebula.  Similar to the LAB\,1 -
LAB\,8 structure, MAMMOTH-1 exists in an extreme overdense region of the universe.

\subsubsection{Photometric centre and photometric principal axis}
\label{sec:phot-centre-posit}

We applied the formalism of image moments \citep[][]{Hu1962,Stobie1980,Stobie1986} to the
adaptive narrow-band image to calculate the photometric centre as well as the angle of the
principal axis, $\theta_\mathrm{PA}$, of the LAB.  In pixel-coordinates $(x,y)$ of the
adaptive narrow-band image $I_\mathrm{xy}$ the photometric centre
$(\overline{x},\overline{y})$ is defined as
\begin{equation}
  \label{eq:xycen}
  (\overline{x},\overline{y}) = \left ( \frac{\sum_{xy} I_{xy} x}{\sum_{xy}
      I_{xy}} \, , \,  \frac{\sum_{xy} I_{xy} y}{\sum_{xy}
      I_{xy}} \right ) \;\text{,}
\end{equation}
and the  angle of the principal axis is defined as 
\begin{equation}
  \label{eq:pa}
  \theta_\mathrm{PA} = \frac{1}{2} \arctan \left ( \frac{2 \,
      \overline{xy}}{\overline{x^2} - \overline{y^2}} \right ) 
\end{equation}
with
\begin{equation}
  \label{eq:3}
  \left ( \, \overline{x^2}, \overline{y^2} \, \right  ) = \left ( \frac{\sum_{xy} I_{xy} x^2}{\sum_{xy}
      I_{xy}} - \overline{x}^2 \, , \, \frac{\sum_{xy} I_{xy} y^2}{\sum_{xy}
      I_{xy}} - \overline{y}^2 \right ) 
\end{equation}
and
\begin{equation}
  \label{eq:1}
  \overline{xy} = \frac{\sum_{xy} I_{xy} xy}{\sum_{xy} I_{xy}} -
  \overline{x}\,\overline{y} \;\text{.}
\end{equation}
For these calculations we only considered pixels in the narrowband image $I_{xy}$ that
have a corresponding pixel above a S/N of six in the maximum S/N map.  Moreover, the
definition of the angle of the principal axis in Eq.~(\ref{eq:pa}) is such that 0$^\circ$
corresponds to the axis from S to N, and that the angle increases
anticlockwis to the east.  The in this way obtained photometric centre,
converted to celestial coordinates (J2000), is located at
RA$=$22$^\mathrm{h}$17$^\mathrm{m}$25.94$^\mathrm{s}$,
Dec$=$+00$^\circ$12\arcmin{}39.338\arcsec{}, and for the angle of the principal axis we
find $\theta_\mathrm{PA}= 20.9^\circ$ east of north.

We show the position of the photometric centre in Figure~\ref{fig:adapt_nb} by a cyan
cross.  As can be seen, the photometric centre is located slightly west to the ``bubble''.
We also indicate the principal axis in Figure~\ref{fig:adapt_nb} by a dashed cyan line.
Figuratively speaking, the principal axis is the axis along which the blob appears most
elongated.  Formally, it describes axis along which the variance in flux is maximised.
For a light distribution of elliptical shape, the so defined principal axis would be
oriented along the major axis of the ellipse.  Thus, our $\theta_\mathrm{PA}$ measurement
is comparable to the measurements of LAB position angles via ellipse fitting in
\cite{Erb2011}.  We discuss the alignment between principal axis and gas kinematics in
Sect.~\ref{sec:large-scale-gas}.

\subsection{Moment based maps of the Ly$\alpha$ line profile}
\label{sec:spectr-morph-lyalpha}

\begin{figure*}
  \centering
  \includegraphics[width=\textwidth]{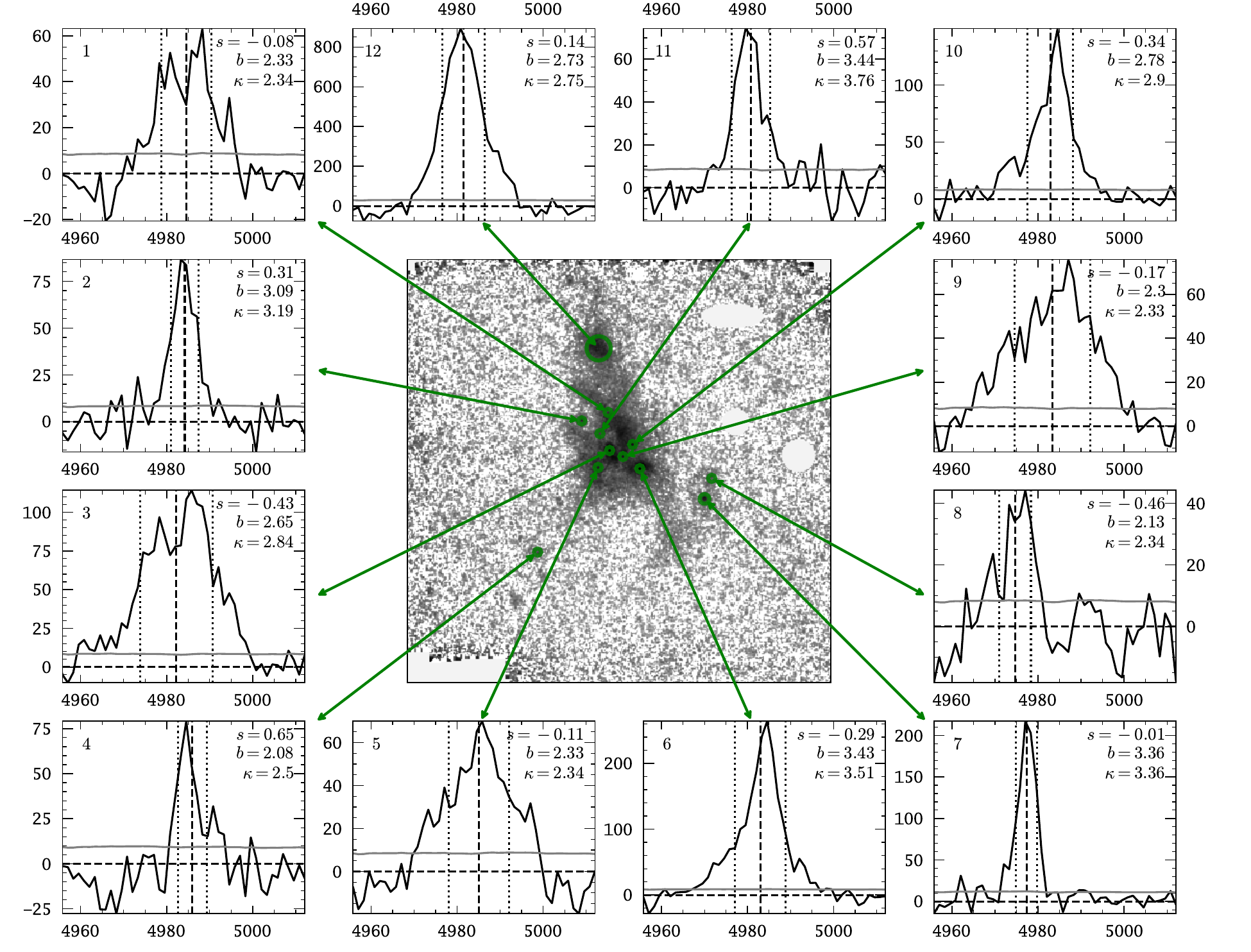}
  \caption{Examples of the variety and complexity of the Ly$\alpha$ line profiles
    encountered in LAB\,1.  All profiles are extracted in circular apertures of
    1.2\arcsec{} diameter, except for LAB\,8 where a 4\arcsec{} diameter aperture was
    used.  The image in the centre is the adaptive narrow-band image shown in
    Figure~\ref{fig:adapt_nb}, but in a logarithmic scale.  Green circles represent the
    extraction apertures with lines connecting to the individual panels that display the
    profiles.  Four of the twelve panels show Ly$\alpha$ profiles at the position of known
    galaxies: LAB1-ALMA3 in panel 5, SSA22a-C11 in panel 6, LAB1-ALMA1 and LAB1-ALMA2 in
    panel 9, and SSA22a-C15 (LAB\,8) in panel 12. For each profile the wavelength axis (in
    \AA{}) is fixed and centred on $z_\mathrm{Ly\alpha}=3.1$, but the axis displaying the
    intensity (in erg\,s$^{-1}$cm$^{-2}$\AA{}$^{-1}$) is scaled to encompass the maximum
    flux value of each profile.  We also indicate in each panel the flux-weighted central
    moment (Eq.~\ref{eq:1mom}, dashed line), and the non-parametric measure for
      the width of the line obtained from the second flux-weighted moment (from
    Eq.~\ref{eq:kmom} with $k=2$, dotted lines).  Moreover, we display in the upper right
    corner of each panel the non-parametric descriptive measures skewness $s$
    (Eq.~\ref{eq:skew}), kurtosis $\kappa$ (Eq.~\ref{eq:kurtosis}), and bi-modality $b$
    (Eq.~\ref{eq:bim}) -- see text for details.  All non-parametric descriptive statistics
    are computed by considering only the range of connected positive spectral bins blue-
    and red-wards of the peak. }
  \label{fig:exspec}
\end{figure*}

\begin{figure*}
  \centering
\includegraphics[width=\textwidth]{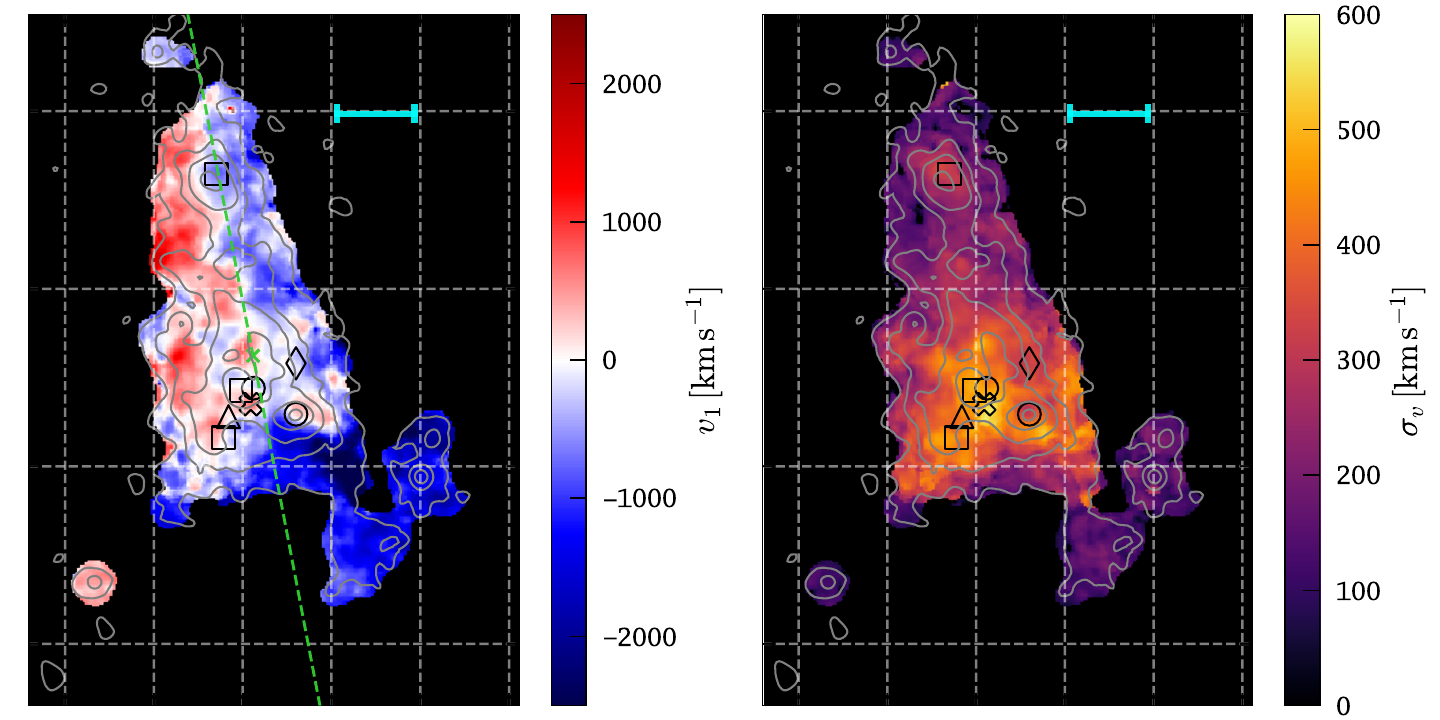}
\caption{Apparent line of sight-velocity (\emph{left panel}) and apparent velocity
  dispersion (\emph{right panel}) as measured from Ly$\alpha$ using the first and second
  flux weighted moments (Eq.~\ref{eq:2} and Eq.~\ref{eq:4}).  Before the moment-based
  analysis was carried out, each layer of the datacube was spatially smoothed with a
  $\sigma = 0.8$\arcsec{} Gaussian kernel.  In each spaxel only spectral bins above a S/N
  threshold of 4 in the \texttt{LSDCat} S/N datacube were used in the summations in
  Eq.~(\ref{eq:1mom}) and Eq.~(\ref{eq:kmom}).  Moreover, the displayed map shows only
  spaxels that have a maximum $\mathrm{S/N}>6$.  Thin grey contours indicate
  surface-brightness levels
  $\mathrm{SB}_\mathrm{Ly\alpha} = [200 , 100 , 50 , 25 , 8.75]\times
  10^{-19}$\,erg\,s$^{-1}$cm$^{-2}$\,arcsec$^{-2}$ as measured in the adaptive narrow-band
  image (Figure~\ref{fig:adapt_nb}).  The positions of confirmed galaxies within the blob
  are indicated with the same symbols as in Figure~\ref{fig:ocas}.  The photometric centre
  and the principal axis of the blob (see Sect.~\ref{sec:adapt-lyalpha-image} and
  Figure~\ref{fig:adapt_nb}) are indicated by a green cross and a green dashed line,
  respectively.  The horizontal cyan line in the upper right of each panel
  indicates a projected proper distance of 50 kpc. } \vspace{0.5em}
  \label{fig:moms}
\end{figure*}

\begin{figure*}
  \includegraphics[width=\textwidth,trim=20 20 10 20,clip=true]{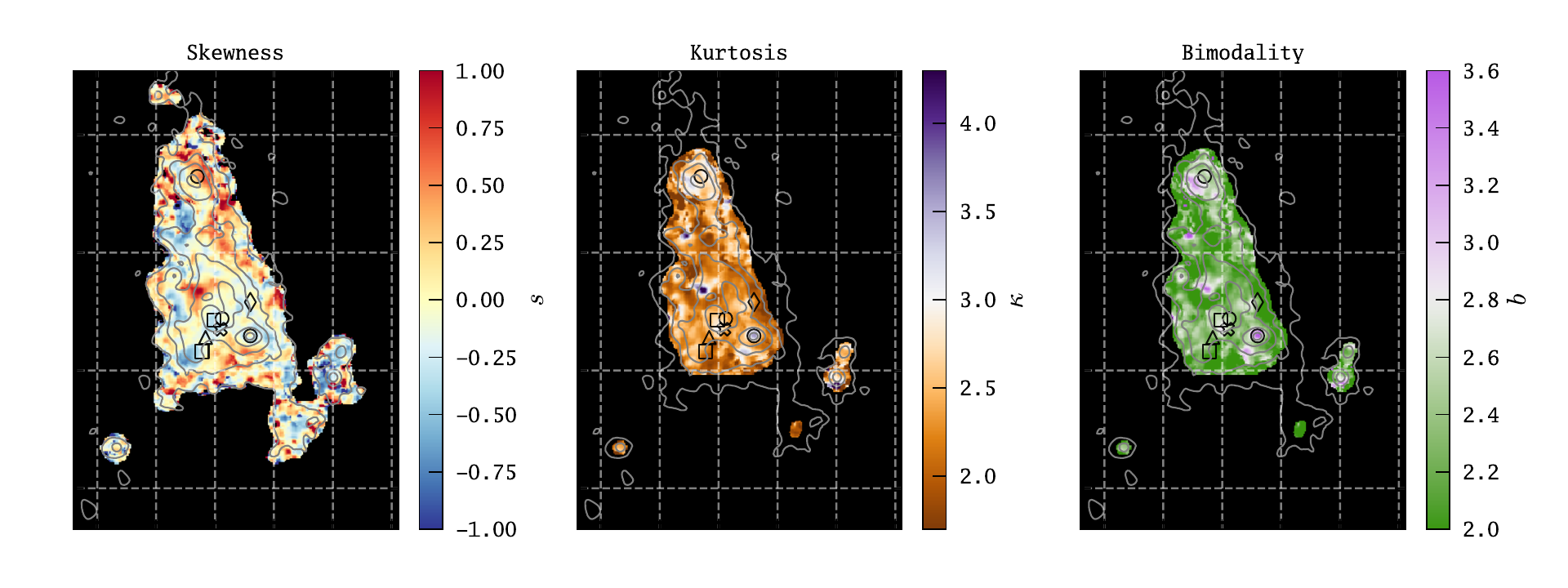}
  \caption{Maps of higher-moment based non-parametric measurements for visualising the
    varying complexity of the Ly$\alpha$ profiles throughout the blob.  We show maps of
    skewness $s$ (\emph{left panel}), kurtosis $\kappa$ (\emph{centre panel}) and
    bimodality $b$ (\emph{right panel}), as defined in Eq.~(\ref{eq:skew}),
    Eq.~(\ref{eq:kurtosis}), and Eq.~(\ref{eq:bim}), respectively (see text for details).
    The displayed map in the left panel (middle and right panel) shows only spaxels that
    have a maximum $\mathrm{S/N}>6$ (maximum $\mathrm{S/N}>13$).  Skewness $s$ quantifies
    the asymmetry around the first central moment $m_1$ (Eq.~\ref{eq:1mom}), with $s=0$
    (yellow) indicating symmetric profiles, while $s<0$ (shades of blue) indicate that the
    profile shows a larger tail towards the blue and $s>0$ (shades of red) indicate that
    the profiles show a larger tail towards the red. Kurtosis quantifies the strength of
    wings (symmetric profiles) or tails (asymmetric profiles), with $\kappa=3$
    (white) indicating that the wings of the profile are comparable to a Gaussian wings,
    while $\kappa>3$ (shades of purple) indicate larger tail extremity of the profiles
    (shades of brown), and $\kappa < 3$ indicate that there is less power in the tails
    compared to a Gaussian .  Bimodality $b$ is an attempt to quantify whether the
    profiles are double component ($b \lesssim 2.6$, green colours) or single component
    ($b \gtrsim 3$, shown in violet) profiles \citep{Remolina-Guti2019}.  However, given
    the presence of noise and finite spectral resolution a clear discriminatory power
    between single component and double component is not given by this measure in range
    $2.6 \lesssim b \lesssim 3$ (light green and light violet colours).  Contours and
    symbols are the same as in Figure~\ref{fig:moms}.}
  \label{fig:hm}
\end{figure*}

By visually inspecting the Ly$\alpha$ spectral profile as a function of position with the
QFitsView software\footnote{The QFitsView software is publicly available via the
  Astrophysics Source Code Library: \url{http://ascl.net/1210.019}.}
\citep{Davies2010,Ott2012}, we find that the line profile complexity varies strongly
throughout the blob.  We illustrate this by showing a selection of representative profiles
in Figure~\ref{fig:exspec}.  As can be seen, in some region the profiles appear very broad
and with a dominating peak (e.g. panel 12 in Figure~\ref{fig:exspec}), while
other regions are characterised by clearly double- (e.g. panel 3) or even
multi-component profiles (e.g. panel 5 or panel 9).  The isolated LAEs
in the outskirts (example in panel 4, see also Figure~\ref{fig:laes}), or in the
shell-like region (panel 7) show narrower Ly$\alpha$ profiles.

The varying complexity of the Ly$\alpha$ profiles as a function of position prohibit
parametric fits of a simple model to the spaxels of the datacube in order to create maps
of, e.g., the velocity centroid ($v_r$) or line-width ($\sigma_v$).  Such an analysis was
presented for the much shallower SAURON data of LAB\,1 \citep{Bower2004,Weijmans2010}, but
the increased sensitivity and resolution of our MUSE data warrant a different approach.
We thus resort on a moment-based non-parametric analysis. Our method is rooted in
descriptive statistics \citep[e.g.][]{Ivezic2014}, but we need to account for two
differences when describing spectroscopic line profiles instead of statistical data with
such an ansatz.  First, formal validity of the summarising parameters is only given when
positive values are considered in the calculations.  Second, the presence of noise and the
usage of small sets of input values can lead to non-trivial biases in moment-based
quantities, especially if the analysed profiles are of low S/N.  To ensure positive values
and in order to minimise low-S/N biases we first applied three preprocessing steps to the
continuum-subtracted datacube:
\begin{enumerate}
\item We used the 3D mask that was already used for the creation of the adaptive
    narrow band image in Sect.~\ref{sec:maximum-sn-image}.  We recall that this 3D mask
    was constructed by thresholding the matched-filtered datacube with $\mathrm{S/N > 4}$.
    Voxels that do not fulfil this criterion will be set to 0 in the analysis.
\item We smoothed each layer of the flux datacube with a circular 2D Gaussian
    ($\sigma$ 0.8\arcsec{}).  This is done to reduce the spaxel-to-spaxel noise in the
    final maps, especially in low surface-brightness regions.  This value significantly
    improved the S/N ratio of the Ly$\alpha$ profiles in the filaments and low
    surface-brightness regions of the blob.
\item We used a 2D mask by thresholding the maximum S/N map shown in
    Figure~\ref{fig:maxsn} to exclude spaxels were only a very small number of spectral
    bins would contribute to the resulting moments.  After visual inspection we set this
    ``display threshold'' to $\mathrm{S/N}_\mathrm{max}=6$, i.e.\, the analysed regions are
    exactly the regions that we regarded as confident detections in
    Sect.~\ref{sec:maximum-sn-image}.
\end{enumerate}
After these preparatory steps we created a 2D array of the central flux-weighted moment
(first moment) from the processed datacube voxels $F_{xyz}$,
\begin{equation}
  \label{eq:1mom}
  m_1^{xy} = \frac{\sum\nolimits_z z \, F_{xyz}}{ \sum\nolimits_z F_{xyz} } \;\text{,}
\end{equation}
as well as 2D arrays $m_k^{xy}$ of the $k$-th flux-weighted moments,
\begin{equation}
  \label{eq:kmom}
  m_k^{xy} = \frac{\sum_z (z - m_1^{xy})^k \, F_{xyz}}{\sum_z F_{xyz}} \;\text{,}
\end{equation}
for $k=2$, $k=3$ and $k=4$.  In Eqs.~(\ref{eq:1mom}) and~(\ref{eq:kmom}), as well as in
the following equations below, $x$ and $y$ denote the indices of spatial axes of the flux
datacube $F$, while $z$ indexes the spectral direction.

The first moment map resulting from Eq.~(\ref{eq:1mom}) directly translates into a
line-of-sight velocity map
\begin{equation}
  \label{eq:2}
  v_1^{xy} = c \times \left (
    \frac{\lambda_\mathrm{vac}(m_1^{xy})}{\lambda_\mathrm{Ly\alpha}} - z_\mathrm{LAB1} - 1
  \right ) \; \text{,}
\end{equation}
where $c$ is the speed of light, $\lambda_\mathrm{vac}(m_1^{xy})$ is the non-linear
translation between MUSE spectral pixel coordinate and vacuum wavelength\footnote{We use
  the air-to-vacuum wavelength conversion that has been adopted in the Vienna Atomic Line
  Database \citep{Ryabchikova2015}:
  \url{https://www.astro.uu.se/valdwiki/Air-to-vacuum\%20conversion}. } and
$z_\mathrm{LAB1}$ is the systemic redshift of LAB\,1.  For this translation we fix the
systemic redshift of LAB\,1 to $z_\mathrm{LAB1}=3.1$, in agreement with known redshifts of
the galaxies within the blob.  The so created $v_1^{xy}$ map is shown in the left panel of
Figure~\ref{fig:moms}.  There we also show the photometric centre and the principal axis
that were computed from the adaptive narrow band image as described in the previous
section.  We point out that the principal axis is oriented orthogonal to the direction of
the apparent large scale velocity gradient.  This feature will be further discussed in
Sect.~\ref{sec:morpho-kin}.

By taking the square root of second moment map (Eq.~\ref{eq:kmom}, with $k=2$) we compute
a map that provides a measure of the width of the spectral profiles:
\begin{equation}
  \label{eq:4}
  \sigma_v^{xy} = c \times \frac{\lambda_{\mathrm{vac}} \left (m_1^{xy} + \sqrt{m_2^{xy}}/2
    \right ) - \lambda_{\mathrm{vac}} \left ( m_1^{xy} - \sqrt{m_2^{xy}}/2  \right )
  }{\lambda_\mathrm{vac}(m_1^{xy})} \; \text{.}
\end{equation}
This $\sigma_v$ map is shown in the right panel of Figure~\ref{fig:moms}.  We
express the width of the profiles $\sigma_v$ in km\,s$^{-1}$, but caution that this
measurement can not be directly interpreted as velocity dispersion as it is often done for
non-resonant emission lines.  For example, double- or multiple peaked profiles generally
have larger second moments than single peaked profiles.  Moreover, radiative transfer
effects are also known to broaden the single-peaked Ly$\alpha$ line profiles when compared
to non-resonant emission lines (see also Sect.~\ref{sec:detect-ionh-lambd}).  We thus call
the second moment based measure ``apparent velocity dispersion''.  Lastly, the spectral
resolution of MUSE also has an effect on the apparent velocity dispersion.  Given the
complexity of the profiles and the non-parametric nature of our measurement the effect of
broadening the profiles via convolution with the spectrograph's line spread function is
not easily quantifiable.  The measured instrumental width for MUSE at 4980\AA{} is
$\sigma_\mathrm{inst} \approx 75$\,km\,s$^{-1}$ \citep{Bacon2017}.  As a figure of merit
estimate, this translates into resolution corrections $\sigma_\mathrm{corr}$ of
-34\,km\,s$^{-1}$, -20\,km\,s$^{-1}$, -15\,km\,s$^{-1}$, and -10\,km\,s$^{-1}$ for
apparent velocity dispersions of 100\,km\,s$^{-1}$, 150\,km\,s$^{-1}$, 200\,km\,s$^{-1}$,
and 300\,km\,s$^{-1}$, respectively, if the observed line profiles and the line spread
function are well approximated by a Gaussian profile i.e.
\begin{equation}
  \label{eq:sigmac}
  \sigma_\mathrm{corr} = \sigma_v - \sqrt{\sigma_v^2 - \sigma_\mathrm{inst}^2}\;\text{.}
\end{equation}

Creating such moment-based maps of line-of-sight velocity and apparent velocity dispersion
is common in the analysis of synthesised datacubes from radio-interferometric
21\,cm observations of galaxies \citep[e.g.][Section 10.5.4]{Thompson2017}.  It now is
also routinely used in the analyses of extended Ly$\alpha$ nebulae surrounding quasars
\citep[e.g.][]{Borisova2016,Arrigoni-Battaia2018,Arrigoni-Battaia2019,Drake2019}.
Moreover, recent theoretical work by \cite{Remolina-Guti2019} and \cite{Smith2019} resort
on a moment-based analysis in the analysis of Ly$\alpha$ profiles from Ly$\alpha$
radiative transfer simulations.  However, in order to create maps that characterise the
varying complexity of the Ly$\alpha$ profile as a function of position in the blob, we use
measurements involving higher-moments.  \cite{Remolina-Guti2019} suggest to use the
skewness $s$, the kurtosis $\kappa$, and the bimodality $b$.  These measurements will be
detailed in the following.  To provide a visual guide on how to interpret these quantities
we also display their values next to the example profiles from the blob shown in
Figure~\ref{fig:exspec}.

We calculate a map of the Ly$\alpha$ profile skewness $s$ via
\begin{equation}
  \label{eq:skew}
  s^{xy} = m_3^{xy} / \left ( m_2^{xy} \right )^{3/2} \;{.}
\end{equation}
The so defined skewness ($-1 \leq s \leq 1$) quantifies the asymmetry of the spectral
profile with respect to $m_1$.  If $s\simeq 0$ the profile is symmetric around $m_1$
(panel 1 in Figure~\ref{fig:exspec}), while for $s>0$ a tail is found redwards of $m_1$
(e.g. panels 4 and 11 in Figure~\ref{fig:exspec}) and for $s<0$ a tail is bluewards of
$m_1$ (e.g. panels 6, 8, and 10 Figure~\ref{fig:exspec}).  We show our computed map for
$s^{xy}$ in the left panel of Figure~\ref{fig:hm}.

We note that other definitions than Eq.~(\ref{eq:skew}) have been used in the literature
to quantify the asymmetric Ly$\alpha$ line-profile morphology of LAEs.  For example,
\cite{Shimasaku2006} quantified the observed skewness in spectral profiles from Lyman
$\alpha$ emitting galaxies by multiplying the definition given in Eq.~(\ref{eq:skew}) with
a measure of the width of the line, however, we prefer to not entangle those two
quantities.  Other authors \citep[][]{Mallery2012,U2015} quantified skewness $s$ by
fitting a skewed Gaussian profile.  But this approach does not capture the complex
Ly$\alpha$ spectral profiles seen here in LAB\,1.  Additionally, \cite{Childs2018} showed
recently that the skewness values derived from fitting an asymmetric Gaussian do not
accurately capture the true skewness of Ly$\alpha$ profiles in the presence of finite
spectral resolution and background noise. Two other alternative definitions have been put
forward by \cite{Dawson2007}.  These authors quantify asymmetry either via the ratio of
flux blue- and redwards of the peak or via the ratio of the widths than encompass 90\% of
the flux blue- and redwards of the peak.  However, given that the profiles in LAB\,1
sometimes show multiple peaks at substantial spectral distance (e.g. panel 1 and 3 in
Figure~\ref{fig:exspec}), quantifying asymmetry around the higher peak would exaggerate
the skew measure compared to the visual perception of symmetry in those profiles.

We obtain a map of the kurtosis of the Ly$\alpha$ profiles via
\begin{equation}
  \label{eq:kurtosis}
  \kappa^{xy} = \frac{m_4^{xy}}{\left ( m_2^{xy} \right )^2} \geq 1 \text{.}
\end{equation}
Kurtosis quantifies how much flux is in the wings of the profiles in comparison to the
wings of Gaussian profile (i.e. their tail extremity).  For $\kappa = 3$ the tails are
comparable to the Gaussian profile, while profiles with $\kappa > 3$ show more pronounced
tails (e.g. panels 6 and 11 in Figure~\ref{fig:exspec}), while $\kappa < 3$ indicates the
absence of pronounced tails (e.g. panels 1, 5, and 9 in Figure~\ref{fig:exspec}).  Of
course, only wings that are significantly above the noise can contribute to this
statistic.  As a corollary, regions of low S/N are biased towards to low kurtosis values.
We avoid these biases by increasing the display threshold to
  $\mathrm{S/N}_\mathrm{max} = 13$.  We show the resulting map for $\kappa^{xy}$ in the
centre panel of Figure~\ref{fig:hm}.

Following \cite{Remolina-Guti2019} we calculate a map of the bi-modality of the
Ly$\alpha$ line profiles using
\begin{equation}
  \label{eq:bim}
  b^{xy} =  \kappa^{xy} - \left ( s^{xy}
  \right )^2  \geq 1 \;\mathrm{.}
\end{equation}
\cite{Remolina-Guti2019} introduced this quantity to discriminate whether their Ly$\alpha$
radiative transfer models result in single- or double component profiles.  We point out
that this measure is not a formal statistical test for bi-modality, but it can capture the
visual appearance of the Ly$\alpha$ profile morphologies.  We find that for
$b \lesssim 2.6$ profiles appear mostly to have clearly distinct double component
structures (e.g. panels 1 and 4 in Figure~\ref{fig:exspec}, but see panel 5 and 9), while
profiles with $b \gtrsim 3$ appear single peaked (e.g. panel 2, 6, and 7 in
Figure~\ref{fig:exspec}).  However, some $b \gtrsim 3$ profiles may also have a
subdominant second component, that is mainly contributing to the kurtosis (e.g. panel 11
in Figure~\ref{fig:exspec}).  In the range $2.6 \lesssim b \lesssim 3$, however, the
discriminatory power of $b$ appears not strong, and visual inspection of those profiles
indicates a high complexity with possible multiple components or peaks (see e.g. panels 3
and 10 in Figure~\ref{fig:exspec}).  Despite its potential lack of accuracy, qualitatively
$b$ captures the visual complexity of the profiles, with higher values indicating simple
single component profiles and lower values indicating more complex profiles, and with the
lowest values often corresponding to the presence of double component profiles.  Moreover,
since the $\kappa$ is biased towards low values in regions of low S/N, also $b$ will be
biased low in those regions.  Thus, we hide the biased regions by setting the
  display threshold to $\mathrm{S/N}_\mathrm{max} = 13$.  We show our computed map for
$b^{xy}$ in the right panel of Figure~\ref{fig:hm}.

It can be seen that most of the blob shows low values of $b$ indicative of double
component Ly$\alpha$ profiles.  This impression is also on par with our visual inspection
of the line profile variations throughout the blob.  Moreover, in the central high
surface-brightness region of LAB, where also the broadest profiles are observed, we obtain
$b$ values in the intermediate range -- these profiles often appear to exhibit a
high-degree of complexity.  Lastly, only a few small island regions can be characterised
by high values of $b$.  These regions show clearly distinct single peaked profiles, often
with very pronounced tails.  We will describe and discuss the here derived and presented
maps from the moment based analysis further in Sect.~\ref{sec:morpho-kin}.

\subsection{Newly discovered faint LAEs at $z\approx 3.1$ in
  proximity to LAB\,1}
\label{sec:newly-disc-isol}

\begin{table}
  \caption{Newly detected faint $z=3.1$ LAEs around LAB1.}
  \centering \small
  \begin{tabular}{ccccccc}
    \hline \hline \noalign{\smallskip}
    ID & RA                                             & Dec & $R_\mathrm{kron}$ &$\log F_\mathrm{Ly\alpha}$ & $\Delta v$    & $\sigma_v$ \\ \hline
    1  & 22$^\mathrm{h}$17$^\mathrm{m}$27.08$^\mathrm{s}$  & 0$^\circ$12\arcmin{}12.2\arcsec{}  & 0.6\arcsec{} & -17.2  & +3690 & 127 \\
    2  & 22$^\mathrm{h}$17$^\mathrm{m}$28.01$^\mathrm{s}$  & 0$^\circ$12\arcmin{}13.7\arcsec{}  & 0.7\arcsec{} & -17.1  & -1445 & 94\\
    3  & 22$^\mathrm{h}$17$^\mathrm{m}$26.83$^\mathrm{s}$  & 0$^\circ$12\arcmin{}20.1\arcsec{}  & 1.5\arcsec{} & -16.7  & +804 & 222 \\
    4  & 22$^\mathrm{h}$17$^\mathrm{m}$26.48$^\mathrm{s}$  & 0$^\circ$13\arcmin{}05.0\arcsec{}  & 1.3\arcsec{} & -16.6  & -275 & 105 \\ \hline \hline
  \end{tabular}
  \tablefoot{$F_\mathrm{Ly\alpha}$ is the Ly$\alpha$ line flux in erg\,s$^{-1}$cm$^{-2}$
    measured within a $2.5\times R_\mathrm{kron}$ aperture on the adaptive narrow-band
    image, $\Delta v$ is the velocity difference in km\,s$^{-1}$ with respect to $z=3.1$,
    and $\sigma_v$ is the measured width of the line in km\,s$^{-1}$ (not
      corrected for instrumental dispersion).  $\Delta v$ and $\sigma_v$ have
    been computed using the first and second flux-weighted moments (see
    Sect.~\ref{sec:spectr-morph-lyalpha}).}
  \label{tab:laes}
\end{table}

\begin{figure}
  \centering
 \includegraphics[width=0.49\textwidth]{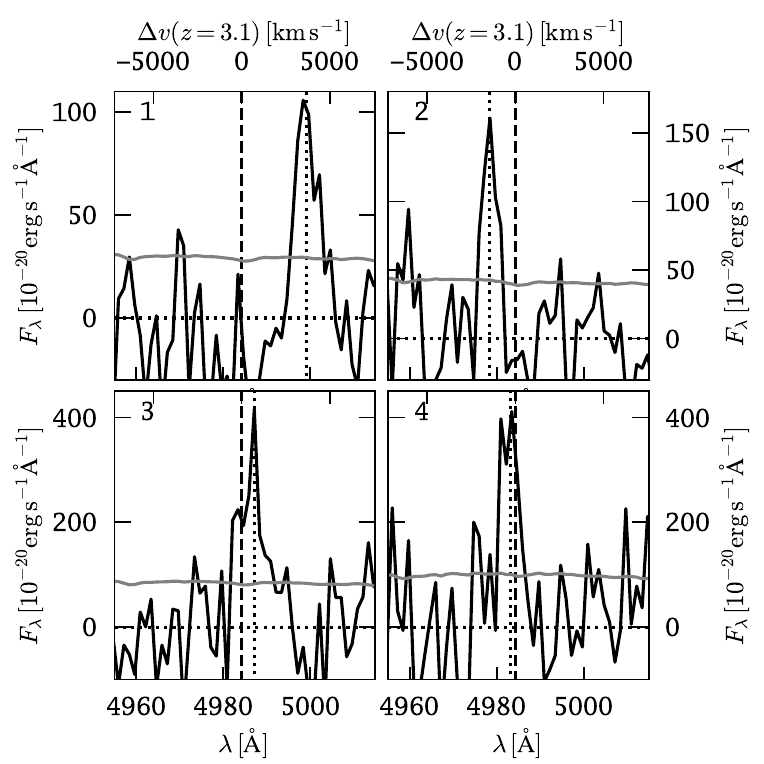}
 \caption{Spectral profiles of the newly discovered LAEs 1 -- 4 (as labelled in
   Figure~\ref{fig:maxsn}, clockwise from top-left to bottom-right). Spectra
   (black lines) have been extracted within a circular aperture of radius
   $R_\mathrm{kron}$ (see Table~\ref{tab:laes}).  The propagated error spectrum
   from the variance cube in this aperture is shown as a grey line.  The
   vertical dashed line indicates $z_\mathrm{Ly\alpha}=3.1$, whereas the
   vertical dotted lines indicate the measured redshifts from the profiles (see
   Table~\ref{tab:laes}). }
  \label{fig:laes}
\end{figure}

As mentioned in Section \ref{sec:adapt-lyalpha-image}, our S/N map revealed four
detections that are not embedded in the extended Ly$\alpha$ radiation from the blob.  We
labelled those sources 1 -- 4 in Figure~\ref{fig:maxsn}.  These sources are detected with
S/N$>6$ in the \texttt{LSDCat} cross-correlated datacube.  Formally there exists one more
detection with S/N$>6$ at $z=3.1$ close to the eastern border of Figure~\ref{fig:maxsn},
however this detection turned out to be an artefact near the edge of our field of view.

The coordinates, Kron-radii, and fluxes of the newly detected LAEs are listed in
Table~\ref{tab:laes}.  These measurements have been obtained with the \texttt{LSDCat}
software \citep{Herenz2017}.  In Figure~\ref{fig:laes} we show the spectral profiles of
the detections.  These 1D spectra have been extracted within a circular aperture of radius
$R_\mathrm{kron}$.  No other lines are detected at these positions and thus we are
confident that the sources are LAEs in physical proximity to the blob.
Additionally, two of the line-profiles (LAE 3 \& LAE 4) are reminiscent
of the characteristic red-asymmetric line-profiles seen typically in LAEs
\citep[e.g.][]{Dawson2007,Yamada2012}.  At $z=3.1$ the range of the measured fluxes is
$\log F_\mathrm{Ly\alpha} [\mathrm{erg}\,\mathrm{s}^{-1}\mathrm{cm}^{-2}] = -17.2 \dots
-16.6$, which corresponds to Ly$\alpha$ luminosities
$\log L_\mathrm{Ly\alpha}[\mathrm{erg}\,\mathrm{s}^{-1}] = 41.7 \dots 42.3$.  Hence those
galaxies occupy the faint-end ($L_\mathrm{Ly\alpha} < L_\mathrm{Ly\alpha}^*$) of the LAE
luminosity function \citep{Drake2017a,Drake2017,Herenz2019} and are thus below the
detection limit of classical narrow-band imaging surveys.

From the spectral profiles we measure the LAEs redshifts using the first flux-weighted
moment (Eq.~\ref{eq:1mom}).  These redshifts are indicated as a vertical dotted
lines in Figure~\ref{fig:laes}.  We list the velocity difference $\Delta v$ with respect
to $z=3.1$ in Table~\ref{tab:laes}.  The two galaxies 1 and 3 south-east of the blob show
large positive $\Delta v$.  In fact, their redshifts appear to be a continuation of the
overall west-to-east line-of-sight velocity gradient seen in the blob.  For such large
values of $\Delta v$ radiative transfer effects are unlikely the main cause for the
redshift offsets.  We speculate, that the peculiar motion of those galaxies are driven by
the gravitation potential of LAB\,1's dark matter halo.  The peculiar motion of the
compact sources embedded in the northern part of the south-western shell-like structure
(labelled as ``knots'' in Figure~\ref{fig:maxsn}) could also be explained by this
scenario.  Moreover, the small blue-shift of our LAE 4 to the north of LAB\,8 appears
consistent with a smooth continuation of the overall blob velocity field.  However, the
base of the blob's filament which points towards the LAE 3 shows blue-shifts and thus
deviates from a smooth velocity-field continuation.  As we will discuss in more detail in
Sect.~\ref{sec:disc}, such small scale modulations of a velocity field could be
interpreted as peculiar motions of individual galaxies or filamentary cooling flows.
Lastly, the eastern-most galaxy (2) is significantly blue-shifted and does not follow any
trend.  This galaxy might thus be at a larger distance from LAB\,1's halo and thus not
subject to its gravitational potential.

We quantify the line widths of the LAEs from the square root of the second flux-weighted
moment (Eq.~\ref{eq:4}).  As already discussed in Sect.~\ref{sec:spectr-morph-lyalpha},
the moment-based $\sigma_v$ measurement is not readily corrected for the instrumental
resolution.  Nevertheless, compared to the complexity of the line profiles seen in the
blob, the profiles of the isolated LAEs appear relatively simple, hence the prescription
for the figure-of-merit estimate in Eq.~(\ref{eq:sigmac}) provides a valid approximation
here.  The so corrected Ly$\alpha$ line-widths are 102\,km\,s$^{-1}$, 57\,km\,s$^{-1}$,
208\,km\,s$^{-1}$, and 73\,km\,s$^{-1}$ for our LAEs 1, 2, 3, and 4, respectively.  These
line-widths are within the range of typical $\sigma_v$ values obtained for a sample of
brighter LAEs in the SSA22 field \citep[][mean$=108$\,km\,s$^{-1}$,
median$=84$\,km\,s$^{-1}$]{Yamada2012}.  Moreover, \cite{Yamada2012} also found that the
line-widths for some less extreme LABs in SSA22 are significantly broader than those of
the isolated LAEs.  We thus may consider these faint LAEs as an extension of the known
SSA22 LAE population that exists below the detection limits of narrow-band selected
samples.

\subsection{Detection of \ion{He}{ii} $\lambda$1640 emission}
\label{sec:detect-ionh-lambd}

\begin{figure*}
  \vspace{-1.2em}
  \includegraphics[width=\textwidth]{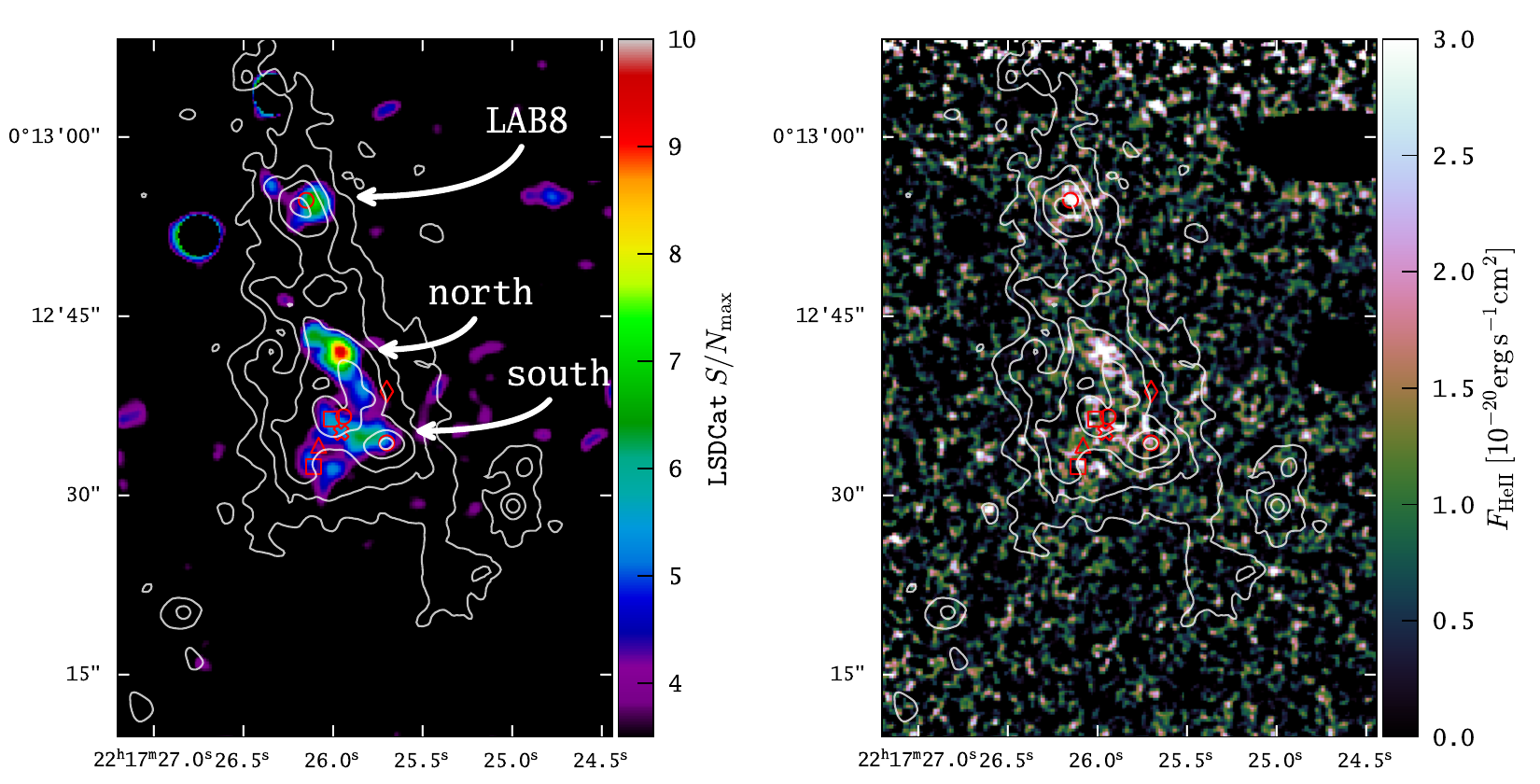} \\ \vspace{0.5em}
  \includegraphics[width=\textwidth]{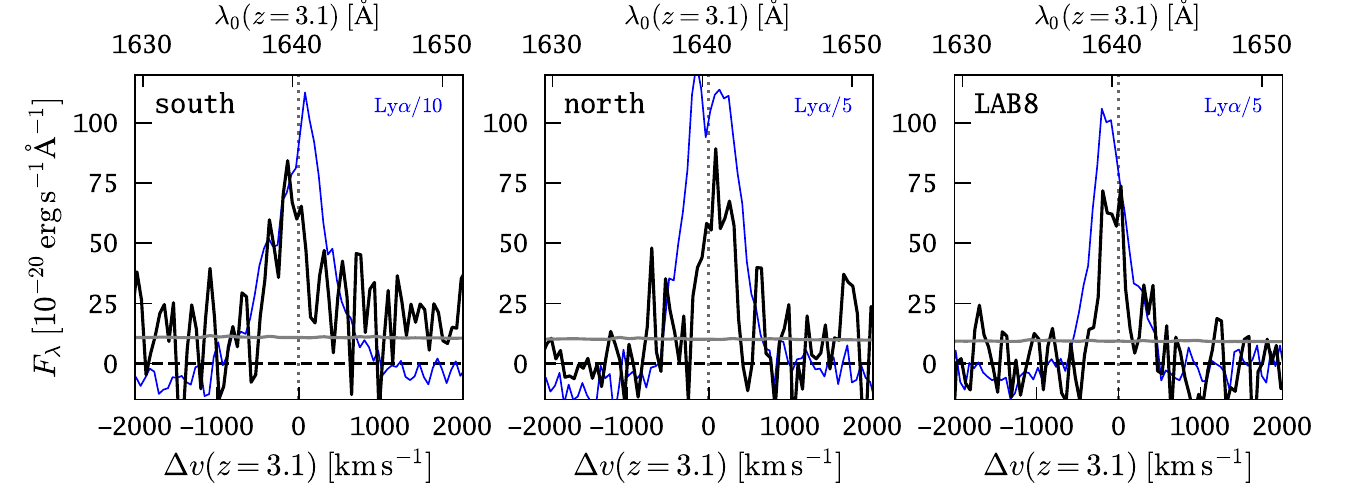}
  \vspace{-1.5em}
  \caption{Detection of extended \ion{He}{ii} $\lambda 1640$ fuzz in three distinct
    regions within LAB\,1. \emph{Top Left}: Map of the maximal \ion{He}{ii} S/N after
    cross-correlation with a 3D Gaussian template (see
    Sect.~\ref{sec:adapt-lyalpha-image}).  We show the maximum from the S/N datacube
    between $\lambda_{\mathrm{min}} = 6719$\AA{} and $\lambda_{\mathrm{max}} = 6729$\AA{}
    (i.e. $\pm 5$\AA{} around \ion{He}{ii} at $z=3.1$).  We label the three \ion{He}{ii}
    emitting regions ``south'', ``north'', and ``LAB8'' as indicated.  \emph{Top Right:}
    Adaptive \ion{He}{ii} narrow-band image.  The creation of this image followed a
    similar procedure as for the adaptive Ly$\alpha$ narrow-band image (see
    Figure~\ref{fig:adapt_nb} and Sect.~\ref{sec:adapt-lyalpha-image}), except that here
    the used S/N analysis threshold is 3, and the default band-width for non-detections
    above this threshold is set to 3 datacube layers.  The adaptive narrow-band image was
    additionally smoothed with a $\sigma=1$px (0.2\arcsec{}) Gaussian kernel.  Grey
    contours indicate Ly$\alpha$ surface-brightness levels
    $\mathrm{SB}_\mathrm{Ly\alpha} = [200 , 100 , 50 , 25 , 8.75]\times
    10^{-19}$\,erg\,s$^{-1}$cm$^{-2}$\,arcsec$^{-2}$ as measured in the adaptive
    narrow-band image (Figure~\ref{fig:adapt_nb}).  In addition to the foreground galaxies
    in the west where bright continuum emission corroborated the median-filter subtraction
    continuum removal, we also masked out \ion{O}{iii} line emission from a $z=0.3$ galaxy
    (at the northern edge of the image), and H$\beta$ emission from a $z=0.4$ galaxy (at
    the north-western edge of the image), as these highly significant emission line
    coincide with the $z=3.1$ \ion{He}{ii} emission. \emph{Bottom panels}: \ion{He}{ii}
    spectral profiles extracted from the three regions ``south'' (\emph{bottom left
      panel}), ``north'' (\emph{bottom centre panel}) and ``LAB\,8'' (\emph{bottom right
      panel}).  The black line is the spectrum and the grey line shows the propagated
    error from the variance cube.  The bottom axis is given in velocities and the top axis
    indicates rest-frame wavelength, both for $z=3.1$.  The vertical dotted line indicates
    the rest-frame wavelength of \ion{He}{ii} (1640.42\AA{}).  Extraction was performed by
    defining apertures connected regions of pixels above a S/N threshold of 6 in the
    maximum S/N map.  For comparison we also show the Ly$\alpha$ profiles scaled by a
    factor of 10 (5) for the ``south'' (``north'' and ``LAB8'') regions as blue lines (for
    those profiles only the bottom velocity axis is relevant).  }
  \label{fig:heii_maxsn}
\end{figure*}

\begin{table*}
  \caption{Properties of \ion{He}{ii} $\lambda1640$ emitting regions within the
    Ly$\alpha$ blob}
  \label{tab:heiiprop}
  \centering
  \begin{tabular}{cccccccccc}
    \hline \hline \noalign{\smallskip}
    Name   & RA   & Dec            & Area & $F_\mathrm{HeII}$ &  $\Delta v (z_\mathrm{Ly\alpha})$  & $\sigma_v^\mathrm{HeII}$ & $\sigma_v^\mathrm{Ly\alpha}$ & \ion{He}{ii}/Ly$\alpha$ & \ion{C}{iv}/Ly$\alpha$ \\
    {}     & [J2000] & [J2000]     & $\square$\arcsec{}      &  [$10^{-18}$\,erg/s/cm$^{2}$]   & [km/s]    & [km/s]  &  [km/s]  & {}                      &  {}                      \\
    \hline
    south  & 22$^\mathrm{h}$17$^\mathrm{m}$25.89$^\mathrm{s}$  & +00$^\circ{}$12\arcmin{}35.3\arcsec{}  & 13.9 & 10.5$\pm$1.8             &  $+$13$\pm$68        &   289$\pm$55  & 464$\pm$9 &   0.06$\pm$0.01        &  $\leq 0.06$   \\
    north  & 22$^\mathrm{h}$17$^\mathrm{m}$26.01$^\mathrm{s}$  & +00$^\circ{}$12\arcmin{}42.2\arcsec{}  & 12.6 & \phantom{1}7.2$\pm$1.1   &  $-$42$\pm$51        &   152$\pm$24  & 368$\pm$4 &   0.07$\pm$0.01        &  $\leq 0.10$   \\
    LAB8   & 22$^\mathrm{h}$17$^\mathrm{m}$26.12$^\mathrm{s}$  & +00$^\circ{}$12\arcmin{}54.1\arcsec{}  & 8.8  & \phantom{1}6.7$\pm$0.9   &  $-$109$\pm$22       &   207$\pm$44  & 298$\pm$6 &   0.11$\pm$0.02        &  $\leq 0.11$   \\
    \hline \hline
  \end{tabular}
  \tablefoot{The area of the \ion{He}{ii} emitting regions is defined as the connected
    area of spaxels with $\mathrm{S/N}>6$ in the maximum S/N map
    (Figure~\ref{fig:heii_maxsn}).  $\Delta v(z_\mathrm{Ly\alpha})$ is the relative
    velocity offset between Ly$\alpha$ and \ion{He}{ii} emission:
    $\Delta v (z_\mathrm{Ly\alpha}) = c \times (z_\mathrm{Ly\alpha} - z_\mathrm{HeII})/(1
    + z_\mathrm{HeII})$}
\end{table*}

We detect low-SB \ion{He}{ii} emission from three distinct regions within the Ly$\alpha$
blob.  We visualise this detection in Figure~\ref{fig:heii_maxsn}, where we show the
resulting maximum S/N map from the \texttt{LSDCat} 3D cross-correlated datacube
(Sect.~\ref{sec:maximum-sn-image}).  Here we evaluated the S/N datacube between
$\lambda_{\mathrm{min}} = 6719$\AA{} and $\lambda_{\mathrm{max}} = 6729$\AA{} (i.e.
$\pm 5$\AA{} around \ion{He}{ii} at $z=3.1$).  We also show in Figure~\ref{fig:heii_maxsn}
an adaptive narrow-band image for \ion{He}{ii} constructed similarly as the
adaptive Ly$\alpha$ image in Sect.~\ref{sec:adaptive-narrow-band}.  Due to the faintness of
the emission we lowered the extraction threshold a bit, i.e. here the image was extracted
by using voxels with $\mathrm{S/N}>3$ from the S/N datacube.  We provide a visual
representation of the noise level in regions not containing any detected signal in the
same manner as we did for the adaptive Ly$\alpha$ image, i.e. we sum $\mathrm{S/N}<3$
spaxels by $\pm 2.5$\AA{} (4 spectral bins) around 6724\,\AA{}.

We label the three regions where \ion{He}{ii} is detected with $\mathrm{S/N} > 6$
``south'', ``north'', and ``LAB8'' in Figure~\ref{fig:heii_maxsn}.  The ``south'' region
is located south-west in proximity the ALMA continuum sources LAB1-ALMA1 and LAB1-ALMA2
while being slightly north-east of the Lyman break galaxy SSA22a-C11.  The ``north''
\ion{He}{ii} region can not be associated with any known source in LAB\,1.  This region is
co-spatial with the feature labelled ``northern high SB region'' in Fig.~\ref{fig:maxsn}.
It is also co-spatial with the region labelled R2 in \cite{Weijmans2010} and
\cite{McLinden2013}.  Lastly, the ``LAB8'' region appears in proximity to the Lyman
break galaxy SSA22a-C15.  We report in Table~\ref{tab:heiiprop} positions of those
regions.  These positions are S/N-weighted, i.e. calculated according to
Eq.~(\ref{eq:xycen}) with $I_{xy}$ replaced by the pixel values of the maximum S/N map
$\mathrm{SN}_{xy}$ and only considering pixels satisfying $\mathrm{SN}_{xy} \geq 6$ for
each region.  All \ion{He}{ii} peaks are within regions of Ly$\alpha$ surface-brightness
above $5\times10^{-18}$erg\,s$^{-1}$cm$^{-2}$arcsec$^{-2}$.  But the morphological
features seen in \ion{He}{ii} are considerably different compared to the morphology of
Ly$\alpha$ above this surface brightness limit.
 
In order to extract 1D spectra from those regions we created apertures consisting of
contiguous regions with $\mathrm{S/N}>5.5$ in the maximum S/N image.  The corresponding
areas of these \ion{He}{ii} morphology-matched apertures are also listed in
Table~\ref{tab:heiiprop}.  The in those regions extracted spectral profiles of the
\ion{He}{ii} emission are shown in the bottom panels of Figure~\ref{fig:heii_maxsn}.  We
furthermore compare in this figure the \ion{He}{ii} spectral profiles to the Ly$\alpha$
profiles extracted in the same regions.

We measured the \ion{He}{ii} fluxes in those regions by summing the spectral profiles
shown in Figure~\ref{fig:heii_maxsn} over their full width at zero intensity.  The error
on the fluxes was obtained from $10^4$ Monte-Carlo (MC) realisations of the profile, where
we used the propagated variances as input for adding noise to each spectral bin.  The so
obtained flux \ion{}{} measurements are listed in Table~\ref{tab:heiiprop}.  Given the
areas of the extraction apertures, the measured fluxes correspond to \ion{He}{ii}
surface-brightness values of 7.6$\times10^{-19}$erg\,s$^{-1}$cm$^{-2}$arcsec$^{-2}$,
5.7$\times10^{-19}$erg\,s$^{-1}$cm$^{-2}$arcsec$^{-2}$, and 7.6$\times10^{-19}$
erg\,s$^{-1}$cm$^{-2}$arcsec$^{-2}$ for region ``south'', ``north'', and ``LAB8'',
respectively.  The measured \ion{He}{ii} surface-brightness is $\sim 4-5$ times fainter
than the upper limits reported for a previous attempt to detect \ion{He}{ii} emission in
LAB\,1 using narrow-band imaging \citep{Arrigoni-Battaia2015}.

To quantify the \ion{He}{ii}/Ly$\alpha$ flux ratios we measured the Ly$\alpha$ fluxes over
their full width at zero intensity within the three \ion{He}{ii} emitting regions.  We
obtained $F_\mathrm{Ly\alpha} = 1.7\times10^{-16}$erg\,s$^{-1}$cm$^{-2}$,
$1\times10^{-16}$erg\,s$^{-1}$cm$^{-2}$, and $6.1\times10^{-17}$erg\,s$^{-1}$cm$^{-2}$,
corresponding to \ion{He}{ii}/Ly$\alpha$ $0.06\pm0.01$, $0.07\pm0.01$, and $0.10\pm0.02$,
for region ``south'', ``north'', and ``LAB8'', respectively (see also
Table~\ref{tab:heiiprop}).  The errors on those ratios have also been computed from $10^4$
MC simulations for the individual \ion{He}{ii} and Ly$\alpha$ flux measurements.  From the
previous non-detection of \ion{He}{ii} \cite{Arrigoni-Battaia2015} determined upper limits
\ion{He}{ii}/Ly$\alpha$$<0.11$ for LAB\,1 and \ion{He}{ii}/Ly$\alpha$$<0.22$ for LAB8,
thus the measured \ion{He}{ii}/Ly$\alpha$ ratios are a factor of two lower.

Using the first flux-weighted moment (Eq.~\ref{eq:1mom}) on the extracted profiles, we
determined the relative redshift offset between \ion{He}{ii} and the Ly$\alpha$
$\Delta v (z_\mathrm{Ly\alpha}) = c \times (z_\mathrm{Ly\alpha} - z_\mathrm{HeII})/(1 +
z_\mathrm{HeII})$.  Again, the errors have been computed from $10^4$ MC realisations.

A significant offset between Ly$\alpha$ and \ion{He}{ii} is only detected for the LAB\,8
region, where Ly$\alpha$ appears modestly blue-shifted.  Blue-shifted Ly$\alpha$ emission
with respect to non-resonant nebular lines is atypical for normal Ly$\alpha$ emitting
galaxies, that show often $\Delta v (z_\mathrm{Ly\alpha}) \gtrsim 200$\,km\,s$^{-1}$
\citep[e.g.][]{Rakic2011,Song2014}, although few exceptions exist
\citep[e.g.][]{Trainor2015}.  Interestingly, our non-detections of velocity offsets
between Ly$\alpha$ and \ion{He}{ii} around C11 and LAB\,8 is consistent with the
non-detection of offsets between Ly$\alpha$ and [\ion{O}{iii}] for those systems by
\cite{McLinden2013}.  Moreover, \cite{Prescott2015b} also report similar low
  velocity offsets between \ion{He}{ii} and Ly$\alpha$ in a \ion{He}{ii} detected LAB at
  $z=1.67$.  These authors also compile data from the literature to show that LABs exhibit
  generally lower kinematic offsets between Ly$\alpha$ and optically thin lines and the
  here presented measurements on \ion{He}{ii} emitting regions in LAB1 corroborate this
  fact.

Lastly, it is visible in Figure~\ref{fig:heii_maxsn} that in all three regions the
Ly$\alpha$ profile appears broader compared to \ion{He}{ii}.  To quantify this we measured
the width of the \ion{He}{ii} and Ly$\alpha$ lines, $\sigma_v^\mathrm{HeII}$ and
$\sigma_v^\mathrm{Ly\alpha}$, using the second flux-weighted moment (Eq.~\ref{eq:kmom},
with $k=2$).  The measurements obtained are reported in Table~\ref{tab:heiiprop}.  We find
that Ly$\alpha$ is 1.6$\times$, 2.4$\times$, and 1.4$\times$ broader than \ion{He}{ii} in
the ``south'', ``north'' and ``LAB8'' region, respectively.  These values are in
  agreement with the broadening for Ly$\alpha$ with respect to \ion{He}{ii} observed in a
  $z=1.67$ LAB by \cite{Prescott2015b}.  As pointed out by these authors, the absence of
  significant velocity offsets between Ly$\alpha$ and \ion{He}{ii} and the presence of
  line broadening in Ly$\alpha$ with respect to \ion{He}{ii} are consistent with a
  scenario where a significant fraction of the extended Ly$\alpha$ emission is produced in
  situ.  We will discuss this physical interpretation further in
  Sect.~\ref{sec:ionh-emiss-from}.

\subsection{Non-detection of the \ion{C}{iv} $\lambda\lambda$1548,1550 doublet}
\label{sec:ionciv}

\begin{figure*}
  \sidecaption
  \includegraphics[width=12cm, trim=20 0 10 0, clip=True]{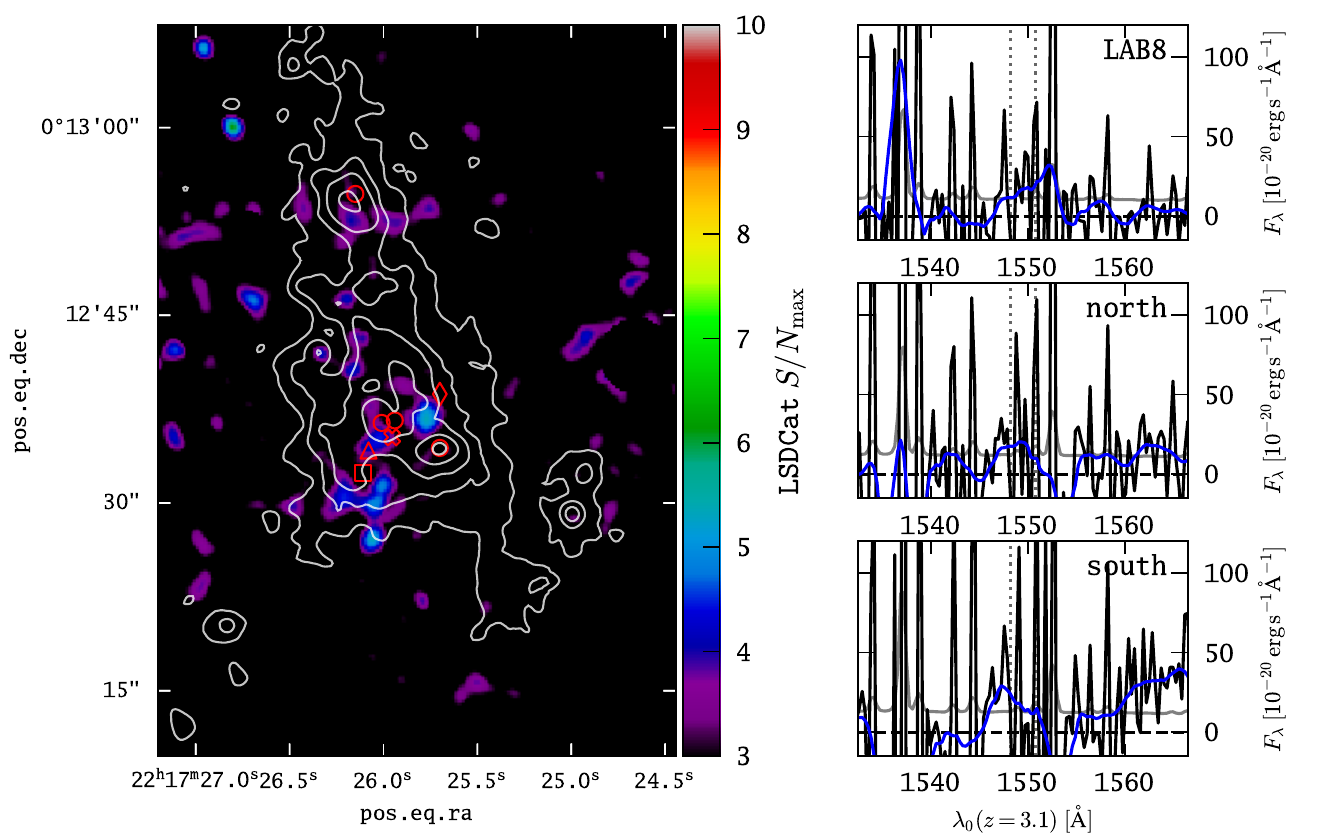}
  \caption{Non-detection of significant signal from \ion{C}{iv} emission within
    LAB\,1.  \emph{Left panel}: Maximum S/N after cross-correlation with a 3D
    Gaussian template (see Sect.~\ref{sec:adapt-lyalpha-image}).  Here the S/N
    cube was evaluated between $\lambda_{\mathrm{min}} = 6345$\AA{} and
    $\lambda_{\mathrm{max}} = 6360$\AA{}.  \emph{Right panels}: Spectral
    extractions in the regions ``south'', ``north'' and ``LAB\,8'', where
    significant \ion{He}{ii} emission was detected
    (Figure~\ref{fig:heii_maxsn}).  In each of those panels the black curve
    shows the extracted spectrum, while the grey curve shows the propagated
    noise from the variance datacube.  Vertical dotted lines indicate the
    rest-frame wavelengths of the \ion{C}{iv} doublet.  We also show a smoothed
    version (Gaussian window, $\sigma=3\,\mathrm{px}$) of the spectrum (blue
    curve), which suppresses the high-frequency and high-amplitude oscillations
    from systematic sky-subtraction residuals in this wavelength range. }
  \label{fig:civ}
\end{figure*}

We detect no significant signal from the \ion{C}{iv} doublet (1548.203\,\AA{} and
1550.777\,\AA{}) in the MUSE data of LAB\,1.  This can be seen in Figure~\ref{fig:civ}
where we show the maximum S/N map for \ion{C}{iv} from evaluation of the 3D
cross-correlated S/N datacube (see Sect.~\ref{sec:adapt-lyalpha-image}).  Here the S/N
cube was evaluated between $\lambda_{\mathrm{min}} = 6345$\AA{} and
$\lambda_{\mathrm{max}} = 6360$\AA{}.  While there are a few patches within the blob that
indicate possible signal at $\mathrm{S/N} \approx 5$, this spectral region of the datacube
is highly contaminated by sky-subtraction residuals.  As these systematic residuals are
not accounted for in the variance datacube, our S/N estimates are biased high.  These
sky-subtraction residuals appear especially pronounced at the edges of the individual
observational datasets (see exposure map in Figure~\ref{fig:exp_map}).  These
edge-residuals are apparent as a patchy horizontal stripe of $\mathrm{S/N}\approx 4$
values slightly south of the photometric centre of LAB\,8, as well as a vertical stripe on
the western edge of the displayed region.  A few pixels in the maximum S/N map show values
above 5 in regions within the blob, but the extracted spectra in those regions do not show
signal.  We show in Figure~\ref{fig:civ} the spectral region of interest around \ion{C}{iv}
for the spectra extracted within the three \ion{He}{ii} emitting regions
(Sect.~\ref{sec:detect-ionh-lambd}).  In order to suppress the high-frequency sky-noise
and residuals we also smoothed those spectra with a $\sigma=3.6$\,\AA{} wide Gaussian.
Interestingly, the smoothed spectrum -- shown as a blue line in Figure~\ref{fig:civ} --
indeed appears to exhibit a small bump at the position of the $\lambda1549$ line
in all three regions.  However, from the noise in those three regions (shown as a grey
line in Figure~\ref{fig:civ}), it is obvious that these features are not significant and
can be only considered as a hint of \ion{C}{iv} emission.

To provide upper limits on \ion{C}{iv} emission we perform a source insertion and recovery
experiment. For this experiment we assume that the \ion{C}{iv} emission would be
co-spatial with the \ion{He}{ii} emitting regions. As noted in
Sect.~\ref{sec:detect-ionh-lambd}, each of the three \ion{He}{ii} emitting region is
defined by a cluster of connected spaxels that have $\mathrm{S/N}>5.5$.  For simplicity we
assume constant surface brightness within each region.  Furthermore, we model the spectral
profile of the $\lambda\lambda 1548,1550$ doublet by two Gaussians, where the dispersion
is set to the average width measured from the \ion{He}{ii} line:
$\langle \sigma_v^\mathrm{HeII} \rangle = 215$\,km\,s$^{-1}$ and we fix the ratio to 1.7
between $\lambda 1548$ and $\lambda 1550$.  We implant the so generated \ion{C}{iv}
emitting regions at surface-brightness levels from
$10^{-19}$erg\,s$^{-1}$cm$^{-2}$arcsec$^{-2}$ to
$9 \times 10^{-19}$erg\,s$^{-1}$cm$^{-2}$arcsec$^{-2}$ in steps of
$10^{-19}$erg\,s$^{-1}$cm$^{-2}$arcsec$^{-2}$ prior to median-filter subtracting the fully
reduced datacube.  After median-filter subtracting each datacube with artificial
\ion{C}{IV} sources we ran the 3D cross-correlation procedure from \texttt{LSDCat}.  From
this experiment we found that we would detect the \ion{C}{iv} emission significantly in
all three regions at surface-brightness levels
$\geq 8 \times 10^{-19}$erg\,s$^{-1}$cm$^{-2}$arcsec$^{-2}$.  Given the areas of the three
\ion{He}{ii} emitting regions, our surface-brightness limit corresponds upper limits in
\ion{C}{iv} flux of $1.1\times 10^{-16}$erg\,s$^{-1}$cm$^{-2}$,
$1.0\times 10^{-16}$erg\,s$^{-1}$cm$^{-2}$, and $6.1\times 10^{-16}$erg\,s$^{-1}$cm$^{-2}$
for region ``south'', ``north'', and ``LAB 8'', respectively.  Given the measured
Ly$\alpha$ fluxes in those regions (see Sect.~\ref{sec:detect-ionh-lambd}), this
corresponds to upper limits in \ion{C}{iv}/Ly$\alpha$ ratios of 0.06, 0.1, and 0.11 for
region ``south'', ``north'', and ``LAB 8'', respectively.  Our upper limits on \ion{C}{iv}
translate into upper limits of $<1$, $<1.4$, $<1$ of \ion{C}{IV} / \ion{He}{ii} in region
``south'', ``north'', and ``LAB\,8'', respectively.

Our upper limit for \ion{C}{iv} emission within the \ion{He}{ii} emitting zones appears,
at first sight, less constraining than the previous upper limit of
7.4$\times 10^{-19}$erg\,s$^{-1}$cm$^{-2}$arcsec$^{-2}$ from narrow-band observations
\cite{Arrigoni-Battaia2015}.  However, the comparison has to be treated with caution, as
the previous estimate assumed constant \ion{C}{iv} emission over the whole area of the
blob ($\sim$200\,arcsec$^2$). But, we assumed here that \ion{C}{IV} emission is confined
to the three \ion{He}{ii} emitting regions ($\lesssim$10\,arcsec$^2$).  The assumption of
overlapping \ion{C}{IV} and \ion{He}{ii} emission appears a good first-order approximation
in scenario where \ion{C}{IV} and \ion{He}{ii} act as a coolant of shock heated gas, as
for both lines the maximum emissivity is obtained in a similar gas phase of $T\sim10^5$K
\citep{Cabot2016}, where \ion{C}{iv} is driven through collisional excitations of C$^{3+}$
(ionisation potential 47.9\,eV) while \ion{He}{ii} is originating from the recombination
cascade of collisionally ionised He$^{2+}$ (ionisation potential 54.4\,eV).  Nevertheless,
the \ion{C}{iv} emissivity also peaks in a region of slightly lower density
($n_\mathrm{H}\sim2-5$\,cm$^{-3}$) compared to \ion{He}{ii}
($n_\mathrm{H}\sim5-20$\,cm$^{-3}$), thus \ion{C}{iv} could potentially also be more
extended \citep{Cabot2016}.  This effect could be further enhanced by the resonant nature
of \ion{C}{iv} \citep[see also][]{Berg2019}.  On the other hand, if photo-ionisation from
a AGN produces C$^{4+}$ (ionisation potential 64.4\,eV) that emits \ion{C}{iv} as a result
of recombinations, the C$^{4+}$ zone would be confined within the He$^{2+}$ zone.  Lastly,
we will show below (Sect.~\ref{sec:ionh-emiss-from}) that under certain conditions
\ion{C}{IV} might be even brighter than \ion{He}{ii}.  For this reason some detections of
extended \ion{C}{IV} without corresponding \ion{He}{II} around quasars have been reported
\citep{Borisova2016,Travascio2020}.

\section{Discussion}
\label{sec:disc}

\subsection{On interpreting the moment maps of Ly$\alpha$ from LAB\,1 as a tracer of gas
  kinematics}
\label{sec:interpr-first-moment}

Before interpreting flux-weighted moment maps from Ly$\alpha$ as a tracer of gaseous
motions, we need to asses how Ly$\alpha$ radiative transfer effects may have influenced
these measurements.  Ly$\alpha$ velocity fields found to exhibit coherent
  large-scale kinematics have been observed around known
\citep[e.g.][]{Arrigoni-Battaia2018,Martin2019} or suspected \citep{Prescott2015b}
quasars.  Details on the density distribution of the gas within those systems are still
not settled, but it has been asserted that the strong ionisation field from the
quasar produces a large fraction of Ly$\alpha$ photons in situ throughout the nebula via
recombinations \citep[e.g.][]{Martin2014,Martin2019,Cantalupo2014,Cantalupo2019}.
Nevertheless, even small residual neutral \ion{H}{i} fractions
($\lesssim 10^{-4}$) within the ionised halo will result in high Ly$\alpha$
optical depths.  Still, we suspect that the radiative transfer modulation of the
Ly$\alpha$ profile for an extended intrinsic Ly$\alpha$ radiation field is less dramatic
compared to the effect of an integrated spectrum within a compact
Ly$\alpha$ emitting source.  The reason is that the profile measured at each
position in the nebula only reflects small local velocity offsets between point of
emission and surface of last scattering \citep{Prescott2015b}.  At typical
  spectral resolving powers then only a broadening of the Ly$\alpha$ line may be
  observed.  This expectation appears consistent with our results for the \ion{He}{ii}
emitting regions in LAB\,1.  Moreover, systemic literature redshifts of the
galaxies associated with LAB\,1 do not show significant offsets with respect
to those determined from Ly$\alpha$ \citep{McLinden2013,Kubo2016,Umehata2017}.
From this we conclude that the first-moment map traces the gas-kinematics close to the
embedded sources.  For those galaxy near zones we then can assume that significant
amounts of Ly$\alpha$ photons must be produced in situ.

For interpreting the entire first-moment map as a tracer of the underlying gas kinematics,
especially in regions far from known galaxies in LAB\,1, we can not resort to the argument
above.  The detection of polarised Ly$\alpha$ emission, both in imaging polarimetry
\citep{Hayes2011} and spectro-polarimetry \citep{Beck2016}, shows that significant
fractions of the observed Ly$\alpha$ photons must have scattered into the line of sight
after being emitted from a central source.  According to radiative transfer theory the
observed polarisation signal in Ly$\alpha$ stems from scattered photons within the wing of
the absorption profile \citep[e.g.][]{Dijkstra2008,Eide2018}.  Hence, while these photons
preserve information regarding the kinematics of the surface of last scattering, they
appear red- or blue-shifted (depending on the kinematics of the scattering medium) with
respect to the overall kinematics of the gas.  Therefore, in a scenario where only
scattered Ly$\alpha$ photons are observed at large distance to the embedded sources, the
measured first-moments would be biased towards higher or lower velocities compared to the
true gas kinematics.  This could potentially increase the observed amplitude of the
observed velocity gradient.

However, the above scenario also requires that significant amounts of Ly$\alpha$
photons are released by the dust-rich ALMA 850$\mu$m detected galaxies in LAB\,1
  \citep[see also][]{Geach2016}.  The possibility of this process is indicated by the
detection of Ly$\alpha$ emission from nearby ultra-luminous infrared galaxies
\citep{Martin2015}.  Feedback driven outflows powered by star-formation is believed to
create the required Ly$\alpha$ escape channels.  Evidence for the required feedback
effects at the positions of the galaxies comes from our higher-moment analysis discussed
below (Sect.~\ref{sec:morpho-kin}), but also from the spectro-polarimetry measurements by
\cite{Beck2016}.  These authors find that the polarisation signal is increased in the
wings of the Ly$\alpha$ profile, especially near the embedded galaxies.  According to
recent Ly$\alpha$ radiative transfer models of \cite{Eide2018} this is an expected
polarisation signature in the presence of outflowing gas.  Moreover, the Ly$\alpha$ escape
channels produced by gaseous outflows can potentially also act as escape channels for
ionising photons from the embedded systems, thus a significant fraction of those photons
might indeed be available to power the emission from the blob via recombination.
Empirical evidence for such a process have also been observed in the nearby universe
\citep{Herenz2017b,Bik2018,Menacho2019}.  Thus, in addition to Ly$\alpha$ scattering 
far from the embedded galaxies, we also expect that also in LAB\,1 a significant
fraction of Ly$\alpha$ radiation is produced in-situ at larger distances from the known
embedded sources.  Hence, the observed emission from the galaxy far-zones is likely
superposition of scattered and in-situ produced Ly$\alpha$ emission.  This potentially
mitigates biases in the first moment maps that would result from pure Ly$\alpha$
scattering.

To summarise, we presented qualitative arguments in favour of interpreting large-scale
coherent features in the first-moment map from Ly$\alpha$ as tracers of the gaseous motion
in the system.  These qualitative arguments are, as of yet, not tested for Ly$\alpha$
radiative transfer simulations in realistic Ly$\alpha$ blob environments.  As we will also
discuss further below, radiative transfer effects are, however, expected to strongly
influence the Ly$\alpha$ line-width and the derived moments of higher order.  We suggest
that the here presented non-parametric moment-based measurements of the line profiles
present a starting point to summarise the complexity of the encountered profiles in LAB
environments. Guided by radiative transfer simulations then these measurements may
  help to disentangle scattering processes from in-situ Ly$\alpha$ photon production.

\subsection{Combined analysis of Ly$\alpha$ line profile morphology and imaging polarimetry}
\label{sec:comb-analys-lyalpha}

\begin{figure*}
\centering
\includegraphics[width=0.9\textwidth]{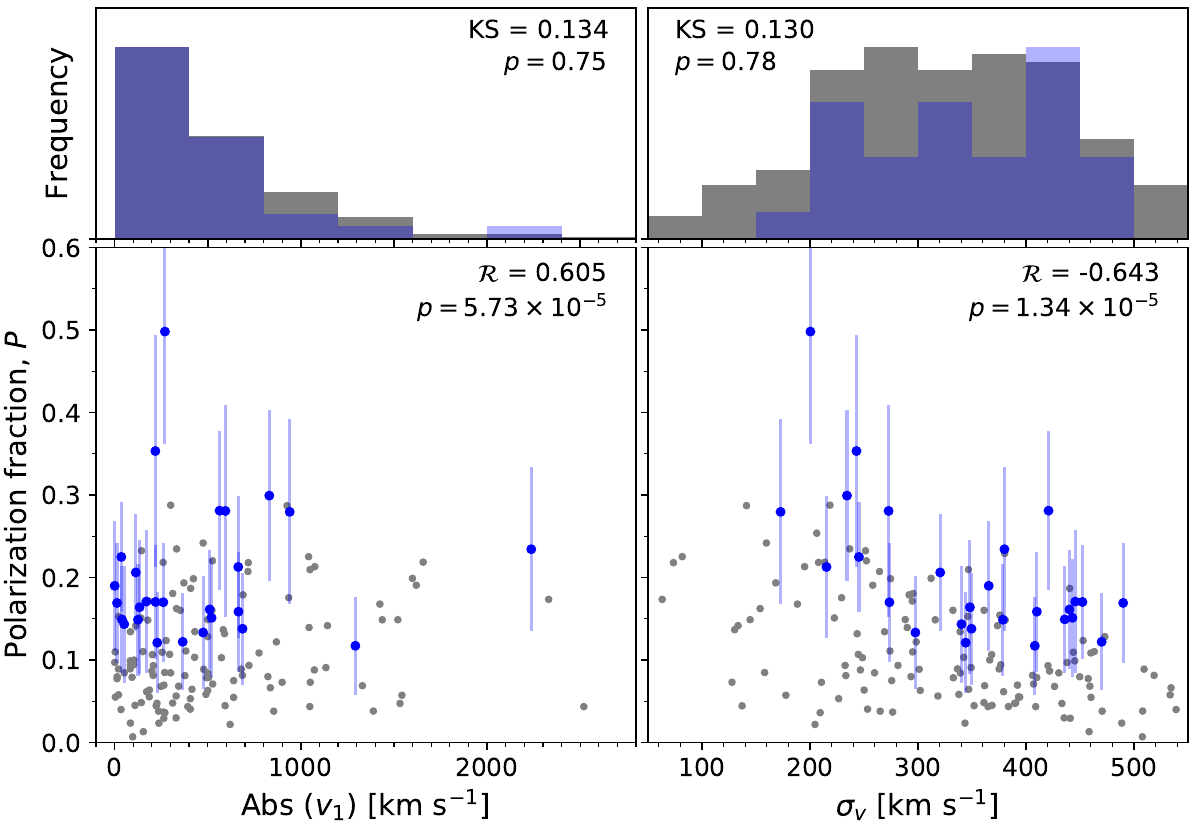}
\caption{A region-by-region comparison of the polarisation fraction ($P$) derived from
  VLT/FORS2 observations, with the kinematic measurements of the Ly$\alpha$ emission
  derived in this paper.  The \emph{left panel} displays $P$ against the absolute value of
  the first moment (Eq.~\ref{eq:2}), while the \emph{right panel} shows $P$ against the
  second moment (Eq.~\ref{eq:4}).  Gray points show every spatial element for which $P$
  could be measured at better than 5~per cent (see Figure~2 of \citealt{Hayes2011}).  Blue
  points show the regions in which $P$ is detected at significance better than 2-$\sigma$.
  Spearman's rank correlation coefficient, $\mathcal{R}$, and the associated $p$-value
  (the likelihood with which a correlation of at least this magnitude arises purely by
  chance) are shown in each figure -- $p<0.05$ is generally considered significant.  The
  histograms above each plot shows the distribution of properties from the lower
  abscissae, where the grey and blue colouring follows that of the lower figures.  The
  values from the two-sample KS-statistic, and the associated $p$-value, are presented in
  each panel.  These $p$-values of almost 1 show that the polarised spaxels are consistent
  with being drawn from the same distribution, and that the trends seen in the blue points
  of the lower figures are not the result of biased selection.}
\label{fig:polfig}
\end{figure*}

\citet{Hayes2011} studied LAB1 using the FORS2 polarimeter at VLT in order to spatially
map the polarimetric properties of the Ly$\alpha$ emission, including the Stokes vectors
($Q$ and $U$), the polarised light fraction ($P$) and angle of the linear polarisation
vector ($\chi$).  The resulting maps of $P$ and $\chi$ were qualitatively consistent with
theoretical predictions by \citet{Lee1998}, \citet{Rybicki1999}, and \citet{Dijkstra2008},
and implied a geometry in which Ly$\alpha$ is produced inside, or close to the central
regions of the highest Ly$\alpha$ surface brightness (at the positions of the
later-discovered ALMA sources 1 and 2) and scattered at large impact parameters, from
vectors in the plane of the sky to the sightline of the observer.  However the narrowband
imaging polarimetry data contained no information regarding the kinematic properties of
the Ly$\alpha$ emission, and we could not study where in the Ly$\alpha$ line profile the
polarisation signal is imparted.  To remedy this \cite{Beck2016} conducted spectroscopic
polarisation observations with the same instrument, to study $P$ and $\chi$ as a function
of frequency.  Those observations showed that the wavelength dependence of $P$ is small
near line centre, but rises towards the line wings.  This is again qualitatively
  consistent with the numerical predictions of \citet{Dijkstra2008}.  Those slit
data contained the necessary frequency information, but only at eight positions identified
along the 1\arcsec{} wide slit.  Moreover, those studies were limited to the
regions of highest surface brightness (\citealt{Beck2016}, their Figure~3).

In this Section we use the 3D spectroscopy provided by MUSE to shed further light on the
origin of the polarisation pattern.  We perform a hybrid study, and contrast the spatially
known (but spectrally-unknown) measurements of $P$ and $\chi$ with the Ly$\alpha$
kinematics using 3D spectroscopy.  \cite{Hayes2011} used Voronoi tessellation to locally
enhance the SNR in the images taken in individual beams (`ordinary \& extraordinary'), and
we apply the same binning patterns to the MUSE data: we aligned the FORS2 Ly$\alpha$
intensity images with the MUSE data, and propagate the recovered astrometric information
into the maps of $P$ and $\chi$.  For every Voronoi tessellated cell we then extract the
corresponding spaxels from the MUSE datacube, and compute the moments described in
Section~\ref{sec:spectr-morph-lyalpha}.

Figure~\ref{fig:polfig} shows the results, where we display the absolute velocity offset
from $z_\mathrm{Ly\alpha}=3.1$ (left panel) and the velocity dispersion (right panel).  We
show the absolute value of the first moment because the polarisation signal is imparted by
wing scatterings, and scatterings in both the blue and red wings manifest in the same $P$
\citep[the quantity measured in our polarimetry data][]{Dijkstra2008}.  We find a positive
correlation between $P$ and the absolute line-of-sight velocity, and also an
anti-correlation between $P$ and the velocity dispersion.  The $p$-values of
$10^{-5} - 10^{-4}$ suggest that correlations of this magnitude or larger are very
unlikely to have arisen by a random process.  That Ly$\alpha$ should exhibit higher
degrees of polarisation for narrow lines is a curious result, and may stem from several
effects.

The polarisation data show a high $P$, and a pattern in $\chi$ that is close-to
tangentially aligned with a central source, and follows contours of surface brightness at
smaller scales \citep[][Figure 2]{Hayes2011}.  To impart such a pattern, scattering in the
wing of the redistribution profile needs to be promoted, so as to in turn enhance
scattering through the $2P_{3/2}$ excited level \citep{Dijkstra2008,Eide2018}.  $P$ will
instead decrease if too many core (resonance) scattering events occur (through the
$2P_{1/2}$ level).  Such an enhancement of wing events cannot be the case for a static
medium, where the gas would absorb Ly$\alpha$ close to line-centre (core).  Some velocity
offset is needed, which implies that high $P$ should avoid $|\Delta v_\mathrm{los}| = 0$
(the original motivation for the outflowing shell models of \citealt{Dijkstra2008}).
Moreover, even if large bulk velocity offsets are present, the velocity dispersion cannot
be significantly larger than the absolute $|v_\mathrm{los}|$: the consequence of a broad
velocity range of absorbing material will always leave some high column density gas close
to line-centre, and maintain significant core scattering.  In a complicated, 3-dimensional
environment with multiple sources of Ly$\alpha$ and many velocity components of scattering
gas, highly polarised regions should favour larger velocity shifts and narrower lines, and
we would expect $P$ to correlate with $|v_\mathrm{los}|$ and anti-correlate with the line
width.  This is precisely what Figure~\ref{fig:polfig} shows and the combination of 2D
polarimetric data and 3D spectroscopic data supports a scenario in which preferential
scattering outside the Doppler core is enhanced by velocity differences.

\subsection{Insights into LAB\,1 from the spatially resolved
  line-profile analysis}
\label{sec:morpho-kin}

\subsubsection{Large-scale gas kinematics from the first-moment map}
\label{sec:large-scale-gas}

We concluded in Sect.~\ref{sec:interpr-first-moment} that the first-moment map from
Ly$\alpha$ (Figure~\ref{fig:moms}, left panel) can be used to trace the large scale
kinematics of the gas within the blob.  As apparent from that map, the blob shows a
coherent large-scale velocity gradient from receding velocities on its eastern side to
approaching velocities on its western side.  This large scale velocity gradient
encompasses LAB\,1 and the northern neighbour LAB\,8.  Moreover, the newly detected
shell-like region in the south-west (SW) of the blob is completely blue-shifted, thus
seemingly follows the large scale E--W velocity gradient.  Furthermore, also within the
shell a gradient from NW to SE is seen.  The velocity field is mostly coherent over
$\sim 150$\,kpc and shows significant perturbations only on smaller ($\lesssim 30$kpc)
scales.

We quantify the amplitude of the large-scale velocity shear as
$v_\mathrm{shear} = ( v_\mathrm{max} - v_\mathrm{min})/2 = 1304$\,km\,s$^{-1}$, where
$v_\mathrm{max}$ and $v_\mathrm{min}$ are the lower and upper 5th percentile of the
observed first moments throughout the blob (this choice of percentiles is robust against
outliers while sampling the true extremes of the distribution).  Coherent large scale
velocity fields of such large amplitude appear to be rare amongst extended Ly$\alpha$
nebulae around quasars at redshifts similar to LAB\,1
\citep{Borisova2016,Arrigoni-Battaia2019}.  Nevertheless, a few comparable examples do
exist: object 3 (Q0042-2627) in \cite{Borisova2016}, the QSO UM\,287 studied in
\cite{Martin2015a} and \cite{Martin2019}, as well as object 13 from
\cite{Arrigoni-Battaia2019} that was described in detail in \cite{Arrigoni-Battaia2018}.
Moreover, two of the four QSOs in the recent $z\sim6$ sample by \cite{Drake2019} show
coherent kinematics, while the other two objects show highly irregular and disturbed
velocity fields. 

We hypothesise that the preponderance of disturbed velocity fields around QSOs could be
related to large-scale AGN feedback effects.  As an additional hypothesis we state that
the large fraction of chaotic velocity fields around QSOs is driven by the short exposure
times that are typically used in those studies (e.g. 1\,h in \citealt{Borisova2016};
45\,min in \citealt{Arrigoni-Battaia2019}).  These short exposures result in lower S/N for
low-SB Ly$\alpha$ emission in the outskirts of the halos and potentially even miss the
faintest regions, where large scale kinematics may become most apparent.  Comparing our
observations with previous IFS observations of LAB\,1 by \cite{Bower2004} and
\cite{Weijmans2010}, which additionally have a coarser spatial resolution (0.4\arcsec{}
per spaxel, and seeing FWHM$\approx$1.5\arcsec{}) compared with our data (0.2\arcsec{} per
spaxel, and seeing FWHM$\approx$0.9\arcsec{}), we indeed find that their velocity fields
represent just a coarser and more noisy representation of the here observed velocity field
in the high SB regions of the blob.  Nevertheless, \cite{Bower2004} described the overall
velocity structure of LAB\,1 as chaotic with a lack of velocity shear.  And undeniably,
the velocity field of LAB\,1 within the high-SB regions
($\mathrm{SB}_\mathrm{Ly\alpha} \gtrsim 10^{-18}$erg\,s$^{-1}$cm$^{-2}$arcsec$^{-2}$)
appears complicated.  As noted by \cite{Weijmans2010} the observed complexity in this
region can be broken down into several sub-regions that show coherent velocity shear and
some of these regions are associated to known galaxies within the LAB.  These small-scale
perturbations, now observed at higher spatial resolution in the MUSE data, might be
related coherent motions of the gas in the individual sub-halos.  But they also could be
caused by kinematic perturbations due to inflowing cold gas, or they could be driven by
feedback effects (see Sect.~\ref{sec:small-scale-gas} below).

The most striking feature of the newly revealed large-scale velocity field is that the
shear is observed perpendicular to the morphological principal axis
(Figure~\ref{fig:moms}, left panel).  The existence of such a large-scale coherent
velocity field of high-amplitude appears in qualitative agreement with theoretical
predictions of hydro-dynamical simulations in a $\Lambda$CDM framework.  In particular it
is a natural outcome of the filamentary nature of the cosmic web and the predicted
existence of dense cold-gas streams that can penetrate deeply into dark-matter potential
wells without being shock-heated by the gravitational potential \citep[see review
by][]{Stewart2017b}.  However, the perpendicular alignment between the velocity field and
the morphological principal axis implies that the angular momentum vector of the gas in
LAB\,1 is aligned parallel with the morphological principal axis.  Simulations in the
$\Lambda$CDM framework predict that the angular momentum vector and the elongation of the
dark matter halos -- including their gaseous and stellar contents -- show correlated
properties with the surrounding large-scale structure of the cosmic-web \citep[e.g.,][and
references therein]{Forero-Romero2014,Libeskind2018,Codis2018,Wang2018}.  These studies
predict that the angular-momentum of high-mass halos ($\gtrsim10^{12}$M$_\odot$) is
expected, on average, to be orthogonal to the major axis of the halo, and the halo's major
axis is expected to be aligned parallel with the connecting filament of the cosmic web.
Thus, the observed alignment between morphological principal axis, which we naively
interpret as the direction of the major cosmic web filament \citep[as suggested
by][]{Erb2011}, renders LAB\,1 as an outlier from the theoretically expected average for
high-mass halos.

We speculate that the deviation from the theoretically expected norm in LAB\,1 is caused
by the complexity of the surrounding large-scale structure.  We regard the observed
multi-filamentary morphology of the blob, with one filament extending to the SE in the
direction of the two newly identified LAEs (labelled ``filament towards LAE\,3'' in
Figure~\ref{fig:maxsn}), another filament extending to the north (``filament to LAE\,4''
in Figure~\ref{fig:maxsn}), and another filament connecting to the shell-like region in SW
(``shell-connecting filament'' in Figure~\ref{fig:maxsn}) as evidence for such a complex
environment.  In fact, the SSA\,22 proto-cluster, which hosts LAB\,1 and LAB\,8, is one of
the most-massive structure known in the high-redshift universe (see \citealt{Kikuta2019}
for another extreme overdensity at $z=2.84$).  In such a high-$z$ high-density peak of the
universe a complex interwoven network of filaments is expected.  Indeed, recent MUSE
observations ($t_\mathrm{exp} \approx 4$\,h) by \cite{Umehata2019} have revealed such a
cold-gas network in Ly$\alpha$ at the core of SSA\,22, located 4.8\,\arcmin{} (2.19\,Mpc)
north-east of LAB\,1.  Interestingly, also this structure shows a mix of perpendicular and
parallel aligned large-scale velocity gradients with respect to the filamentary structure
\citep[][their Fig.~3]{Umehata2019}.

The complexity of angular momentum alignments in dense cosmic regions was studied in
simulations by \cite{Lee2018}, who also showed that in these environments the orientation
of angular momentum vector is expected to change frequently and significantly.  In such a
scenario the observed alignment of LAB\,1's angular momentum can be reconciled as an early
evolutionary stage of a massive galaxy.

\subsubsection{Small-scale kinematical perturbations in the first-moment map}
\label{sec:small-scale-gas}

As mentioned before, the coherent velocity field of LAB\,1 shows also various
perturbations on smaller scales.  One prominent perturbation occurs at the position of the
Lyman break galaxy SSA22a-C11, where the velocity shears from north-west to
  south-east, i.e. almost perpendicular to the apparent large-scale motion.  The coherent
small-scale velocity shear at the position of SSA22a-C11 was already noted by
\cite{Weijmans2010}.  In this region the velocity field varies from
$\sim450$\,km\,s$^{-1}$ to $\sim - 1500$\,km\,s$^{-1}$ just within 2\arcsec{} or 15\,kpc
in projection.  It appears as if Ly$\alpha$ traces the velocity field of this individual
bright galaxy that is not aligned with the blob's large scale velocity field.  Such
misalignments of sub-halos are also in qualitative agreement with the theoretical studies
mentioned in the previous section \citep[e.g.][]{Codis2018}.  However, especially given
the large amplitude on small physical scale, this disturbance of the velocity field may
also be caused by feedback effects (galactic outflows, as proposed by
\citealt{Weijmans2010}).  Less prominent disturbances are apparent near the sub-mm
detected sources LAB1-ALMA1 and LAB1-ALMA2.  Here, as we will discuss
below in Sect.~\ref{sec:interpr-high-order}, the analysis of the higher order moment map
will provide arguments in favour star-formation driven winds and/or outflows.

While the small-scale perturbations may be caused by individual velocity fields of the
embedded galaxies and/or star-formation driven winds or outflows, we also have to consider
the alternative (but not competing) hypothesis that they may in fact represent evidence
for cold-flow multi-filamentary inflows.  Recently, \cite{Martin2019} presented a
quantitative framework to fit the expected kinematic signatures from cold-flow accretion
streams to Ly$\alpha$ intensity weighted velocity maps.  These authors motivate their
parametric ansatz by numerical simulations of a galaxy that exhibits cold
filamentary inflows.  The velocity field of this simulation was found to be optimally
described by a large-scale rotating component that is modulated by radially and
azimuthally varying components.  These modulations are shown to capture the kinematic
perturbations caused by the filamentary cooling flows.  \cite{Martin2019} show that these
parametric models also provide an excellent fit to the observed Ly$\alpha$ velocity fields
of two large extended Ly$\alpha$ nebulae around radio-quiet quasars.  We find that the
qualitative appearance of velocity field modulations caused by the cooling flows in the
\citeauthor{Martin2019} ansatz bear similarity to the small scale perturbations seen in
LAB\,1's first-moment map.  A quantitative treatment with the \citeauthor{Martin2019}
ansatz is, however, beyond the scope of the present analysis.

\subsubsection{Interpretation of higher-order moment maps}
\label{sec:interpr-high-order}

Empirical insights into the Ly$\alpha$ photon production and/or scattering mechanisms
within the blob can be gained from our maps of the Ly$\alpha$ line-width
(Figure~\ref{fig:moms}, right panel), its skewness, its kurtosis $\kappa$, as well as its
bi-modality (Figure~\ref{fig:hm}).  In the following we will discuss several notable
features in those maps.  By visually inspecting the spaxels in each of the discussed
regions we ensured that the mapped values are indeed representative of the observed line
profile morphology.

The highest line-widths ($\sigma_v \gtrsim 500$\,km\,s$^{-1}$) are observed N and S of
LAB1-ALMA1 and LAB1-ALMA2. Towards the north this high dispersion
regions shows a V-shaped morphology that traces the edges of the prominent ``bubble''
feature seen in the narrow-band image (Figure~\ref{fig:adapt_nb}).  While Ly$\alpha$ shows
no pronounced double peaked profile ($b \gtrsim 2.6$) at the position of
LAB1-ALMA1 and LAB1-ALMA2, more pronounced double peaks are observed in
the V-shaped region ($b \sim 2$), with the NE arm showing the most pronounced double peaks
($b \lesssim 1.8$).  Thus, while it is tempting to interpret the broad widths close to the
sub-mm source as gas that is kinematically hotter due to feedback from the galaxies, the
measured large width of the profiles in those regions is rather the result of double
component Ly$\alpha$ profiles, i.e. Ly$\alpha$ radiative transfer.

However, we might witness possible signatures of outflowing gas traced by the Ly$\alpha$
skewness at the position of LAB1-ALMA1, where the Ly$\alpha$ profiles show a pronounced
skew towards the blue ($s\lesssim -0.3$).  Interestingly, the Ly$\alpha$ profiles in some
of the \cite{Martin2015} ULIRGs show profiles with a blue tail.  \cite{Martin2015}
interpret these blue tails as outflow signatures, with the Ly$\alpha$ in the wing being
produced by gas cooling within the outflow.  Also, SE of LAB1-ALMA3 strong blue-skewed
profiles that are indicative of an outflow are observed.  This is corroborating the
inference by \cite{Umehata2017} who argued that the properties of the [\ion{C}{ii}]
158$\mu$m emission from LAB1-ALMA3 are suggestive of an interaction between a galactic
outflow with an intergalactic gas stream.

As can be seen from Figure~\ref{fig:moms} the line-width of Ly$\alpha$ is narrower in the
outskirts of the blob.  Both the shell-like region in the SW, the filament emanating to
the N and connecting to LAB\,8, as well as the filament emanating to the NE are
characterised by $\sigma_v \lesssim 200$\,km\,s$^{-1}$.  However, as traced by the
higher-moment based statistics (Figure~\ref{fig:hm}), even in those regions the observed
Ly$\alpha$ line profiles appear to be quite complex.  For example along the LAB\,1 -
LAB\,8 connecting filament the line profile is strongly skewed to the red
($s \gtrsim +0.5$).  Most prominently this structure emerges as a red band in the skewness
map.  Throughout most of this filament the discriminatory power of the adopted bi-modality
indicator is weak ($2.5 \gtrsim b \gtrsim 3$).  However, visual inspection of the line
profile hints at least in some places at the appearance of a secondary weaker red peak as
cause for the measured red skew.  For the NE filament observationally the situation is
more clear, here the profile is clearly double peaked ($b \lesssim 2.5$, see also example
spectrum in panel 1 of Figure~\ref{fig:exspec}), mostly with a weaker blue component.
Both the red-skewed profiles with a potential secondary peak and the double peaked
profiles are likely caused by radiative transfer.  As we discussed above
(Sect.~\ref{sec:comb-analys-lyalpha}), the strongest polarisation signal arises in regions
of smaller line-width ($\sigma_v \lesssim 250$\,km\,s$^{-1}$), also supportive of
Ly$\alpha$ scattering from a central source.  Hence, the low-line width regions in the
outskirts and in the filamentary regions hint at dense patches of gas where Ly$\alpha$
scattering takes place.  The filamentary morphology of some of these patches is
morphological reminiscent of cooling-flows, but whether the dominant Ly$\alpha$ signal is
produced by cooling or photo-ionisation in those regions remains unclear.  At least the
line profiles in those regions appear consistent with a super-position of in-situ produced
Ly$\alpha$ photons (dominant peak) and scattered Ly$\alpha$ radiation (wing or second
component).

Another peculiar and outstanding feature in our higher moment-based maps is found at the N
edge of the bubble. Here the profile is significantly skewed ($s \gtrsim +0.5$) and
described by a single-component ($b \gtrsim 3$) profile with a pronounced wing
($\kappa \approx 3.8$).  An aperture extracted Ly$\alpha$ spectral profile from this
region is shown in panel 11 of Figure~\ref{fig:exspec}.  The bubble or cavity was already
remarked as a peculiar feature in LAB\,1 by \cite{Bower2004}.  As a possible scenario
\cite{Bower2004} discuss, that the bubble is filled by hot ionised gas.  \cite{Bower2004}
speculate that a potential radio-jet from the central sub-mm sources could have heated
this gas.  In this scenario the skewed Ly$\alpha$ line arises in the region where the
beamed emission is being slowed down by the denser cold-gas.  Observational evidence for
radio-emitting gas influencing the Ly$\alpha$ spectral properties have been presented for
radio loud quasars and radio galaxies
\citep[e.g.][]{Maxfield2002,Humphrey2007,Roche2014,Morais2017}.  However, no extended
powerful radio-lobes are seen in the VLA 10cm radio-continuum images and with
$S_\mathrm{10cm} \approx 7 \,\mu$Jy the nearby compact radio-continuum source is
$\sim 10^3 \times$ weaker compared to typical radio-loud quasars \citep{Ao2017}.
Moreover, \cite{Ao2017} remark a slight radio-excess compared to the expectation from the
sub-mm determined star-formation, but they conclude that star-formation is a significant
contributor in heating the gas.  These observations appear compatible with the alternative
hypotheses from \cite{Bower2004} that the cavity has been blown by a star-formation driven
wind.  The emergence of \ion{He}{ii} emission in the vicinity of the bubble may be linked
to shock heated gas within the compressed shell of the bubble, although photo-ionisation
by a potentially obscured radio-quiet AGN can not be ruled out (see
Sect.~\ref{sec:ionh-emiss-from} below).  Nevertheless, the Ly$\alpha$ profile in this
region also shows a bump on the blue side, and such a profile appears consistent with
theoretically expected Ly$\alpha$ radiative transfer modulations within shock fronts
\citep{Chung2016}.

Lastly, we note that at the position of C11 the Ly$\alpha$ profile is single-peaked
($b>3$), but shows significant tails towards both the blue and red ($\kappa > 3$).
Inspecting the profile from C11 we find that these wings appear to have substructure in
the form of two additional peaks (see also panel 6 in Figure~\ref{fig:exspec}).
This triple-peaked structure is reminiscent of the Ly$\alpha$ profile from a multiply
lensed-galaxy by \cite{Rivera-Thorsen2017}, although not as extreme.
\cite{Rivera-Thorsen2017} showed, that triple peaked Ly$\alpha$ profiles are indicative of
low-covering fractions of neutral gas along the sight-line \citep[see
also][]{Behrens2014a}.  Thus, such galaxies potentially leak ionising radiation along the
sight-line.  Indeed, recently \cite{Rivera-Thorsen2019} confirmed the ionising photon
leakage from their system.  Thus, although C11 is located in an extremely dense
environment, at least in its vicinity most of the gas appears highly ionised, thereby
possibly promoting the escape of ionising radiation.

\subsection{\ion{He}{ii} emission from the LAB: Driven by feedback, cooling flows, or hard
  ionising radiation?}
\label{sec:ionh-emiss-from}

\begin{figure}
  \centering
  \includegraphics[width=0.49\textwidth]{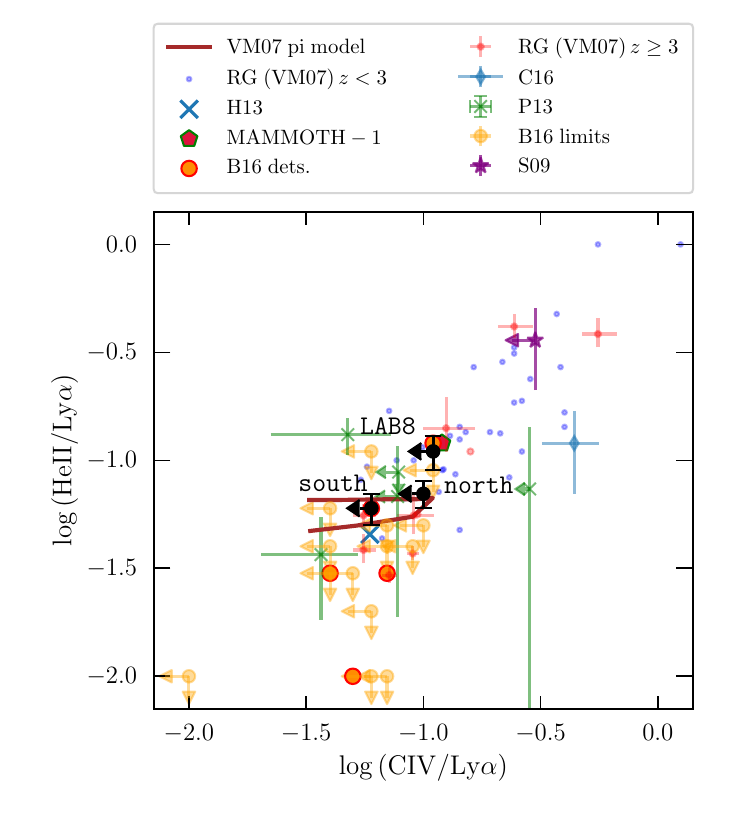}  \vspace{-2.5em}
  \caption{Measured \ion{He}{ii}/Ly$\alpha$ ratios vs. upper limits on
      \ion{C}{iv}/Ly$\alpha$ ratios for the three regions where we detect \ion{He}{ii}
      emission in the LAB (individually labelled black circles and error bars) in
      comparison to ratios or upper limits seen in other high-$z$ sources with extended
      Ly$\alpha$ nebulae.  The comparison includes the following objects from the
      literature: 61 high-$z$ radio galaxies (labelled RG in the legend, with different
      symbols for the $z<3$ and $z\geq 3$ sources) from the compilation by
      \cite{Villar-Martin2007b}; 17 radio-quiet quasars and two radio-loud $z\sim3$
      quasars (labelled B16 -- mostly upper limits in \ion{He}{ii} and \ion{C}{iv} with
      only four sources with \ion{C}{iv} detections, and with the two highest
      \ion{C}{iv}/Ly$\alpha$ ratios coming from the radio-loud quasars) by
      \cite{Borisova2016}; five extended $2 \lesssim z \lesssim 3$ Ly$\alpha$ nebulae
      found in a blind broad-band search (labelled P13) from \cite{Prescott2013}; the
      extreme Ly$\alpha$ nebulae around a $z=2.3$ quasar MAMMOTH-1 from \cite{Cai2017}; a
      $z=2.3$ LAB from \cite{Scarlata2009b} that is detected only in \ion{He}{ii}
      (labelled S09); a faint ($\approx 10^{42}$erg\,s$^{-1}$) extended nebulae found
      behind a gravitational lens (labelled C16) by \cite{Caminha2016}; and an extended
      line emitting nebulae around a $z=2.5$ radio-loud quasar (labelled H13) from
      \cite{Humphrey2013}.  We also show a track of a photo-ionisation model with varying
      ionisation parameter by \cite{Villar-Martin2007b} for a power-law spectral energy
      distribution with $n_\mathrm{H} = 100$\,cm$^{-3}$ and solar metallicity (labelled
      VM07 pi model -- see text for details).  }
  \label{fig:compar}
\end{figure}

\begin{figure}
  \centering
  \includegraphics[width=0.5\textwidth]{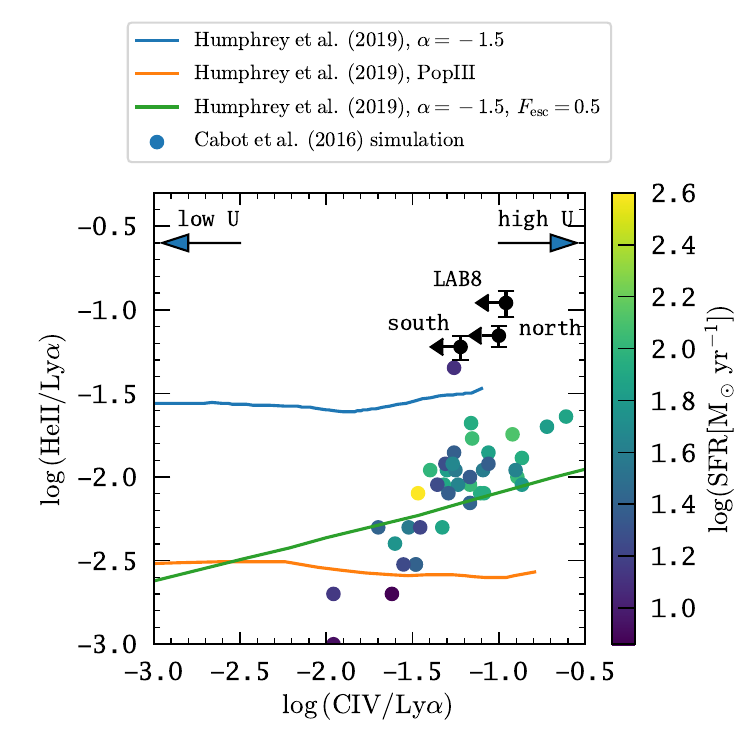}
  \vspace{-2em}
  \caption{Measured \ion{He}{ii}/Ly$\alpha$ ratios vs. upper limits on
    \ion{C}{iv}/Ly$\alpha$ ratios for the three \ion{He}{ii} emitting regions in LAB\,1
    compared to hydrodynamic simulations from \cite{Cabot2016} and tracks of
    photo-ionisation models by \cite{Humphrey2019}.  The individual simulated LABs
      from \cite{Cabot2016} are colour-coded according to the star-formation rate of the
      galaxies within the LAB hosting halos.  The \cite{Humphrey2019}
      photo-ionisation models are computed  for low metallicity
    ($\mathrm{Z}=0.01\mathrm{Z}_\odot$) low-density gas in the circum-galactic medium of
    AGN host-galaxies (blue and green curve -- see text) or Pop-III star-forming galaxies
    (orange curve). For the photo-ionisation models the ionisation parameter
      varies from $U=0.25$ for the highest \ion{C}{iv}/Ly$\alpha$ ratios to
    $U\approx-2.5$ where \ion{C}{iv}/Ly$\alpha = -3$.  }
  \label{fig:chheii}
\end{figure}

\begin{figure*}
  \centering
  \includegraphics[width=0.49\textwidth]{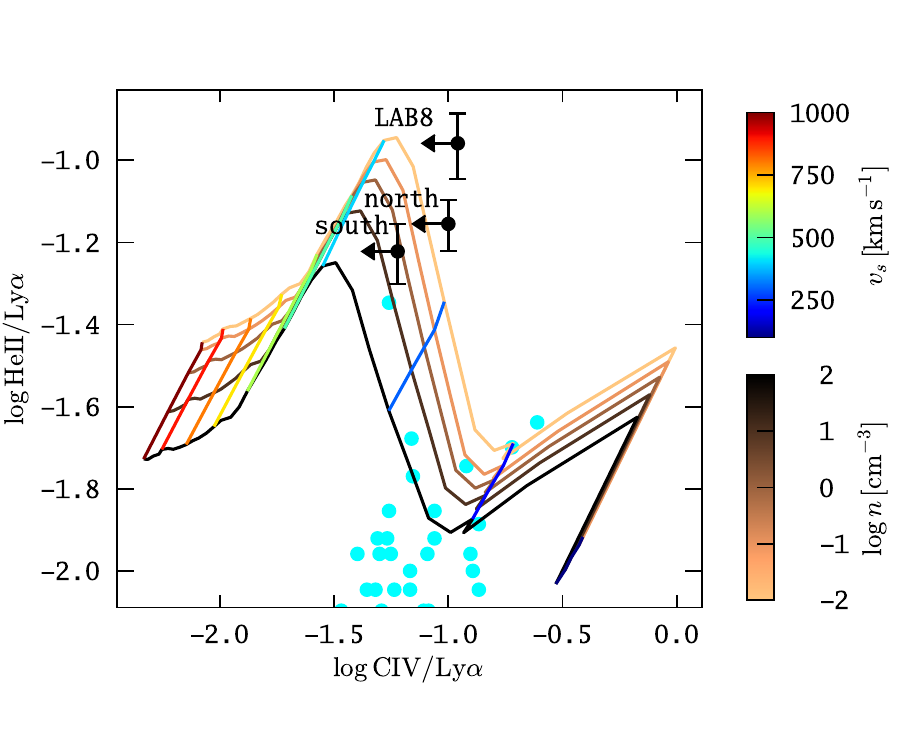}
  \includegraphics[width=0.49\textwidth]{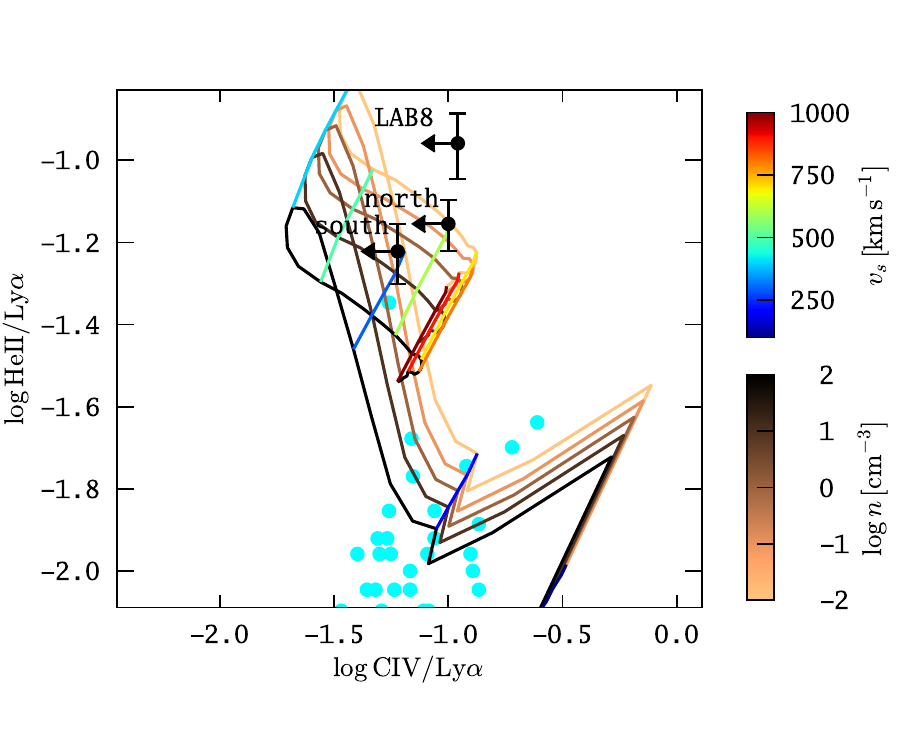}
  \vspace{-2em}
  \caption{Measured \ion{He}{ii}/Ly$\alpha$ ratios vs. upper limits on
    \ion{C}{iv}/Ly$\alpha$ ratios for the three regions where we detect \ion{He}{ii}
    emission in the LAB compared to shock (\emph{left panel}) and shock+precursor
    (\emph{right panel}) models from \cite{Allen2008}.  We show all models for a constant
    magnetic parameter $B/n^{1/2} = 3.23\, \mu$G/cm$^{3/2}$. The grids are colour coded
    according to the density $n_\mathrm{H}$ of the pre-shocked region and the
    shock velocity $v_s$. We also show the predicted line-ratios from ab-initio
    hydro-dynamical simulations of $z=3$ LABs by \cite{Cabot2016}. }
  \label{fig:shockm}
\end{figure*}

In Sect.~\ref{sec:detect-ionh-lambd} we reported on the detection of three distinct
extended \ion{He}{ii} emitting zones with
$\mathrm{SB}_\mathrm{HeII} \sim 6 - 7 \times 10^{-19}$erg\,s$^{-1}$cm$^{-2}$arcsec$^{-2}$.
Extended \ion{He}{II} emission is commonly observed within the extended Ly$\alpha$
emitting regions surrounding high-$z$ radio galaxies
\citep[e.g.,][]{Maxfield2002,Villar-Martin2007b,Villar-Martin2007,Morais2017,Marques-Chaves2019}.
Broad line profiles and alignments with the radio-jet axes indicate that \ion{He}{ii} is,
at least partly, powered by jet-gas interactions in those objects.  Detections of
\ion{He}{ii} emitting zones within the extended Ly$\alpha$ nebulae surrounding radio quiet
quasars or in LABs not directly associated with an AGN appear less frequent and
more difficult to interpret
\citep{Scarlata2009,Prescott2009,Cai2017,Cantalupo2019,Humphrey2019}. Different
explanations have been put forward, e.g. with \cite{Scarlata2009} favouring
cooling-radiation in their system, \cite{Prescott2009} arguing for photo-ionisation by an
AGN or an extremely metal poor stellar population in their object, while \cite{Cai2017}
reason that their extended \ion{He}{ii} zone is powered by shocks.  We now try to
constrain the physical mechanisms responsible for the extended \ion{He}{ii} emission in
the blob.  To this aim we will also consider the upper limits on \ion{C}{IV} emission from
the \ion{He}{ii} emitting zones (Sect.~\ref{sec:ionciv}).

To place our result in context with previous observations (or upper limits) of
\ion{He}{ii} and \ion{C}{iv} emission from high-$z$ sources with extended Ly$\alpha$
emission we compare in Figure~\ref{fig:compar} our measured \ion{He}{ii}/Ly$\alpha$ ratios
and our derived upper limits on \ion{C}{iv}/Ly$\alpha$ to the literature.  The majority of
comparison objects in Figure~\ref{fig:compar} harbour a bright active galactic nucleus.
We observe that that the ratios and upper limits in LAB\,1 appear not unusual with respect
to other extended Ly$\alpha$ objects, i.e, some objects exhibit significantly higher
ratios both in \ion{He}{ii}/Ly$\alpha$ and \ion{C}{iv}/Ly$\alpha$, but also significantly
lower ratios or upper-limits have been reported.  The diversity in ratios observed for
nebulae with known powering source imply that varying physical conditions
(e.g. metallicity, temperature, and density) within the emitting gas significantly
influences the observed ratios.  Additionally, radiative transfer effects in both
Ly$\alpha$ and \ion{C}{iv} \citep[][their Sect. 2.3]{Berg2019} can also bias conclusions
drawn from this diagram \citep[e.g.][]{Caminha2016}.  This diagram has now been frequently
used in the literature for attempting to understand the physical conditions of the ionised
gas.

We show in Figure~\ref{fig:compar} the track of a photo-ionisation model from
\cite{Villar-Martin2007b} that appears to be in good agreement with the
\ion{He}{ii}/Ly$\alpha$ ratios and \ion{C}{iv}/Ly$\alpha$ ratios from our observations.
This track was computed for gas of solar metallicity at a density of
$n_\mathrm{H} = 100$\,cm$^{-3}$ as a function of varying (dimensionless) ionisation
parameter
\begin{equation}
  \label{eq:5}
  U = \dot{Q}/(4 \pi c n_\mathrm{H} r^2)   \; \text{,}
\end{equation}
where $\dot{Q}$ denotes the rate of hydrogen ionising photons produced by the source and
$r$ denotes the distance of the ionised gas cloud to the source.  As spectral energy
distribution of the ionising source \cite{Villar-Martin2007b} assumed an idealised quasar
described by a $F_\nu \propto \nu^{\alpha}$ power-law continuum with slope $\alpha=-1.5$.
Along the displayed track $\log U$ varies from -2.3 (start of the upper branch) to 0 (end
of the lower branch); the inflection point in \ion{C}{iv}/Ly$\alpha$ is at $\log U = -1$
\citep[the model is adopted from Figure 1 of][]{Villar-Martin2007b}.

While the predicted line-ratios show some agreement with our observation, for region
``north'' the assumptions in this model require extremely high photo-ionisation rates:
Region ``north'' is $\sim$ 50 kpc in projection towards LAB-ALMA1, LAB-ALMA2, and
  VLA-LAB1a. Thus, according to Eq.~(\ref{eq:5}), even the smallest $\log U = -2.3$ in
the \cite{Villar-Martin2007b} model corresponds $\dot{Q} \sim 10^{57}$s$^{-1}$ if we
assume a galaxy at the position of LAB-ALMA1, LAB-ALMA2, or VLA-LAB1a as the
ionising source.  This value would be similar to the photo-ionisation rate inferred for
the luminous quasar UM\,287 that powers the ``Slug'' nebulae \citep{Cantalupo2014} and two
orders of magnitude higher than the $\dot{Q} \approx 10^{55}$s$^{-1}$ that would be
required to power the whole blob via photo-ionisation (under case-B assumptions).  While
the photo-ionisation model might be more fitting for the zones ``south'' and
``LAB\,8'', as they are in close vicinity to known galaxies, existing deep
multi-wavelength data does not find strong evidence in favour of such a luminous AGN
\citep{Ao2017}.  Thus, we also explore different mechanism that have been put forward to
explain \ion{He}{ii} in LABs.

The co-spatial occurrence of \ion{He}{ii} emitting gas with embedded galaxies in the blob
is qualitatively consistent with hydrodynamic simulations from \cite{Cabot2016} that do
not contain photo-ionisation by an AGN.  The \cite{Cabot2016} models predict that the main
source of \ion{He}{ii} emission in LABs should be shock-heated gas from supernovae driven
winds. Included in the simulations are both Ly$\alpha$ radiative transfer and
dust-extinction effects, as well as photo-ionisation from a star-forming population at
star-formation rates compatible with typical LABs
($\log \mathrm{SFR} [\mathrm{M_\odot \, yr}^{-1}] \sim 2$).  In Figure~\ref{fig:chheii} we
compare our measured \ion{He}{ii}/Ly$\alpha$ ratios and \ion{C}{iv}/Ly$\alpha$ upper
limits with the results from the 48 simulated LABs by \cite{Cabot2016}.  As can be seen
our measured \ion{He}{ii}/Ly$\alpha$ ratios are significantly above the ratios predicted
by the models, while our upper limits on \ion{C}{iv}/Ly$\alpha$ are compatible with the
models.  The discrepancy between models and data may indicate that, contrary to the
assumption simulations, shock heating is not the dominant mechanism in producing the
observed \ion{He}{ii}/Ly$\alpha$ ratios.  Alternatively, the physical properties of the 46
simulated halos by \cite{Cabot2016} might be significantly different from that of the halo
hosting LAB\,1.

We find a similar discrepancy between the \ion{He}{ii}/Ly$\alpha$ ratios in LAB\,1 and
recent AGN photo-ionisation models by \cite{Humphrey2019}.  As shown in
Figure~\ref{fig:chheii}, these models also predict significantly lower
\ion{He}{ii}/Ly$\alpha$ than what is observed here.  The \cite{Humphrey2019}
photo-ionisation tracks where computed for different ionising sources illuminating
low-metallicity ($Z=0.01Z_\odot$) gas.  Similar to the calculations by
\cite{Villar-Martin2007b} a gas density of $n_\mathrm{H} = 100$\,cm$^{-3}$ was
adopted. The aim of these computations was to find an explanation of the extreme ratios
and/or upper limits found for some nebulae around AGN or radio-galaxies that appeared
incompatible with previous photo-ionisation attempts (see Figure~\ref{fig:compar}).  We
show in Figure~\ref{fig:chheii} the resulting predictions for photo-ionisation by an
idealised quasar (described by a $F_\nu \propto \nu^{\alpha}$ power-law continuum spectral
energy distribution with $\alpha=-1.5$, blue curve), whose ionising continuum also has
been modified by intervening cold-gas that absorbs 50\% of the ionising photons (green
curve) and a fiducial zero-metallicity stellar population (modelled by a $8\times10^4$\,K
black body) for varying ionisation parameters.  In Figure~\ref{fig:chheii} the tracks are
shown from $U=0.25$ for the highest \ion{C}{iv}/Ly$\alpha$ ratios to $U\approx-2.5$ where
\ion{C}{iv}/Ly$\alpha = -3$.  Within the displayed range the \ion{He}{ii}/Ly$\alpha$ ratio
exhibits a weak dependence on $U$, while the \ion{C}{iv}/Ly$\alpha$ ratio rapidly
decreases with decreasing $U$.  The reason for the almost constant \ion{He}{ii}/Ly$\alpha$
ratio is that in the simulated low-metallicity gas the fraction of He$^{++}$ ions remains
saturated for an $U\gtrsim -2.5$ photo-ionisation field.  On the other hand, the large
decrease in \ion{C}{IV}/Ly$\alpha$ for decreasing $U$ is driven by both the reduction in
electron temperature (resulting in lower collision rates for C$^{3+}$) and the decreasing
fraction of $C^{3+}$ ions for smaller values of $U$.  We remark also that for $U < -2.5$
(outside of the displayed range in Fig.~\ref{fig:chheii}, see \citealt{Humphrey2019}) the
\ion{He}{ii}/Ly$\alpha$ ratios will also start to decrease significantly.  As explained
above, for $n_\mathrm{H} = 100$\,cm$^3$ gas in region ``north'' only $\log U \lesssim -4$
would be compatible with a scenario where the whole blob is powered by photo-ionisation,
but also at these extreme ionisation parameters the \cite{Humphrey2019} model are
incompatible with our observations.  Thus, photo-ionisation of low-metallicity gas by an
AGN or Pop-III stellar population under the assumptions made in the \cite{Humphrey2019}
models appears not to be a valid scenario to explain the \ion{He}{ii} emitting zones in
LAB\,1.

It is also apparent in Figure~\ref{fig:chheii} that the photo-ionisation tracks overlap
with the predictions from the ab-initio simulations by \cite{Cabot2016} in which
shock-heated emission dominates.  A similar degeneracy was commented upon by
\cite{Arrigoni-Battaia2015}, who computed a different set of AGN photo-ionisation models.
Similar to \cite{Villar-Martin2007b}, these authors modelled the expected line ratios in
solar-metallicity optically-thick and optically-thin gas.  Their optically-thin models are
compatible with our \ion{He}{ii}/Ly$\alpha$ ratios at very low ionisation parameters
($\log U \lesssim -2.5$), where \ion{C}{iv}/Ly$\alpha \lesssim 10^{-2}$ \citep[Figure 12
in][]{Arrigoni-Battaia2015}, i.e. $\approx 10$ times below our upper limit for this ratio.
As discussed above, such low values of $U$ are not unrealistic for the \ion{He}{ii}
emitting zones within the LAB.  On the other hand, their optically thick models
($N_\mathrm{H}$) would require relatively high ionisation parameters to reproduce our
observed \ion{He}{ii}/Ly$\alpha$ ratios.  For those models then, the predicted
\ion{C}{iv}/Ly$\alpha$ values would be slightly below our upper limits.  Unfortunately,
this region of the \ion{He}{ii}/Ly$\alpha$ vs.  \ion{C}{iv}/Ly$\alpha$ parameter space is
also covered by shock models.  This is illustrated by Figure~\ref{fig:shockm}, where we
compare our measurements and upper limits to the predictions from shock models at solar
metallicity by \cite{Allen2008}.  For those models we use the same parameters motivated by
\cite{Arrigoni-Battaia2015}, and we refer to Sect.~5.2 of their publication for an
in-depth discussion of the motivation and description of the models.

As evident by Figure~\ref{fig:shockm}, both shock and shock+precursor models can reproduce
the observed \ion{He}{ii}/Ly$\alpha$ ratios for \ion{C}{IV}/Ly$\alpha$ ratios that would
be consistent with our upper limits for shock velocities of
$v_\mathrm{s}\sim 250 - 500$\,km\,s$^{-1}$.  While for region ``south'' the observed
line-width in \ion{He}{ii} appears compatible with those velocities, the other two regions
show significantly narrower \ion{He}{ii} emission.  However, especially the emission
stemming from the pre-cursor arises in kinematically more quiescent gas which is
photo-ionised by the hard-UV radiation from the shock.  The measured line-widths might thus
indicate that we observe \ion{He}{ii} emission from the pre-shock phase.

A third interpretation for our \ion{He}{ii} detection is provided in the models from
\cite{Yang2006}.  These models suggest that extended zones of narrow
($\lesssim 250$\,km\,s$^{-1}$) \ion{He}{ii} emission within Ly$\alpha$ blobs are a
signature of gravitational cooling radiation.  We here observe \ion{He}{ii} in close
vicinity to star-forming galaxies (regions ``south'' and ``LAB\,8'') and next to the
expanding bubble within the blob (region ``north'').  The \cite{Yang2006} models
  also predict the highest \ion{He}{ii} and Ly$\alpha$ surface brightness in close
vicinity of the star-forming galaxies within the blobs.  Considering only Ly$\alpha$
emission, a similar conclusion was recently reached by \cite{Trebitsch2016}.  Hence, our
observations do not rule out contributions from cooling radiation within two of our three
detected \ion{He}{ii} patches.  But again, region ``north'' is an exception as it not
close to a known galaxy.

To summarise, the detection of faint extended \ion{He}{II} emission is consistent with all
mechanisms that are suggested to power the Ly$\alpha$ blob -- i.e. cooling radiation,
feedback driven shocks, and/or photo-ionisation from an embedded active galactic nuclei.
For the latter two scenarios we compared \ion{He}{ii}/Ly$\alpha$ ratios and upper limits
on \ion{C}{iv}/Ly$\alpha$ with observed ratios from the literature and
models. \ion{He}{ii}/Ly$\alpha$ and \ion{C}{iv}/Ly$\alpha$ ratios comparable to our
measurements have been found in extended Ly$\alpha$ halos around high-$z$ radio galaxies
and AGNs, and these ratios can be explained by some photo-ionisation models.  However, our
ratios are not as extreme as the ratios that would be expected from quasar
photo-ionisation of extremely metal deficient gas at very low ionisation parameters
\citep{Humphrey2019}.

A potential caveat when comparing our measured line ratios with photo-ionisation models is
that measurements were obtained within apertures that cover a projected area of
$\sim 10^3$kpc$^2$.  We expect realistically that gas within those regions exhibits a
range of different densities and temperatures.  However, the observed
\ion{He}{ii}/Ly$\alpha$ ratios from photo-ionised gas from an embedded quasar will depend
sensitively on the density distribution of the line emitting gas \citep{Cantalupo2019}.
In fact, for any realistic density distribution the aperture integrated
\ion{He}{ii}/Ly$\alpha$ ratio will always be lower than the intrinsic ratios within the
volume covered by the aperture.  Thus, the quantitative comparison between our measured
line ratios and upper limits to photo-ionisation models must be treated with caution, as
these models always assumed constant densities.

\section{Summary and conclusions}
\label{sec:summary-conclusions}

We presented an analysis of 17.2\,h MUSE observations of the prototypical SSA22a LAB\,1 at
$z=3.1$.  Our analysis revealed many previously unknown features of this enigmatic
high-$z$ object and provided detailed look into the early formation stage of a massive
galaxy within an extremely dense environment.  The main results of our study are the
following:
\begin{enumerate}
\item Our IFS data reaches a limiting depth of
  $\approx 6 \times 10^{-19}$erg\,s$^{-1}$cm$^{-2}$arcsec$^{-2}$ in Ly$\alpha$.  This is a
  factor ten deeper than previous IFS observations of LAB\,1 by \cite{Weijmans2010} that
  revealed Ly$\alpha$ down to $5.6 \times 10^{-18}$erg\,s$^{-1}$cm$^{-2}$arcsec$^{-2}$. At
  this unprecedented depth we uncover several hitherto unknown features of the blob.  Most
  prominently we find a filamentary bridge connecting LAB\,1 with its northern neighbour
  LAB\,8 and a shell-like arc towards the SW of LAB\,1.  Potentially another filament is
  emanating to the SE in the direction of two newly identified faint LAEs.  These
  filamentary features are visible both in the sequence of 2.5\,\AA{} narrow-band slices
  (Figure~\ref{fig:channel_maps}) as well as in the adaptive Ly$\alpha$ image
  (Figure~\ref{fig:adapt_nb}).  The newly uncovered structures are morphologically
  reminiscent of cold streams that are predicted to funnel cool gas into the potential
  wells of massive halos. As cooling flows are expected to align with the underlying
  filamentary large scale structure, the presence of multiple filaments pointing into
  various directions may imply that LAB\,1 harbours a node in the cosmic web where a
  complex interwoven network of filaments arrives from various directions.
\item We find a ring-like structure slightly west of the photometric centre that is
  deprived in Ly$\alpha$ emission.  While previous observations hinted already at the
  existence of this feature \citep{Bower2004,Weijmans2010}, the improved spatial
  resolution and sensitivity of our observations allow to slice through this shell in
  velocity space (Figure~\ref{fig:channel_maps}).  The detection of \ion{He}{ii} emission
  at a \ion{He}{ii}/Ly$\alpha$ ratio consistent with expectations for shock heated gas in
  regions directly demarcating the ``bubble'' appears supportive of the hypothesis that we
  are observing a genuine cavity filled by hot ionised gas.  However, the line ratio is
  also consistent with photo-ionisation models that assume an AGN as ionising source, but
  the required photo-ionisation rates for those models appear too high when an AGN is
  assumed at the position of the central most plausible AGN sources
  (Sect.~\ref{sec:ionh-emiss-from}).  Moreover, as seen in our higher-order moment maps
  (Figure~\ref{fig:hm}), the Ly$\alpha$ profile at this position is markedly different
  than in the surrounding regions -- it is strongly skewed with a pronounced red tail, but
  not bi-modal.  Such a profile is reminiscent of the Ly$\alpha$ radiative transfer
  simulations for expanding shells \citep{Verhamme2006,Gronke2016a}, and furthermore the
  faint blue-wing seen in the profile (Figure~\ref{fig:exspec}, panel 11) appears
  consistent with predictions for the Ly$\alpha$ profiles in shock-fronts
  \citep{Chung2016}.  These results may indicate that pressure exerted by the hot ionised
  gas drives an expanding bubble (Sect.~\ref{sec:interpr-high-order}).  Potential
  driving mechanisms for this expansion are a star-formation driven wind or radio emission
  from an obscured or faded AGN associated to the central sub-mm/radio sources.
\item The Ly$\alpha$ profiles observed from the blob exhibit varying degrees of complexity
  (Figure~\ref{fig:exspec}).  We tried to map the varying complexity of the blob by
  utilising a moment-based non-parametric analysis of the profiles
  (Sect.~\ref{sec:spectr-morph-lyalpha}).  This analysis provided us with maps of the line
  of sight velocity (central moment), the width of the line, its skewness, kurtosis, and
  bi-modality (Figure~\ref{fig:moms} and Figure~\ref{fig:hm}).  Especially within the
  central regions of the blob we find broad double- or even triple peaked profiles.
  Unfortunately, however, for these complex broad profiles the used bi-modality measure
  from \cite{Remolina-Guti2019} appeared to be not very precise.  However, it spatially
  pinpoints three compact regions that are characterised single peaked profiles that are
  sometimes skewed towards the blue or the red.  Two of these regions are associated with
  the known Lyman break galaxies SSA22a-C11 and SSA22a-C15, and the third is associated
  with the edge of the expanding bubble.  Interestingly, these single peaked features
  appear in the vicinity of the \ion{He}{ii} emitting zones.  The spatially varying
  complexity of the profiles likely encodes a mix of different Ly$\alpha$ photon
  production mechanisms: in-situ production from ionising sources and gravitational
  cooling, or both simultaneously.  Moreover, projection effects from the line-of-sight
  passing through multiple filaments may also be responsible for the appearance of
  multiple peaks. Single peaked profiles might indicate a small residual fraction of
  neutral gas at line centre, and therefore pinpoint a high degree of ionisation.
  Quantitatively disentangling the production mechanisms and projection effects based on
  the line profiles appears notoriously difficult.  An analysis of realistic Ly$\alpha$
  blob simulations in a cosmological context, as e.g. very recently presented by
  \cite{Kimock2020}, could reveal whether the here presented moment based moment has
  diagnostic power in this respect.
\item We find the highest degrees of Ly$\alpha$ polarisation in regions that exhibit high
  velocity shifts and narrow line profiles (Sect.~\ref{sec:comb-analys-lyalpha} and
  Figure~\ref{fig:polfig}).  These regions are far from known embedded galaxies.  It
  appears that this result is consistent with theoretical expectations for Ly$\alpha$
  scattering from a central source \citep{Eide2018}, although numerical simulations for
  Ly$\alpha$ spectro-polarimetry in complex 3D environments have yet to be
  performed.
\item The line-of-sight velocity field of the blob is characterised by a large-scale
  velocity gradient that is oriented perpendicular to the morphological major axis of the
  blob (Figure~\ref{fig:moms}, left panel).  The observed shearing amplitude is
  $v_\mathrm{shear} \approx 1300$\,km\,s$^{-1}$.  The orientation of the velocity gradient
  implies a parallel alignment between angular momentum vector and major axis. This
  parallel alignment between major axis and angular momentum appears at odds with the
  theoretically expected average for massive halos (Sect.~\ref{sec:large-scale-gas}).  We
  argue that this peculiar alignment reflects the complexity of the dense environment in
  which the blob resides. This leads us to speculate that LAB\,1 is formed at the node of
  multiple intersecting filaments of the cosmic web.  The kinematic interpretation of the
  large-scale velocity field relies on the assumption that the first-moment map from
  Ly$\alpha$ is a good tracer of the large-scale kinematics
  (Sect.~\ref{sec:large-scale-gas}), an assumption that has of yet not been tested against
  radiative-transfer simulations.
\item We detect extended \ion{He}{ii} $\lambda$1640 emission at three disjunct regions in
  the blob (Sect.~\ref{sec:detect-ionh-lambd} and Figure~\ref{fig:heii_maxsn}). The
  \ion{He}{ii} emission from those regions shows a surface brightness of
  $5 - 7\times 10^{-19}$erg\,s$^{-2}$cm$^{-2}$arcsec$^{-2}$.  Two of those regions
  surround known embedded galaxies, and the third region demarcates the expanding bubble.
  Our observations do not reveal \ion{C}{iv} $\lambda$1549 emission from those regions
  (Sect.~\ref{sec:ionciv}).  A comparison between predicted \ion{He}{ii}/Ly$\alpha$ and
  \ion{C}{iv}/Ly$\alpha$ ratios from shock- and photo-ionisation models from the
  literature with our measurements and upper limits is consistent with both
  photo-ionisation from an AGN \citep{Villar-Martin2007b} and fast radiative shocks
  \citep{Allen2008}.  However, the observed ratios were found to be incompatible with
  recent AGN photo-ionisation models by \cite{Humphrey2019}, that assume low-gas
  metallicity, an ionising source partly covered by optically thick neutral gas, or very
  low ionisation parameters.  Moreover, our observed \ion{He}{ii}/Ly$\alpha$ ratios are
  also significantly above the predicted ratios from the hydrodynamic simulations of
  \cite{Cabot2016} that predict feedback driven shocks and gas accretion shocks as main
  source for the observed \ion{He}{ii} emission in LABs.
\end{enumerate}
Our new observations provide the most detailed view of a Ly$\alpha$ blob to date.  Given
the numerous galaxies within the blob and the surrounding dense proto-cluster environment
it appears natural to suspect LAB\,1 as the progenitor of a massive cluster elliptical.
The detection of \ion{He}{ii} emission may hint at the importance of feedback effects in
the early evolutionary stage of such systems, but AGN powering can not be ruled out.
Moreover, we find no clear evidence for gravitational cooling, but we can also not rule
this mechanism out.  We suggest that simulations of Ly$\alpha$ blobs in cosmological
environments should be compared to the non-parametric line profile analysis presented
here.  Moreover, rest-frame optical emission line studies, e.g. with the \emph{James Webb}
Space Telescope, will help to better constrain the powering mechanisms.  Finally, our
study also highlights the potential importance of environmental effects on the kinematic
properties of massive halos that lie at the intersection of multiple filaments of the
cosmic web.  As our study focused on a single object we can not establish any empirical
trends in this respect, hence future studies of LAB samples at sufficient depth in
proto-cluster environments are desirable.

\begin{acknowledgements}
  We thank the anonymous referee for a careful report that contained many suggestions that
  helped to improve this paper.  E.C.H. thanks the extragalactic group at Stockholm
  University for a wonderful time and many insightful discussions during this project.
  E.C.H. also thanks Aaron Smith and Peter Laursen for useful comments during early stages
  of this project. This research made extensive use of the \texttt{astropy} package
  \citep{Astropy-Collaboration2018}.  All figures in this paper were created using
  \texttt{matplotlib} \citep{Hunter2007}.
\end{acknowledgements}

\bibliographystyle{aa}
\bibliography{lab1_paper_bibliography.bib}

\end{document}